\documentclass[12pt]{iopartb}
\usepackage{iopams}
\usepackage{amsmath}
\usepackage{amsfonts}
\usepackage{latexsym}
\usepackage{euscript}
\usepackage[dvips]{graphicx}
\usepackage{lscape}
\begin{document}
\begin{flushright}
DAMTP-2006-11\\
October 2006\\
\end{flushright}
\title{Topological Defects from First Order Gauge Theory Phase
Transitions}
\author{M Donaire}

\address{\it DAMTP, CMS, University of Cambridge\\Wilberforce Road, Cambridge CB3 OWA, United
Kingdom} \ead{M.A.Donaire@damtp.cam.ac.uk}
\begin{abstract}\\
We investigate the mechanism by which topological defects form in
first order phase transitions with a charged order parameter. We
show how thick superconductor vortices and heavy cosmic strings
form by trapping of magnetic flux. In an external magnetic field,
intermediate objects such as strips and membranes of magnetic flux
and chains of single winding defects are produced. At non-zero
temperature, a variety of spontaneous defects of different winding
numbers arise. In cosmology, our results mean that the magnetic
flux thermal fluctuations get trapped in a primordial
multi-tension string network. The mechanism may also apply to the
production of
 cosmic-like strings in brane collisions. In a thin type-I superconductor film, flux strips are found to be
meta-stable while thick vortices are stable up to some critical
value of the winding number which increases with the thickness of
the film. In addition, a non-dissipative Josephson-like current is
obtained across the strips of quantized magnetic flux.
\end{abstract}
\section{Introduction}
The Kibble mechanism \cite{Kibble76} has proved successful in
explaining the formation of topological defects in phase
transitions of field theories which experience spontaneous
breakdown of a global symmetry. It applies to both first and
second order phase transitions. The vacuum (minimum energy
configuration) is degenerate and the system can fall into any
vacuum state in different spatial locations. In first order phase
transitions, where bubbles nucleate spontaneously, it has been
tested theoretically in numerical simulations and analytical
calculations \cite{Srivas92,Srivas92',Ferrera99}. Experimentally
it has been verified in nematic liquid crystals through the
spatial distribution of disclinations
\cite{Chuang91,Bowick94,Digal99}.\\ \indent In second order phase
transitions Kibble's argument was completed by Zurek
\cite{Zurek85} on what is known as Kibble-Zurek mechanism (KZ). It
applies to the generation of topological defects as the system in
question cools down in a finite time. When the transition takes
place, non-equilibrium phenomena make the coherence length finite.
This allows the system to choose different vacuum states in
different locations. The mechanism has been tested experimentally
in the prediction of the vorticity pattern of superfluid $^{3}$He
\cite{Bauerle96,Ruutu96}.\\ \indent In cosmology, the analogues of
disclination lines and superfluid vortices are global cosmic
strings. They may have been generated in the early stages of the
universe \cite{Kibble80,Vilenkin94,Arttu03}. If they exist, Kibble
mechanism is expected to explain their origin.\\ \indent The order
parameters which describe the phases of liquid crystals, $^{3}$He
and field theories with global cosmic strings are neutral. In the
contrary, the order parameter which describes the symmetry of a
gauge theory carries a charge and is coupled to a gauge field. KZ
mechanism is based on the dynamics of a neutral order parameter,
and therefore it is insufficient for describing generally either
the formation of local cosmic strings in cosmological phase
transitions of gauge theories or the origin of vortex lines in
superconductors \cite{Hindmarsh2000}.\\ \indent It is easy to see
this in a simple condensed matter example. When a type-II
superconductor is slowly cooled from the normal to the
superconducting phase in an external magnetic field, an Abrikosov
lattice consisting of vortices of the same sign is formed. This
cannot be explained by Kibble-Zurek mechanism because it predicts
the same number of vortices and anti-vortices. In the same way,
Kibble mechanism is not sufficient to explain the formation of
intermediate structures such as strips of magnetic flux in type-I
superconductors under an external magnetic field
\cite{Schawlow56,Alers57}.\\ \indent For cosmology and particle
physics prospects, gauge theories are more relevant than global
theories. Grand Unified Theories (GUTs) are gauge theories which
predict generically the spontaneous formation of local cosmic
strings, monopoles and other topological defects. In particular,
cosmic strings are tubes of quantized magnetic flux. The value of
the tension of the strings increases with their flux. The
phenomenological effects of these strings are mainly gravitational
and depend quantitatively on their tension. Different cosmological
observations generally attribute different upper bounds to the
tension of cosmic strings \cite{Davis05}. These are average values
which could actually correspond to a distribution of windings as a
result of the evolution of a primordial multi-tension string
network \cite{Donaire05}. Kibble mechanism, on the contrary, gives
rise to a unique tension because all strings have unit winding.\\
\indent In a more theoretical context, recent findings suggest
that cosmic strings may be a generic prediction of super-string
theory \cite{Copeland04} and could form in brane collision
scenarios \cite{Dvali04}. In some brane-inflation models like, for
instance, the $K^{2}LM^{2}T$ model \cite{Kachru03}, D-strings are
produced as a result of the collision of D-branes. The D-strings
are intended to be cosmic strings and the D-branes contain a net
magnetic flux. The resultant D-string network confines that flux
in an analogous manner that a vortex network does in a
superconductor when an external magnetic field is applied.
Therefore, a different mechanism to KZ must be responsible for the
formation of those strings.\\ \indent As it was shown in
\cite{Hindmarsh2000,Rajantie02}, generally in gauge theories KZ
mechanism is only a good approximation in those cases where the
dynamics of the gauge field is negligible. In the opposite limit,
to which the examples above correspond, KZ mechanism is not enough
and a different explanation is needed for the formation of
topological defects. In second order phase transitions an
explanation in terms of the non-equilibrium dynamics of the gauge
field has already been found
\cite{Hindmarsh2000,Kibble03,Donaire04}. Numerical simulations
show the formation of clusters of equal sign vortices.\\ \indent
On the other hand, the possibility for the production of high
winding topological defects was already suggested in
\cite{Zurek96}. It was first argued in \cite{Kibble95} and then
discussed in more detail in \cite{Rajantie02} that they could form
as a result of trapping of thermal fluctuations in first order
phase transitions. We will investigate in this paper the mechanism
by which thick vortices and heavy cosmic strings indeed form when
the underlying symmetry is the $U(1)$ gauge symmetry and the phase
transition is strong first order. It is in this case the joint
dynamics of the gauge field together with the complex phase of the
order parameter -- not necessarily out of equilibrium -- that
gives rise to trapping of magnetic flux. \\ \indent The reader is
warned that throughout this paper we commonly use the term vortex
to refer to both cosmic strings and vortices regardless of the
number of spatial dimensions we deal with. Only where the
dimensionality is of some relevance we will stress the
distinction. Also, we will refer to the regions in a minimum
energy state either as 'vacuum regions' or as 'superconducting
regions'. Likewise, we will refer to the regions in the symmetric
phase either as 'symmetric phase regions' or as 'normal phase
regions'. \\ \indent This paper is organized as follows. Section
$2$ describes the model and the setup of the numerical
simulations. In section 3 we present the mechanism by which
high-winding topological defects form in gauge theories. Section 4
is devoted to the mechanism of formation of strips of magnetic
flux and chains of single-winding defects. In section 5 we explain
the different properties of the vortex lines in a superconductor
film with respect to the vortices in the fully two-dimensional
theory. In section 6 we argue that spontaneous vortices and cosmic
strings  form in the thermal ensemble. Section 7 describes some
mathematical aspects of the $U(1)$ gauge theory which lead to the
derivation of Josephson-like effects as well as to the
reinterpretation of the so-called geodesic rule.
\section{Preliminaries\label{intro}}
\subsection{The model\label{model}}
Let us consider the Ginzburg-Landau (G-L) free energy for
superconductors near the phase transition point:
\begin{equation}\label{aa1}
F[\vec{A},\phi]=\frac{1}{2\mu_{0}}(\vec{\nabla}\times\vec{A})^{2}+\frac{1}{2m}|(-i\hbar\vec{\nabla}+\frac{2e}{c}\vec{A})\phi|^{2}+V(|\phi|),
\end{equation}
where $m$ is the effective mass of a Cooper pair and $\mu_{0}$ is
the magnetic permeability of free space.\\ \indent Let us also
consider the Abelian Higgs model, which can be seen as an
idealized relativistic extension of the above G-L theory. In
natural units,\footnote{Natural units,
$c=\hbar=\mu_{0}=\epsilon_{0}=k_{B}=1$, are used hereafter.} its
Lagrangian is
\begin{equation}\label{a1}
\mathcal{L}=-\frac{1}{4}F_{\mu\nu}F^{\mu\nu}+D_{\mu}\phi
D^{\mu}\phi^{*} - V(|\phi|).
\end{equation}
Greek indices run from $0$ to the number of space dimensions. We
will use the indices $0$ and $t$ to denote time components and
both Latin indices and $x$, $y$, $z$ for the spatial components.
In the following tensorial formulae only the spatial components
apply to (\ref{aa1}).\\ \indent The charged complex scalar field
$\phi\equiv |\phi|\exp{(i\theta)}$ is generically referred to as
'the order parameter'. It is interpreted as the quantum wave
function of Cooper pairs in the G-L theory and as the Higgs field
in a particle physics context. The vector field $A_{\mu}$ is a
$U(1)$ gauge field intended either as the electromagnetic field of
the usual Maxwell theory in the context of superconductivity or as
a $U(1)$ field appearing in the spectrum of a grand unified theory
(not necessarily the photon field). $F_{\mu\nu}$ is the Maxwell
field strength and $D_{\mu}$ is the covariant derivative operator:
\begin{eqnarray}\label{a2}
F_{\mu\nu} & \equiv &
\partial_{\mu}A_{\nu}-\partial_{\mu}A_{\nu},\\D_{\mu} & \equiv &
\partial_{\mu}+ieA_{\mu}.
\end{eqnarray}
In the G-L theory the value of $e$ is twice the electron charge in
agreement with the interpretation of a Cooper pair as a
two-electron bound state. The Lagrangian (\ref{a1}) and the free
energy (\ref{aa1}) are invariant under the $U(1)$ gauge
transformations
\begin{eqnarray}\label{aa2}
\theta & \rightarrow & \theta + \Lambda,\nonumber\\A_{\mu} &
\rightarrow & A_{\mu} - (1/e)\:\partial_{\mu}\Lambda ,
\end{eqnarray}
where $\Lambda$ is a real scalar function of space and time. \\
\indent In the context of Grand Unified Theories, the model
described by the Lagrangian (\ref{a1}) is meant to be just an
idealized toy model. When implementing radiative corrections,
additional couplings of $A_{\mu}$ and $\phi$ to other fields must
be considered. Also first order time derivatives in the effective
equations of motion for both fields, $\phi$ and $A_{\mu}$, emerge
as considering non-zero momenta corrections of order one. \\
\indent Concerning dynamical phenomena in superconductivity, the
time-dependent Ginzburg-Landau equations (TDGL) were derived in
\cite{Gor'kov59,Gor'kov68,Abrahams66} as effective macroscopic
equations from the microscopic BCS theory. A first order time
derivative of $\phi$ emerges as a result of the coupling of the
electrons in the superconducting phase to the electrons in the
normal phase. The equation for $A_{i}$ also contains a first order
time derivative as a result of the finite conductivity in the
normal phase. This is an ohmic current. Thus, the dynamics of both
fields, $\phi$ and $A_{i}$, is approximately diffusive in a
superconducting film. The TDGL equation for $\phi$ does also
include a 'small' second order time derivative. The equation for
$A_{i}$ must  contain a second order time derivative as well. This
is necessary if the gauge field has to satisfy Maxwell equations
out of a superconducting film \cite{Donaire04}.\\ Therefore, we
can explain with analogous equations of motion the dynamics of
both a superconductor -- described by the time-dependent G-L
theory -- and a particle physics phase transition -- described by
an effective field theory. In the temporal gauge, $A_{0}=0$, those
equations read
\begin{equation}\label{a7}
[\partial_{tt} + \sigma\partial_{t} - D_{i}D^{i}+ \frac{\delta
V(\phi)}{\delta \phi^{2}}]\phi=0,
\end{equation}
\begin{equation}\label{a8}
\partial_{tt}A_{i} + \sigma\partial_{t}A_{i} +
\epsilon_{ijk}\epsilon^{k}_{lm}\partial^{j}\partial^{l}A^{m}= j_{i}
,
\end{equation}
where $\sigma$ is the damping rate, $\epsilon_{ijk}$ is the three
dimensional Levi-Civita tensor and summation over spatial indices
applies if repeated in different location on pairs of tensor
symbols. The electric current $j_{i}=
-2e\mathbb{I}m[\phi^{*}D_{i}\phi]$ is a Noether
  current which is conserved by gauge invariance:
$\partial^{\mu}[\phi^{2}(\partial_{\mu}\theta + eA_{\mu})]=0$. The
term   $-\sigma\partial_{t}A_{i}$ is an ohmic current
 $j_{i}^{ohm}$.
\\ \indent Following the original phenomenological Ginzburg-Landau
theory in which the free energy can be expanded analytically in
powers of $|\phi|$ near the phase transition and the fact that many
GUTs predict a radiatively corrected effective potential of a
similar kind, we will work with a scalar potential of the form
\begin{equation}\label{a3}
V(|\phi|)= m_{H}^{2}|\phi|^{2} - \kappa |\phi|^{3} +
\lambda|\phi|^{4} ,
\end{equation}
where $m_{H}$, $\kappa$ and $\lambda$ are in general positive
definite temperature-dependent parameters. A first order phase
transition takes place at some critical temperature $T_c$. Below
$T_{c}$ the coupling parameters must satisfy the inequalities
$9\kappa^{2}-32m_{H}^{2}\lambda
> 0$, $\kappa^{2}-4m_{H}^{2}\lambda>0$ for the existence of a meta-stable minimum at $\phi=0$. The true
minimum is located at $|\phi|_{min}=(3\kappa +
\sqrt{9\kappa^{2}-32 \lambda m_{H}^{2}})/8\lambda$. In what
follows we will refer to $|\phi|_{min}$ as the vacuum expectation
value (v.e.v.) and to the set of absolute minimum energy local
states as vacuum manifold. Those states are characterized by the
conditions
\begin{eqnarray}
|\phi| & = & |\phi|_{min},\nonumber\\
D_{\mu}\phi & = & 0 .
\end{eqnarray}
\indent The transition takes place by means of spontaneous
nucleation of bubbles of true vacuum. These bubbles appear as a
result of the growth of very special and infrequent fluctuations
of $\phi$ \cite{Coleman77,Linde83}. Spontaneous symmetry breaking
(SSB) occurs as the order parameter $\phi$ chooses randomly a
point $\alpha$ on the $U(1)$ vacuum manifold such that
$\phi_{vac}=|\phi|_{min}e^{i\alpha}$ (figure \ref{fig3}(b)). That
way, spatial inhomogeneity in the manner the symmetry breaks down
arises naturally in a first order phase transition.
 This is a requirement for defect formation.\\ \indent Nucleation
of bubbles and expulsion of magnetic field out of them drive the
system out of equilibrium. These processes produce latent heat.
Restoration of equilibrium involves readjustment of the magnetic
flux and steady bubble expansion. It is achieved by means of the
dissipation provided by the damping terms which, at the same time,
remove the latent heat.\\ \indent The damping rates --
conductivity in the case of the gauge field -- determine the time
scale of the dynamics of the fields -- the steady bubble expansion
rate in the case of the scalar field. In principle no a priori
assumption exists as regarding the damping rates. However they are
important phenomenologically since their values determine whether
the fields evolve in adiabatic, diffusive or dissipative regimes.
For the sake of simplicity, we will assume that conductivity is
uniform in the symmetric (normal) phase and we will assign the
same damping rate $\sigma$ to both fields. In doing so we
guarantee conservation of charge by fulfilling Gauss' law
$\vec{\nabla}\cdot\vec{E}=-q$ \cite{Donaire04}.\footnote{Violation
of conservation of charge is not actually a problem in
superconductors phenomenology. In fact the TDGL equations violate
the gauge symmetry explicitly if one considers the system they
describe as isolated.}

\subsection{The gauge choice and the gauge-invariant field ${D_{\mu}\theta}$\label{choice}}
Beyond tree-level, spontaneous breakdown of a local symmetry has
been proved to be impossible without gauge fixing
\cite{Elitzur75}. In the mean-field theory approximation a similar
situation holds. Spontaneous symmetry breaking manifests itself in
the gauge fixing as we will illustrate in section
\ref{trapping}.\\ \indent The fact that topological defects do
form implies that a magnetic field gets confined around a discrete
number of zeros of $\phi$. It results in a multiply-connected
vacuum region. Along any curve passing through one of the zeros of
$\phi$ the complex phase $\theta$ presents a discontinuity of
value $\pi$. These discontinuities cannot be gauged out because
they are physical and so gauge-invariant.\\ \indent Deep in a
simply connected domain of broken phase like a bubble, it is
always possible to choose the unitary gauge or, more generally,
$\theta(x)=\alpha$ uniform. In fact, because the magnetic flux is
expelled by Meissner effect, it is natural to choose $A_{\mu}=0$
there. Conversely, in the symmetric phase region between bubbles
where $\phi=0$, $\theta$ is ill-defined and so is
$\partial_{\mu}\theta$. As the bubbles of broken phase expand and
collide they may form a multiply-connected broken phase region.
The gauge $\theta(x)=\alpha$ can only be maintained if this region
is not to contain topological defects. If topological defects do
form, the unitary gauge is not possible because, for any closed
loop $C$ around a defect, $\oint_{C}\partial_{i}\theta\:
dx^{i}\neq 0$ holds.\\ \indent The above situations illustrate the
fact that neither $\partial_{\mu}\theta$ nor $A_{\mu}$ alone are
good quantities to be tracked in order to explain the formation of
topological defects in gauge theories for they are not
gauge-invariant quantities. In fact, the relevant degrees of
freedom to be considered in the region where topological defects
are to be formed are in the gauge-invariant combination of the
Goldstone modes and the gauge field\footnote{See section
\ref{phase} for further comments on this definition.}
\begin{equation}\label{a40}
D_{\mu}\theta\equiv \partial_{\mu}\theta + eA_{\mu} .
\end{equation}
\subsection{Setup of the numerical simulations\label{setup}}
In the numerical simulations presented in this paper we deal mainly
with two kinds of scenarios. In the first one we show how thick
vortices can be obtained when an external magnetic flux gets trapped
in a multi-bubble collision. At the same time, chains of
single-winding vortices form at the junction of each pair of
bubbles. In the second scenario we show how chains of vortices and
flux strips form in the gap between two parallel superconducting
blocks in the presence of a net magnetic flux.
\subsubsection{Multi-bubble collision picture\\\\}
 Nucleation of bubbles is implemented by setting-up disconnected
spherical bubbles of broken phase. Their initial profile
$|\phi|(r)$ is approximately that of the bounce solution in the
thin bubble wall approximation \cite{Coleman77,Linde83}.
Spontaneous symmetry breaking consists of choosing randomly a
uniform complex phase value in each bubble:
\begin{equation}\label{b01}
\fl
\phi(r)=\sum_{i=1}^{N_{b}}\frac{|\phi|_{min}}{2}\Bigl[1+\tanh{\Bigl(m_{h}(R_{N}-|r-r_{i}|)\Bigr)}\Bigr]\Bigl(\cos{(\alpha_{i})}
+
 i\sin{(\alpha_{i})}\Bigr),
\end{equation}
where $r$ is the space-coordinate vector, $m_{h}^{2}$ is given by
(\ref{mh}) in section \ref{sc}, $N_{b}$ is the number of bubbles,
$\alpha_{i}$ is the initially uniform complex phase of bubble $i$,
$r_{i}$ stands for the position vector of the center of bubble $i$
and $R_{N}$ is the initial radius of the bubbles. $R_{N}$ is large
enough so that bubbles expand. When letting the system evolve with
the equations of motion, the bubbles grow until colliding. The
distance between bubbles is large enough so that the bubble-wall
profile takes its actual form much before they collide. The bubble
expansion rate is controlled by the damping rate $\sigma$
appearing in (\ref{a7}). Soon after bubbles begin to grow the
expansion rate $v$ approaches $v\approx\exp{(-\sigma)}$.\\ \indent
When a uniform magnetic flux is present, nucleation of bubbles
requires the expulsion of the magnetic flux out of them before
they start expanding. We first set up a uniform magnetic field $B$
parallel to the $y$-axis in the gauge $A_{x}= \frac{-B}{2}z$ ,
$A_{z}=\frac{B}{2}x$. To expel flux we let the gauge field  evolve
according to ({\ref{a8}) while the scalar field configuration is
kept fixed at its original form (\ref{b01}). That way, it is by
means of dissipation that the system relaxes until getting to the
minimum energy configuration compatible with both the nucleated
bubbles with profile (\ref{b01}) and the presence of a given
magnetic flux in equilibrium. The optimal damping rate is given by
$\sigma_{relax}\approx 1/l$, where $l$ is the typical length of
the lattice. The time it takes for the magnetic field to relax is
approximately $t_{relax}\approx l$. After relaxation, both $\phi$
and $A_{i}$ are evolved according to equations (\ref{a7}) and
(\ref{a8}). As it was mentioned above, in order to satisfy
conservation of charge, the same damping rate $\sigma$ is used in
both equations.\\ \indent The approach presented here corresponds
to the mean-field zero-temperature effective theory. However, when
a uniform external magnetic field is applied, this field resembles
a long wave-length magnetic field thermal fluctuation.
\subsubsection{Collision of two-superconducting blocks with magnetic flux in
the junction\\\\} This set-up is meant to mimic the collision
between two big bubbles when a magnetic flux gets trapped in the
junction. It is a good approximation as long as the radius of the
bubbles is much larger than the width of the bubble-wall. The
cubic lattice has dimensions $l_{x}\times l_{y}\times l_{z}$,
where $0\leq x\leq l_{x}$, $0\leq y\leq l_{y}$ and $0\leq z\leq
l_{z}$. The collision axis is the $x$-axis, orthogonal to the
walls of the blocks. The gap between the blocks has width $d$ and
its middle point is localized at $x=l_{x}/2$. A uniform magnetic
field parallel to the $y$-axis is confined in the gap. The total
flux is quantized with value $\frac{2\pi}{e}N_{w}$. Periodic
boundary conditions are imposed in the directions $z$ and $y$ and
the axial gauge $A_{x}=0$ is chosen:
\begin{eqnarray}\label{block1}
\fl\phi(x,y,z) & = &
\frac{|\phi|_{min}}{2}\Bigl[1+\tanh{\Bigl(m_{h}(l_{x}/2-d/2-m_{h}^{-1}-x)\Bigr)}\Bigr]\Bigl(\cos{(\alpha_{1})}
+
 i\sin{(\alpha_{1})}\Bigr)\nonumber\\
\fl & + & \frac{|\phi|_{min}}{2}\Bigl[1+\tanh{\Bigl(m_{h}(l_{x}/2-d/2-m_{h}^{-1}-|x-l_{x}|)\Bigr)}\Bigr]\nonumber\\
\fl & \times &\Bigl(\cos{(\alpha_{2} + eBd\:z)} +
 i\sin{(\alpha_{2} + eBd\:z)}\Bigr)\nonumber
\end{eqnarray}
\begin{eqnarray}\label{block2}
\fl A_{x} =  A_{y} = 0\nonumber\\ \nonumber \\ \fl A_{z} = \left\{
\begin{array}{ll}
 0 & \textrm{for $x < l_{x}/2-d/2$}\\
 B(x-l_{x}/2 + d/2) & \textrm{for $l_{x}/2-d/2\leq x \leq l_{x}/2 +
 d/2$}\\
 Bd & \textrm{for $x\geq
 l_{x}/2+d/2$},
 \end{array}\right.\nonumber
 \end{eqnarray}
\\ where $B=\frac{2\pi N_{w}}{e\: d\: l_{z}}$ and $(\alpha_{2}-\alpha_{1})$
is the spontaneous gauge-invariant phase difference (see below)
between both blocks.\\
 \indent In the fully two dimensional
simulations, $l_{y}=0$. In the superconductor film model, the film
thickness $\mu$ satisfies $\mu\ll l_{x},l_{y},l_{z}$. In both
models, we refer to the strip between blocks of width $d$ along
the $x$-axis, length $l_{z}$ along the $z$-axis and thickness
$\mu$ --which is zero in the fully two dimensional model-- as
'junction'. We will show that it is in many aspects similar to a
Josephson junction. In the fully three dimensional model, as the
blocks extend initially all along the $y$-axis, translational
invariance holds at any time in the $y$ direction and the problem
is actually two dimensional. We refer to the $x$ direction as
'transverse direction across the junction' and to the $z$
direction as 'direction along the junction'. In the collision of
two three-dimensional bubbles along the $x$ axis, rotational
invariance in the $y,z$ plane rather than $y$-invariance holds if
no magnetic field is applied. We are however interested in the
situation where a magnetic field is applied in the $y$ direction.
Since we take a planar approximation in which the bubbles are big
enough so that they can be considered as infinite blocks in the
$z,y$ plane, the problem becomes effectively two dimensional.
Therefore, we will maintain the nomenclature we use in the fully
two dimensional model as referring to the directions along (i.e.
the $z$ direction) and transverse to (i.e. the $x$ direction) the
junction.
\subsubsection{The superconductor film model (SC)\\\\}
The superconductor film simulations require extra conditions to
account for the finite film thickness effects. In the first place,
spherical bubbles are replaced with cylinders of length equal to
the thickness of the film $\mu$. More importnatly, vanishing of
the variation of the free energy (\ref{aa1}), $\delta F$, with
respect to arbitrary variations of the scalar field $\phi^{*}$ in
the superconductor film not only implies (\ref{a7}). An additional
surface term with the form
\begin{equation}\label{surfaceterm}
\oint_{\Omega}\delta\phi^{*}(r)D_{i}\phi(r)\: ds^{i}(r)
\end{equation}
must vanish as well. $\Omega$ denotes the boundary of the film and
$ds^{i}$ is the differential vector normal to the surface at point
r. For arbitrary $\delta\phi^{*}$, in order for
(\ref{surfaceterm}) to vanish, it is necessary that the component
of $D_{i}\phi$ normal to the film goes to zero at any point of the
boundary. On the other hand, because the electromagnetic field
does not couple to any charge particle out of the film, purely
Maxwell equations hold in the free space and no dissipative term
(i.e. $\sigma=0$) is to be included there. That is, if the film is
placed in the $xz$-plane with center at $y=0$ and thickness $\mu$,
the additional superconductor film conditions read:
\begin{eqnarray}\label{sccon}
\phi  & = & 0 \qquad\qquad  \textrm{for $y<-\mu/2$ or
$y>\mu/2$}\nonumber\\ \sigma & = & 0 \qquad\qquad \textrm{for
$y<-\mu/2$ or $y>\mu/2$}\nonumber\\ D^{y}\phi & = & 0 \qquad\qquad
\textrm{at $y=-\mu/2$ or $y=\mu/2$} .
\end{eqnarray}
We will be mainly interested in the stability of flux strips and
thick vortices.\\ \indent The setup for the formation of flux
strips is the same as in the previous subsection for the
confinement of magnetic flux in between two superconducting
blocks.\\ \indent Formation of thick vortices of winding $N_{w}>1$
is achieved by confining initially a flux of value
$\frac{2\pi}{e}N_{w}$ in a square of side $d$ equal to the typical
diameter of the vortex, that is
$d\approx\frac{\sqrt{2N_{w}}}{e|\phi|_{min}}$. If the center of
the square is placed at $(x=l_{x}/2, y=0, z=l_{z}/2)$, the initial
conditions are:\\
\begin{eqnarray}\label{block3}
\fl \phi(x,y,z) & = & |\phi|_{min}\qquad\qquad\qquad\qquad\qquad\:
\textrm{for $z>l_{z}/2+d/2$ or $z<l_{z}/2-d/2$},\nonumber\\
\nonumber \\ \fl \phi(x,y,z) & = &
\frac{|\phi|_{min}}{2}\Bigl[1+\tanh{\Bigl(m_{h}(l_{x}/2-d/2-m_{h}^{-1}-x)\Bigr)}\Bigr]
+\nonumber\\ \fl & +
&\frac{|\phi|_{min}}{2}\Bigl[1+\tanh{\Bigl(m_{h}(l_{x}/2-d/2-m_{h}^{-1}-|x-l_{x}|)\Bigr)}\Bigr]\nonumber\\
\fl & \times &\Bigl(\cos{(eBd\:z)} +
 i\sin{(eBd\:z)}\Bigr)\qquad\quad \textrm{for $l_{z}/2+d/2 \geq z\geq
 l_{z}/2-d/2$},\nonumber
\end{eqnarray}
\\
\begin{eqnarray}\label{block4}
\fl A_{x} =  A_{y} = 0, \nonumber\\ \nonumber \\ \fl A_{z}  =
\left\{
\begin{array}{ll} 0 & \textrm{for $z>l_{z}/2+d/2$ or
 $z<l_{z}/2-d/2$}\\ 0 & \textrm{for $x < l_{x}/2-d/2$ and $l_{z}/2+d/2 \geq z\geq
 l_{z}/2-d/2$}\\ B(x-l_{x}/2 + d/2) & \textrm{for $l_{x}/2-d/2\leq x \leq l_{x}/2 + d/2$}\\  & \textrm{and $l_{z}/2+d/2 \geq z\geq
 l_{z}/2-d/2$}\\ Bd & \textrm{for $x\geq l_{x}/2+d/2$ and $l_{z}/2+d/2 \geq z\geq
 l_{z}/2-d/2$},
\end{array}\right.\nonumber
\end{eqnarray}
\\ where  $B=\frac{2\pi N_{w}}{e\: d^{2}}$. As the system evolves, a
stable, meta-stable or unstable thick vortex form. The lattice
size is big enough so that finite size effects are negligible.
Periodic boundary conditions are imposed on the direction $y$
orthogonal to the film.\\ \indent Some simulations in this paper
differ slightly from the ones presented here. We will describe the
corresponding setups in each particular case. A description of the
lattice discretization method is found in \ref{appendB}. The
values of the parameters used in each simulation are compiled in
table \ref{tabfig4}.
\section{Thick vortices and heavy cosmic strings\label{thick}}
\subsection{Global Theory\label{global}}
First suggested by Kibble in \cite{Kibble76}, the formation of
topological defects in a scalar field theory with a global
continuous symmetry can be explained by means of the spontaneous
breakdown of the symmetry and the so-called geodesic rule.\\
\indent In a first order phase transition of the Abelian global
theory, spontaneous symmetry breaking takes place as bubbles
nucleate at arbitrary points of the $U(1)$ vacuum manifold. Since
in the global theory any vacuum configuration requires
$\partial_{\mu}\theta=0$, the value of the complex phase of the
scalar field in each bubble remains uniform as the bubble expands.
When two bubbles coalesce the above condition cannot be satisfied
at the contact area unless both bubbles nucleated with the same
complex phase value. The geodesic rule states that \emph{the
complex phase $\theta$ of the scalar field must interpolate across
the junctions between pairs of bubbles in such a way that it
traces the shortest path in the $U(1)$ manifold}. When several
bubbles nucleated at different points $\{\alpha_{i}\}$ of the
vacuum manifold meet (figure \ref{fig3}(a)), the complex phase
interpolates between each $(\alpha_{i}$, $\alpha_{j})$-pair as
shown in figure \ref{fig3}(b). In the case of a three-bubble
collision the condition for the formation of a single winding
topological defect is
\begin{eqnarray}\label{d00}
\alpha_{1}+\pi<\alpha_{3}<\alpha_{2}+\pi,\qquad
\alpha_{1}<\alpha_{2}<\alpha_{3}.
\end{eqnarray}
This condition is fulfilled with probability $1/4$.
\begin{figure}[htp!]
\begin{center}
\includegraphics[scale=0.50]{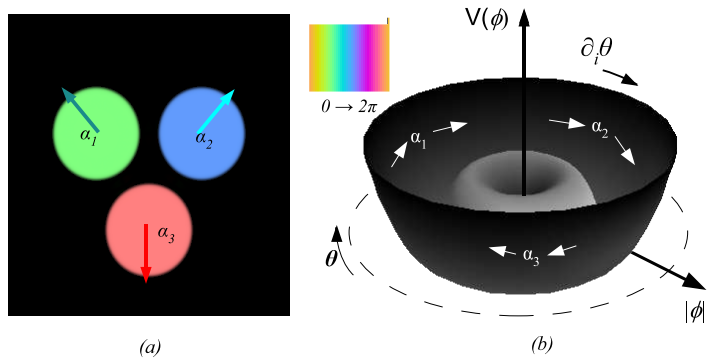}
\end{center}
\caption{\footnotesize{(a) Meeting of three bubbles nucleated at
different points of the vacuum manifold of the global theory,
$\alpha_{1}$,$\alpha_{2}$,$\alpha_{3}$.} \footnotesize{(b) $U(1)$
vacuum manifold. In absence of an external magnetic field the
complex phase follows the geodesic rule interpolating between the
pairs $\alpha_{1}-\alpha_{2}$, $\alpha_{3}-\alpha_{2}$ and
$\alpha_{1}-\alpha_{3}$ in the vacuum manifold. The complex phase
value $\theta$ is depicted using the colour gradient legend at the
top of the figure.}}\label{fig3}
\end{figure}
As the complex phase interpolates according to the geodesic rule,
the gradient energy $\int\:
\partial_{i}\phi\partial^{i}\phi^{*}\: dx$ across the junctions between bubbles, $a<x<b$,
 is minimized:
\begin{equation}\label{ddd0}
\int_{a}^{b} \partial_{i}\phi\partial^{i}\phi^{*} dx\: = \:
\int_{a}^{b}\:
\partial_{i}|\phi|\partial^{i}|\phi|\:dx \quad + \quad \int_{a}^{b}\:
|\phi|^{2}\partial_{i}\theta\partial^{i}\theta\:dx,
\end{equation}
where $\partial_{i}|\phi|\partial^{i}|\phi|$ is the bubble-wall
tension and $|\phi|^{2}\partial_{i}\theta\partial^{i}\theta$ is
the current self-interaction energy density.\footnote{The
one-dimensional integrations in (\ref{ddd0}) are actually either
linear or superficial densities of energy in two or three spatial
dimensions respectively. For brevity, we will refer to them as
energy.} More precisely, what is minimized by following the
geodesic rule is the term associated to $\partial_{i}\theta$, that
is, the current self-interaction energy density.
\subsection{Gauge Theory\label{gauge}}
In the gauge theory $\partial_{\mu}\theta$ is not a
gauge-invariant quantity anymore. The analogue of equation
(\ref{ddd0}) is
\begin{equation}\label{ddd0'}
\int_{a}^{b} D_{i}\phi D^{i}\phi^{*}\: dx\: = \:\int_{a}^{b}
\partial_{i}|\phi|\partial^{i}|\phi|\: dx \quad + \quad \int_{a}^{b}\:|\phi|^{2}D_{i}\theta D
^{i}\theta\: dx,
\end{equation}
where the current self-interaction energy density reads
$\varepsilon(x)\equiv |\phi|^{2}D_{i}\theta D ^{i}\theta$.\\
\indent When two bubbles collide, a different \emph{rule} to the
geodesic follows: \emph{The phase gradient $D_{x}\theta$ and the
gauge-invariant phase difference
$\gamma_{ab}\equiv\int_{a}^{b}D_{x}\theta\: dx$ across the
junction $a<x<b$ between any pair of bubbles decrease in time
until vanishing. In doing so, $A_{i}$ evolves (and so does
$\partial_{i}\theta$) satisfying the conservation of flux
requirement.} That way, the magnetic flux is expelled, a gradient
in the complex phase, $\partial_{x}\theta$, is generated and the
current self-interaction energy gets to its minimum value. At the
same time the order parameter takes the v.e.v. across the
junction.\\ \indent In a multiply-connected broken phase region
the above rule applies to any given pair of vacuum sectors along
each loop $\mathcal{C}$ enclosing a simply-connected symmetric
phase area. Thus, $\oint_{\mathcal{C}}D_{i}\theta\: dx^{i} = 0$
holds when  $\mathcal{C}$ is contained in the surrounding vacuum
region. Continuity in $\theta$ implies
$\oint_{\mathcal{C}}\partial_{i}\theta\: dx^{i} = 2\pi N_{w}=
-\oint_{\mathcal{C}}eA_{i}dx^{i} = -e\Phi_{E.M.}$.\\ \indent In
absence of magnetic field the geodesic rule is a good
approximation. In fact it is equivalent to the gauge rule above if
the gauge $A_{0}=0$, $A_{x}(t=0)=0$ is taken, where $x$ stands for
the direction along the path traversing from one to another
bubble. Hence the formation of a single vortex between the three
bubbles of figure \ref{fig4} can be explained by Kibble mechanism.
\begin{figure}[htp!]
\begin{center}
\includegraphics[scale=0.70]{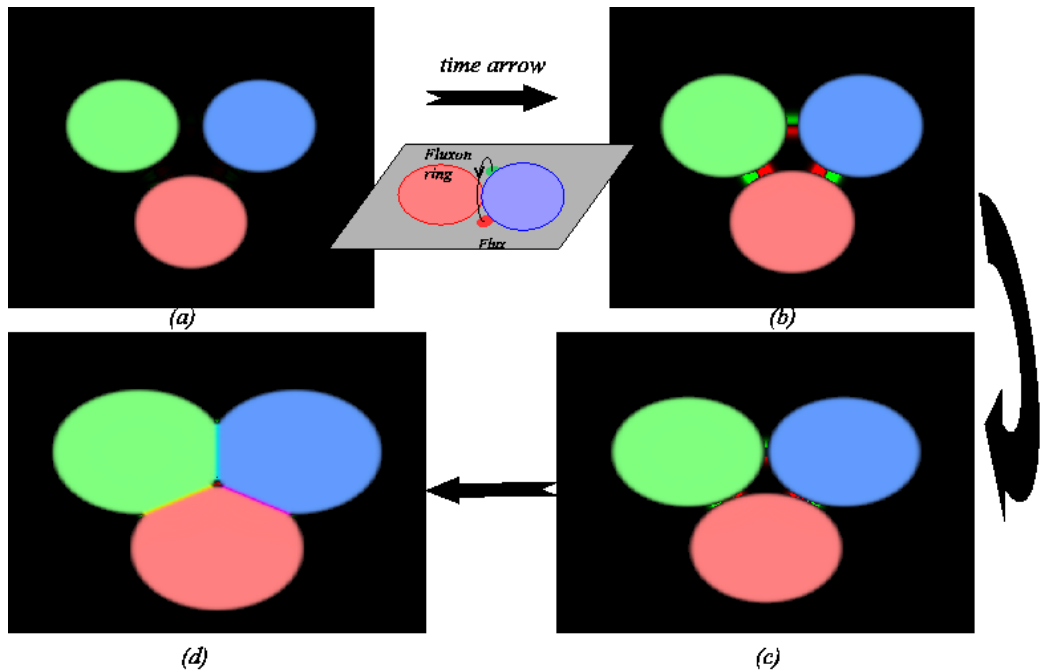}
\end{center}
\caption{\footnotesize{Fully two-dimensional model simulation.
Snapshots of a three-bubble collision. Red colour is used for
negative values of the magnetic field and green colour for
positive values.}}\label{fig4}
\end{figure}
When the three bubbles of figure \ref{fig4}(a) coalesce a fluxon
generates per bubbles pair in figure \ref{fig4}(b). If the
condition (\ref{d00}) is fulfilled, those fluxons add up
constructively at the center of the collision giving rise to a
single winding vortex like the one in figure \ref{fig4}(d).
Further discussion about the geodesic rule in the gauge theory in
absence of an external magnetic field can be found in section
\ref{geodesic}.\\ \indent The difference between the gauge rule
and the geodesic rule in the gauge theory manifests itself in the
presence of an external magnetic field. Let us consider for
instance the phase transition of an infinite superconductor film
in an external magnetic field. Because the total flux is
conserved, the resultant vortex pattern must consist of a lattice
of vortices of the same sign.
\begin{figure}[htp!]
\begin{center}
\includegraphics[scale=0.50]{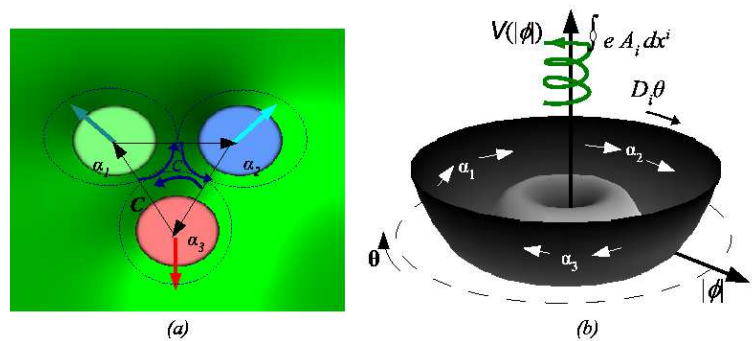}
\end{center}
\caption{\footnotesize{(a) Meeting of three bubbles nucleated at
different points of the vacuum manifold
$\alpha_{1}$,$\alpha_{2}$,$\alpha_{3}$ in the presence of an
external magnetic flux (in green). (b) Vector connection
$D_{i}\theta$ on the vacuum manifold in the gauge
theory.}}\label{figp}
\end{figure}
On the contrary, the randomness of the complex phase pattern
obtained by SSB would give rise to the same number of positive and
negative vortices according to Kibble's argument. Hereafter we
will refer to the above rule in gauge theories as
\emph{geodesic-Meissner rule}.\\ \indent A typical scenario for
the formation of a high winding topological defect is a
multi-bubble collision in presence of a magnetic flux. As the
bubbles of figure \ref{fig6} grow, $D_{i}\theta$ goes to zero
inside them and the magnetic flux is expelled. That is the
Meissner effect. Ultimately, when the bubbles meet and the
geodesic-Meissner rule is applied to each two-bubble junction, the
flux in the symmetric phase region enclosed by the bubbles gets
trapped and a thick vortex forms at the center of the collision.
The main contribution to the total flux in the vortex comes from
the trapping of part of the magnetic flux which is at the moment
bubbles nucleate, $t_{N}$, inside the triangle with contour curve
$C$ whose vertices are the centers of the bubbles (figure
\ref{figp}(a)). An upper bound is in fact given by
$\Phi^{max}_{F.T.}\approx\oint _{C}A^{i}(t_{N})dx_{i}$, where
$F.T.$ stands for Flux Trapping. A lower bound is
$\Phi^{min}_{F.T.}\approx\oint _{C'}A^{i}(t_{N})dx_{i}$ where $C'$
is the contour curve of the area enclosed by the bubbles when they
meet at some time $t>t_{N}$ (figure \ref{figp}(a)). The precise
amount depends on kinematic effects, i.e. on the bubble expansion
rate and the dynamics of the magnetic field. The upper bound
$\Phi^{max}_{F.T.}$ is saturated when either the bubbles expand at
a rate close to the speed of light or the dynamics of the magnetic
field is diffusive. In both cases the magnetic flux expelled from
the bubbles gets stuck to the bubble walls (within a layer of
thickness $\sim 1/\sigma$ in the latter case) all the way until
they meet. This effect appears signaled by the rings in bright
green color surrounding the bubbles of figure \ref{fig17}. In the
opposite limit, when the bubbles expand adiabatically and the
magnetic field escapes from bubbles at the speed of light,
$\Phi^{min}_{F.T.}$ is a better approximation. In general, the
precise value of $\Phi_{F.T.}$ depends on the bubble expansion
rate and the damping rate of the magnetic field. In addition,
order-one numerical factors are determined by the actual geometry
of the collision. The quantitative study will be reported in a
separate article \cite{Me2}.\\ \indent The snapshots of the
two-dimensional simulation in figure \ref{fig6} correspond to a
soft bubble collision in an external magnetic field. After the
three bubbles collide gently, i.e. $v\ll1$, a thick vortex gets
formed at the very center. Its magnetic flux $\Phi_{E.M.}$ lies in
the range $\Phi^{min}_{F.T.}<\Phi_{E.M.}<\Phi^{max}_{F.T.}$
because $\sigma$ is chosen so that the dynamics of $|\phi|$ is
close to adiabatic while the dynamics of the magnetic field is
diffusive. It is easy to calculate its value by simply counting
the times the complex phase increases $2\pi$ units around the
vortex.\\ \indent The contribution of Kibble mechanism through SSB
is negligible in the presence of an intense magnetic field, that
is, for $\Phi_{F.T.}\gg 2\pi/e$.
\begin{figure}[htp!]
\begin{center}
\includegraphics[scale=0.70]{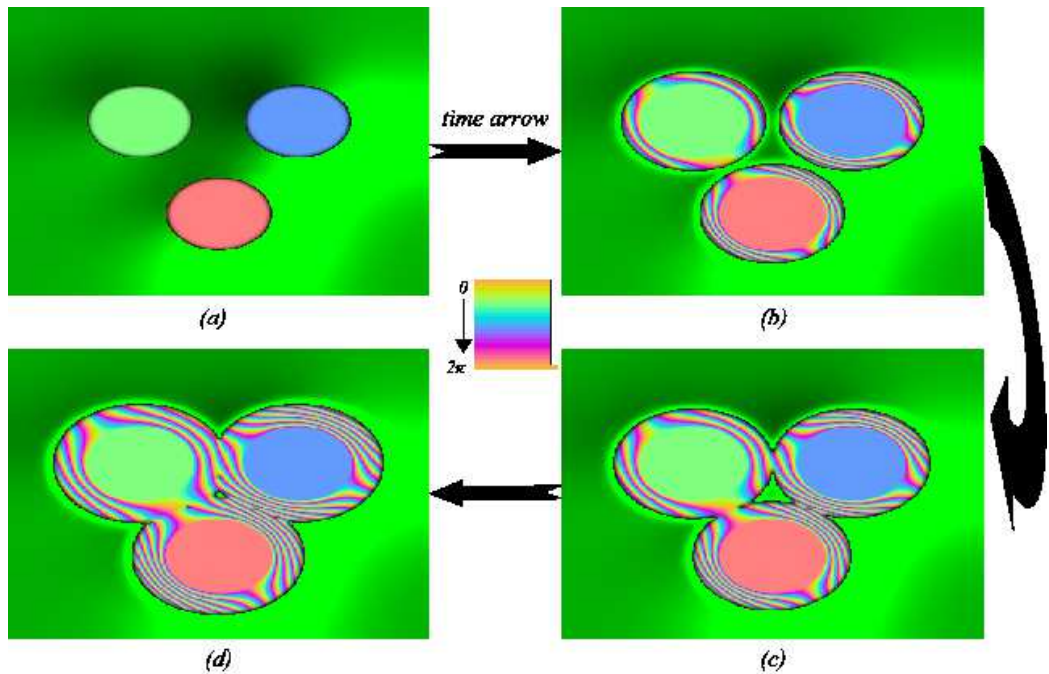}
\end{center}
\caption{\footnotesize{Fully two-dimensional simulation. Snapshots
of a three-bubble collision in the presence of an external
magnetic flux (in green). The colours in the bubbles stand for the
complex phase value of $\phi$ according to the colour legend. In
(d), a thick winding-five vortex gets formed at the center of the
collision.}}\label{fig6}
\end{figure}
Let us consider a fixed value for the initial magnetic field in
the setup of figure \ref{fig6}(a). If one chooses randomly the
initial $\{\theta_{i}\}$ values of the complex phases of the three
bubbles the winding number of the vortex that gets formed at the
center of the collision takes just one of the three possible
values ${\frac{e}{2\pi}\Phi_{F.T.}\pm 1}$. Only if the
contributions due to Kibble mechanism and Flux Trapping were
additive, according to the probability for condition (\ref{d00})
to be satisfied, the probability $P$ of getting each value would
be:
\begin{eqnarray}\label{probas1}
P(\frac{e}{2\pi}\Phi_{F.T.})& = & 3/4,\nonumber\\
P(\frac{e}{2\pi}\Phi_{F.T.}+1)& = & 1/8,\nonumber\\
P(\frac{e}{2\pi}\Phi_{F.T.}-1)& = & 1/8.
\end{eqnarray}
However we will show at the end of section \ref{trapping} that
Kibble mechanism and Flux Trapping are not actually additive.\\
\indent If the collision is violent, i.e. $v\sim1$, as it is the
case in figure \ref{fig17}, the flux trapped corresponds roughly
to $\Phi^{max}_{F.T.}$. The non-equilibrium dynamics is relevant,
reason why the flux trapped is not only confined in the form of a
thick vortex at the very center of the collision. On the contrary,
chains of single-winding vortices form at the junctions.
\begin{figure}[htp!]
\begin{center}
\includegraphics[scale=0.75]{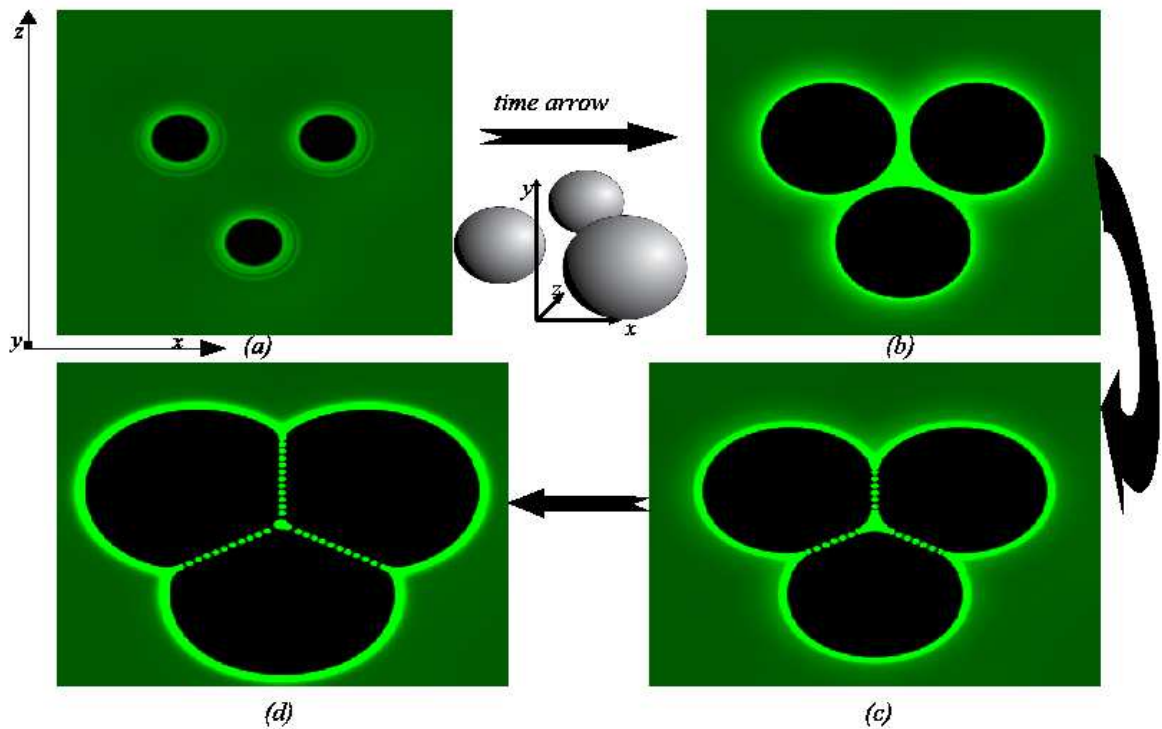}
\end{center}
\caption{\footnotesize{Fully three-dimensional model simulation.
Snapshots of the cross-section of a violent three-bubble collision
in the presence of an external magnetic flux. The cross-section is
taken on the collision plane. Information about the complex phase
is omitted. The broken phase bubbles appear in black while the
magnetic flux density is depicted in green colour gradient. A
thick cosmic string form at the center while chains of single
winding strings form at the junctions.}}\label{fig17}
\end{figure}
\newpage
\section{Magnetic flux strips and single vortex chains\label{trapping}}
By causality, when two bubbles collide in an external magnetic
field, a finite amount of magnetic flux gets trapped at the
junction. Let us assume in first approximation that small
structure effects --i.e. effects at  scales comparable to the
bubble wall thickness $\sim m_{h}^{-1}$-- can be  neglected in the
dynamics of the bubbles. That way the bubbles expand isotropically
at constant rate $v$ at all times, even after their walls overlap.
Let us assume the less favorable situation for magnetic flux to
get trapped. That is, let us assume that the magnetic flux
consists of massless magnetic particles which are expelled from
the surface of the bubbles at the speed of light as if they were
light-rays. That way we are neglecting several effects. First, we
are assuming no interaction between the magnetic field and the
scalar field at small scales of the order of $\sim m_{h}^{-1}$.
Second, we are ignoring the long range magnetic field interaction,
which we will show later on is present in the superconducting
model. Third, we are neglecting any diffraction effect. All these
effects contribute positively to the trapping of magnetic flux
otherwise. As a consequence, the estimate of the flux trapped that
we proceed to calculate disregarding these effects is a lower
bound.\\
\begin{figure}[htp!]
\begin{center}
\includegraphics[scale=3.55]{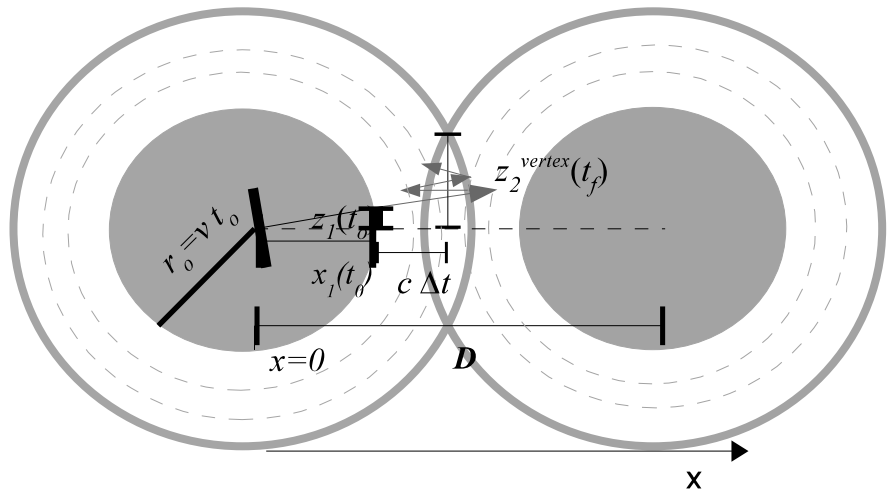}
\end{center}
\caption{\footnotesize{Trapping of a 'magnetic ray' by causality.
A 'magnetic particle' is expelled at the speed of light $c$ from
the surface of the left bubble at $(x_{1}(t_{0}),z_{1}(t_{0}))$.
It gets trapped by causality at some point $(x=D/2,z)$ along the
junction between the two bubbles. The centers of the bubbles are
separated a distance $D$ and both bubbles expand at constant rate
$v$.}}\label{MagOpt}
\end{figure}
 \indent Let us take two identical bubbles separated a distance
 $D$, nucleated at time $t=0$ and expanding at constant rate $v$
 as shown in figure \ref{MagOpt}. A magnetic particle abandons the surface
 of the bubble on the left at height $z_{1}(t_{0})$ from the
 collision axis when the bubble radius is $r_{0}=vt_{0}$. The ray
 experiences several reflections as it strikes several times on each bubble.
 These reflections may, or may not, follow Snell's law. In the case
 they do, the process is referred to in \cite{Me2} as \emph{magnetic optic approach}. Nevertheless,
 in the worst of the cases the exit angle is arbitrarily large
 at each reflection. However, the distance the magnetic particle travels in a time
 interval $\Delta t = t - t_{0}>0$ cannot exceed $c(t-t_{0})$.
 If $t >D/2v$, bubbles meet and overlap during that time interval such that the vertex
 of the overlap is located at height $z_{2}^{vertex}(t)$.
 Because the speed of the vertex along the vertical axis $z$ is
 initially superluminal, it is possible that
 for some time interval $T_{f}$,
\begin{equation}\label{causal1}
c(t_{f}-t_{0})\leq\|(x_{1}(t_{0}),z_{1}(t_{0}))-(D/2,z_{2}^{vertex}(t_{f}))\|,\quad
t_{f}\in T_{f},
\end{equation}
where $(x_{1}(t_{0}),z_{1}(t_{0}))$ stands for the coordinates of
the point from which the ray is first expelled and
$(D/2,z_{2}^{vertex}(t_{f}))$ stands for the coordinates of the
 vertex. If inequality (\ref{causal1}) is satisfied for
 some $t_{f}\in T$ then the light-ray must have got trapped at
 some point of the junction ($x=0$) where bubbles overlap.\\
 \indent  The vertex height, $z_{2}^{vertex}$, evolves in time as
\begin{equation}\label{Vertexloc}
 z_{2}^{vertex}(t)=\sqrt{v^{2}t^{2}-D^{2}/4},\qquad t\geq D/2v.
\end{equation}
The maximum height $z_{2}$ at which a photon expelled at
$(x_{1}(t_{0}),z_{1}(t_{0}))$ can get along the junction at time
$t>t_{0}$ is
\begin{equation}\label{z2maxloc}
\fl z_{2}(t)|_{(x_{1}(t_{0}),z_{1}(t_{0}))}=
 z_{1}(t_{0})+\sqrt{c^{2}(t-t_{0})^2 -
 (x_{1}(t_{0})-D/2)^2},\quad t\geq
t_{0}+(1/c)(D/2-x_{1}(t_{0})).
\end{equation}
From (\ref{z2maxloc}) one reads that the rate at which $z_{2}$
grows is initially infinite. However, from the analysis of its
second derivative, it tends asymptotically to $c$  faster than the
first derivative of (\ref{Vertexloc}) tends to $v$. As $c\geq v$,
inequality (\ref{causal1}) may get saturated at time $t_{f}^{Max}$
for some pair $(x_{1}(t_{0}),z_{1}^{Max}(t_{0}))$ such that\\
\begin{eqnarray}\label{causal3}
z_{2}(t_{f}^{Max})|_{(x_{1}(t_{0}),z_{1}^{Max}(t_{0}))}& = &
z_{2}^{vertex}(t_{f}^{Max}),\nonumber\\
\frac{dz_{2}(t)|_{(x_{1}(t_{0}),z_{1}^{Max}(t_{0}))}}{dt}|_{t_{f}^{Max}}
& = & \frac{dz_{2}^{vertex}(t)}{d t}|_{t_{f}^{Max}}.
\end{eqnarray}\\
Therefore, the flux in the differential area
$z_{1}^{Max}(x_{1})dx$ gets trapped. We have solved the equations
above numerically discretizing the $x$-axis with space step
$\delta x=D/2N$. We search for $z_{1}^{Max}$ for each
$x_{1}^{i}=i\times\delta x$, $i\in[0,N]$. Starting with a uniform
initial magnetic field $B_{0}$, the flux trapped by causality is
given by
\begin{equation}\label{dicricasual}
\Phi_{E.M.}^{Caus.trapped}=4B_{0}\sum_{i=0}^{N}z_{1}^{Max}(x_{1}^{i})\delta
x.
\end{equation}
 Figure \ref{magno} shows the causality bound estimate
(curve (b)) in comparison to the actual data (curve (a)) of the
simulation of a two bubble collision in an initially uniform
magnetic field $B_{0}$. Quantization of the magnetic flux has been
taken into account in the causality bound computation. A damping
rate $\sigma$ is used to slow down the bubble expansion rate $v$
through (\ref{a7}). However, no damping term is included in the
equation for $A_{i}$. That way the dynamics of $A_{i}$ is neither
dissipative nor diffusive.
\begin{figure}[htp!]
\begin{center}
\includegraphics[scale=0.64]{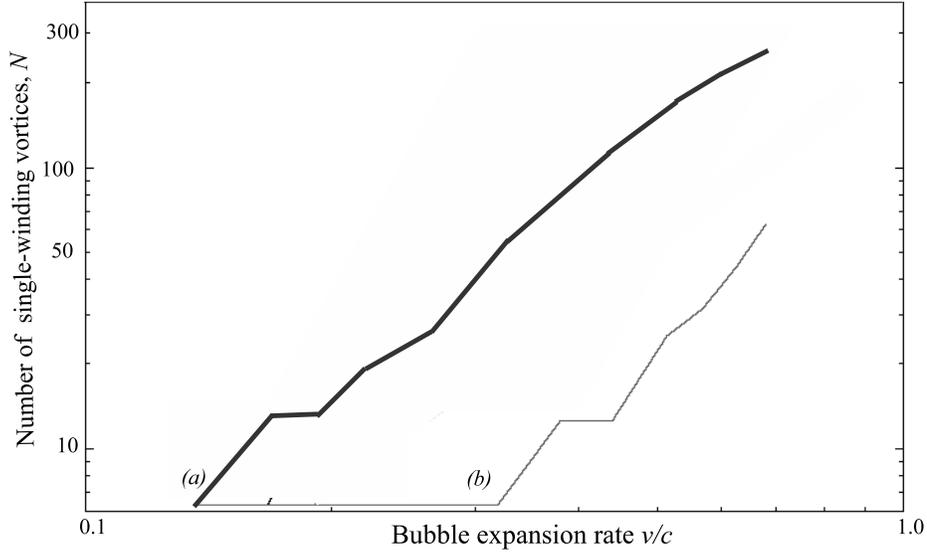}
\end{center}
\caption{\footnotesize{Number of single-winding vortices trapped
in a two-bubble collision. The bubbles expand at constant rate
$v$. Initially, there exists a uniform magnetic field $B=7.76\cdot
10^{-3}/e$ and the distance between the centers of the bubbles is
$D=300$ in lattice spacing units. Curve (a) corresponds to the
actual fully two-dimensional simulation. Curve (b) corresponds to
the causality bound estimate.}}\label{magno}
\end{figure}
\\ \indent Having shown that magnetic flux gets necessarily trapped at the junction
by kinematical reasons (modulo quantization effects), we aim to
explain how it gets confined in the form of  vortex chains like
those in figure \ref{fig17}(d). This involves the study of small
structure effects, that is, the dynamics of $|\phi|$ and
$D_{i}\theta$ at the junctions.\\ \indent Let us simplify the
problem by considering that the collision area of two big bubbles
can be approximated by two parallel superconducting blocks
separated by a gap of width $d$ in the transverse direction $x$ in
which a net magnetic flux is placed. For the sake of simplicity we
will work with the fully two-dimensional theory. Periodic boundary
conditions (pbc) are imposed in the direction $z$ along the
junction. The total flux in the gap is quantized. Its value is
$\frac{2\pi}{e}N_{w}$, where $N_{w}$ is an integer. The reason for
imposing pbc is to confine the flux as it would happen when two
big bubbles coalesce (by dynamical reasons in the latter case).\\
\indent Given two disconnected broken phase blocks we  introduce a
uniform magnetic flux in between. With no loss of generality we
will choose the gauge $A_{x}=0$ at $t=0$ as shown in figure
\ref{fig199}.
\begin{figure}[htp!]
\begin{center}
\includegraphics[scale=0.60]{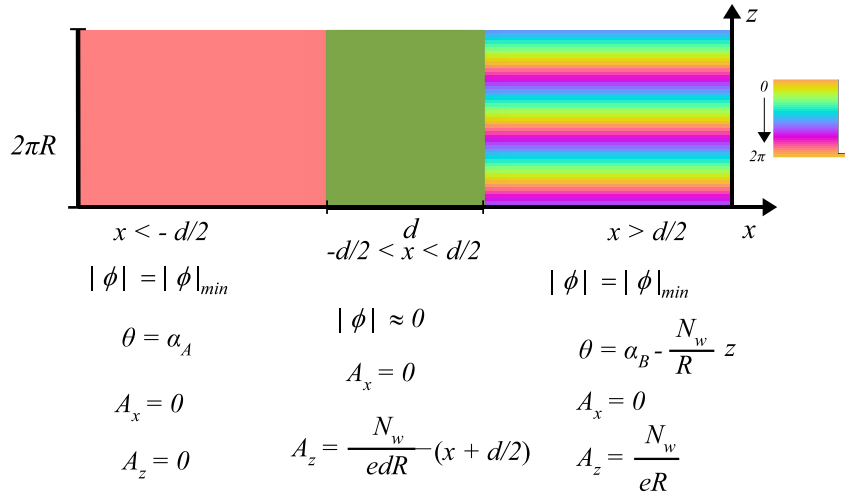}
\end{center}
\caption{\footnotesize{Two superconducting blocks separated by a gap
of symmetric phase. The magnetic flux in the gap (in green) is
$\frac{2\pi}{e}N_{w}$, with $N_{w}=3$. The gauge is chosen in such a
way that $D_{i}=0$ in the superconducting blocks. The complex phase
value in the gauge below is depicted according to the color gradient
legend.}}\label{fig199}
\end{figure}
In this gauge the field $A_{i}$ is continuous and so are
$\partial_{i}\theta$ and $D_{i}\theta$ in the areas of broken
phase. Furthermore, the gauge-invariant phase difference
$\gamma_{ab}(z)$ across the junction when the blocks meet at time
$t=0$ is equivalent to the complex phase difference
$\Delta\theta_{ab}(z)$. That is
\begin{eqnarray}\label{e2o}
\gamma_{ab}(z)& = & \int_{a}^{b}D_{x}\theta(z)\: dx\nonumber\\& = &
\int_{a}^{b}\partial_{x}\theta(z)\:dx\:\:=\:\:\Delta\theta_{ab}(z)\qquad\textrm{in
the gauge $A_{x}=0$}.
\end{eqnarray}
Therefore, according to the gauge choice in figure \ref{fig199},
\begin{equation}\label{e2}
\gamma_{ab}(z)\:= (\alpha_{B}-\alpha_{A})\:\:-(N_{w}/R)\: z,
\end{equation}
where $(\alpha_{B}-\alpha_{A})$ is the \emph{spontaneous phase
difference} and $-(N_{w}/R)\: z$ is the \emph{induced phase
difference}.\footnote{See section \ref{phase} for further
explanation about this decomposition.} After fixing the gauge
there exists still the freedom in choosing
$\Delta\alpha_{ab}\equiv(\alpha_{B}-\alpha_{A})$. This freedom
reflects the invariance of the system along the junction
($z$-invariance) when the blocks are separated. The equality
$\gamma_{ab}=\Delta\alpha_{ab}$ holds in absence of magnetic
field. Thus, the arbitrary choice of $\Delta\alpha_{ab}$
corresponds to the spontaneous breakdown of the $U(1)$ symmetry as
announced in section \ref{choice}. The spontaneous difference
$\Delta\alpha_{ab}$ is physical and manifests itself as blocks
overlap and the $z$-invariance gets broken. At the moment the
blocks get in contact, the physical local gauge-invariant
quantities  are not translation invariant anymore along the
junction due to the presence of the magnetic field (see
\ref{appendA} for a proof).\\ \indent The scalar field profile
falls off exponentially within a typical length equal to the
inverse of its mass $1/m_{h}$. Consequently, it is necessary that
the size of the gap between the blocks $d$ is less than
$d_{s}\approx 2/m_{h}$ so that the blocks overlap and interact
with each other.\\ \indent If $d\gg d_{s}$, the blocks behave as
fronts of big bubbles whose expansion is driven by the potential
$V(|\phi|)$ and the bubble-wall tension
$\partial_{i}|\phi|\:\partial^{i}|\phi|$. The $z$-invariance of
 $|\phi|$ and $B$ along the junction persists and no current traverses the gap.
 Consequently no vortex can form. On the other hand the trapped
flux gives rise to a repulsive force which, if $d>d_{s}$, can
balance the expansion of the blocks towards each other. If one
neglects the bubble-wall tension in first approximation it is easy
to calculate the critical value for the magnetic field necessary
to prevent the blocks from approaching each other beyond some
distance greater than $d_{s}$. At zero-temperature the equilibrium
configuration corresponds to the minimum energy configuration
compatible with a net magnetic flux in the junction. In the fully
two-dimensional theory the magnetic field $B$ does not present
long-range interaction in the symmetric phase and the equilibrium
configuration corresponds to a magnetic flux density uniformly
distributed in the junction. Therefore, for a fixed value of the
magnetic flux, the magnetic energy depends only on the size of the
junction. The equilibrium distance $d_{eq}$ between the two
superconducting blocks of figure \ref{fig199} is the distance at
which the expansion of the blocks is compensated by the magnetic
pressure,
\begin{equation}
d_{eq}=\frac{\Phi_{E.M.}}{2\pi R\sqrt{2V_{min}}},
\end{equation}
where $2\pi R$ is the length of the blocks along the junction,
$\Phi_{E.M.}$ is the total flux and $V_{min}$ stands for
$|V(|\phi|_{min})|$. Requiring $d_{eq}=d_{s}$ we calculate the
minimum value for the magnetic flux so that the blocks never
overlap and the flux strip remains straight, that is,
\begin{equation}\label{e5}
\Phi_{E.M.}^{strip,min} = \frac{4\pi}{m_{h}}R\sqrt{2 V_{min}}.
\end{equation}
\\ \indent If the value of $\Phi_{E.M.}$ in the gap is less than
$\Phi_{E.M.}^{strip,min}$ the small scale structure effects are
relevant and the dynamics of both $|\phi|$ and $D_{i}\theta$
cooperate towards the formation of topological defects. A detailed
analysis of the generation of Josephson-like currents can be found
in \cite{Me}. In what follows we just make use of the essential
results there.\\
 \indent At the moment the blocks get in contact,
$\gamma_{ab}(z)=\int_{-\chi/2}^{\chi/2}D_{x}\theta(x,z)\: dx \sim
\pi$ for some $z$, where $\chi$ is the typical length of the
region where $D_{x}\theta\neq 0$ and $\partial_{x}|\phi|\approx
0$, $\chi \ll d$. This implies that $D_{x}\theta(z)$ is a function
of $\gamma_{ab}(z)$ and the equations of motion incorporate
non-local effects and non-linear terms in $D_{x}\theta$. In
particular, a Josephson-like coupling shows up (see also sections
\ref{joseph} and \ref{geodesic}). The quantity to be studied is
$\gamma_{ab}(z)$ which, right at the moment the blocks get in
contact, follows a perturbed sine-Gordon equation in $1+1$
dimensions \cite{Clem90},
\begin{equation}\label{gordon}
\partial_{tt}\gamma_{ab}-\partial_{zz}\gamma_{ab} +
\sigma\partial_{t}\gamma_{ab} +
2e^{2}|\phi|_{0}^{2}\sin{\gamma_{ab}} = 0,
\end{equation}
where $|\phi|_{0}$ is the $z$-independent factor of the norm of
the scalar field at the time the blocks meet, i.e.
$|\phi|^{2}_{|x|<\chi/2}\simeq |\phi|_{0}^{2}\times f(z)$. In
\cite{Me} $|\phi|_{0}^{2}$ is referred to as 'reduced Cooper pairs
density'. For $\chi \ll d$, it is
\begin{equation}\label{phinot}
|\phi|_{0}\sim (1/2)|\phi|_{min}e^{-d/2}.
\end{equation}
Therefore, initially, the vortex formation dynamics corresponds to
that of one-dimensional sine-Gordon solitons. At this stage the
current self-interaction energy density depends on $z$ as
\begin{equation}\label{e5'}
\varepsilon_{ab}(z)\approx \frac{2}{\chi}|\phi|_{0}^{2}\Big ( 1 -
\cos{[\gamma_{ab}(z)]}\Big )=
\frac{4}{\chi}|\phi|_{0}^{2}\sin^{2}{[\gamma_{ab}(z)/2]}.
\end{equation}
At $|\phi|_{0}$ fixed, equation (\ref{e2}) implies that
$\varepsilon_{ab}$ gets maximum at those points ${z_{n}}$
satisfying $\gamma_{ab}\small{(}z_{n}\small{)} = (2n+1)\pi$, which
are
\begin{equation}\label{e6}
z_{n}  =  \frac{R}{N_{w}}[\alpha_{B}-\alpha_{A}-(2n+1)\pi],\quad
0\leq n < N_{w},
\end{equation}
where the linear flux density is expressed in terms of the winding
number, $\frac{d\Phi_{E.M.}}{d\:z}=\frac{N_{w}}{eR}$.
\\ \indent When the blocks get in contact, $\phi|_{x<0}$ and $\phi|_{x>0}$ interfere at the middle of the
gap $x=0$ giving rise to an interference pattern along the
junction as
\begin{equation}\label{interf}
|\phi|^{2}_{x=0}\approx
2|\phi|_{0}^{2}\Bigl(1+\cos{[\gamma_{ab}(z)]}\Bigr)=4|\phi|_{0}^{2}\cos^{2}{[\gamma_{ab}(z)/2]}.
\end{equation}
Such an interference is constructive at the points ${z_{n'}}$ such
that $\gamma_{ab}(z_{n'})=2\pi n'$ (equal colours according to the
color gradient legend in figure {\ref{fig199}). On the contrary it
is totally destructive if (\ref{e6}) is satisfied. As a result
$N_{w}$ minima of $|\phi|$ turn up at the points
$\{(x=0,z=z_{n})\}$ when the blocks first meet. (see figure
{\ref{fig20}(b)).\\
 \indent In \cite{Me} it is shown that $|\phi|_{0}$
is not $z$-invariant anymore either as $|\phi|$ follows the
equation of motion (\ref{a10}). We will just use here a minimum
energy principle to argue for the formation of vortices.
In view of formula (\ref{interf}), the potential (\ref{e5'}) gets
minimum at the same points where $|\phi|$ is maximum and vice
verse. The points $\{(x=0,z=z_{n})\}$ are in fact centers of the
orbits of the current flow (figure {\ref{fig20}(d)) and attractors
of magnetic flux. The current flow $\sim\vec{D}\theta$ turns
around the minima of $|\phi|$ while the magnetic field, $B=(1/e)\:
\vec{\nabla}\times \vec{D}\theta$, goes straight into them. They
are seeds for the formation of a chain of single winding vortices
if an additional condition holds (see below). Equation
(\ref{gordon}) must be intended as a good starting point to
evaluate the initial evolution of $\gamma_{ab}$. Afterwards,
additional non-linear terms enter and the problem becomes
two-dimensional in space.\\ \indent It might still happen that the
magnetic repulsion stops the vortex formation process.
Unit-winding vortices are rotationally symmetric with a radius
$r_{v}$ that scales with the London penetration length
$\lambda_{L}$ which is equivalent to the inverse of the photon
mass \cite{Hindmarsh..}: $r_{v}\approx
1/m_{\gamma}=(\sqrt{2}e|\phi|_{min})^{-1}$. Therefore the
inequality $\pi R\gtrsim N_{w}r_{v}$ must be satisfied in order to
accommodate $N_{w}$ single vortices in a chain of length $2\pi R$.
This condition imposes an additional constrain on the maximum
value of the flux trapped for a chain of single-winding vortices
to form:
\begin{equation}\label{e7}
\Phi^{chain,max}_{E.M.}=\frac{2\pi^{2}R}{e\: r_{v}}\approx
2^{3/2}\pi^{2}R|\phi|_{min}.
\end{equation}
In the intermediate situation the magnetic flux lies in the range
$\Phi_{E.M.}^{strip,min}>\Phi_{E.M.}>\Phi^{chain,max}_{E.M.}$. In
that case the $N_{w}$ minima of $|\phi|$ do form but single
vortices do not. Instead, periodic flux 'ripples' appear along the
junction between blocks. Figure \ref{fig19} shows the three
typical cases.
\begin{figure}[htp!]
\begin{center}
\includegraphics[scale=0.72]{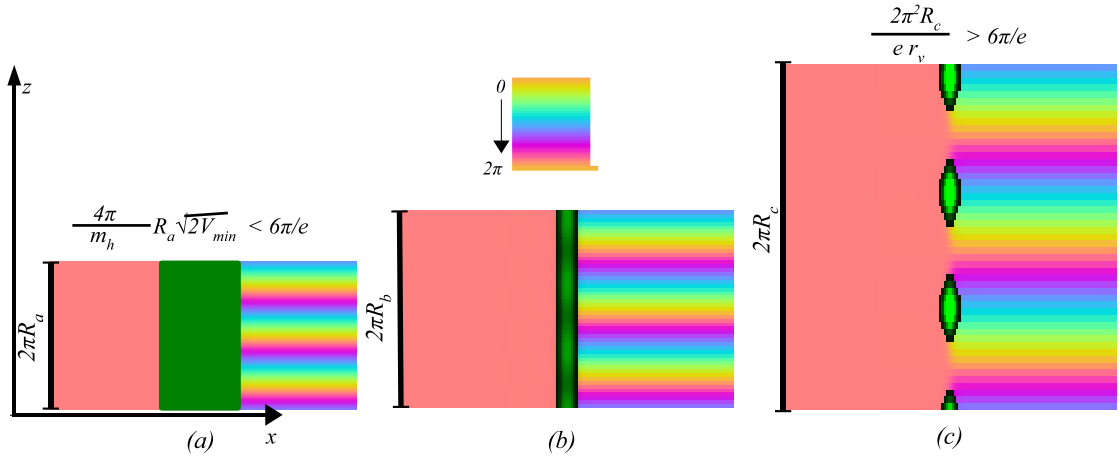}
\end{center}
\caption{\footnotesize{Fully two-dimensional model simulation.
Typical magnetic flux structures at the junction between two
superconducting blocks. The lengths of the junctions $2\pi R_{()}$
increase from left to right while the magnetic flux is $6\pi/e$ and
the spontaneous phase difference is $3\pi/4$ in all the cases. (a)
Flux strip. (b) Flux 'ripples'. (c) Vortex chain.}}\label{fig19}
\end{figure}
The stability/meta-stability of the configurations \ref{fig19}(b)
and \ref{fig19}(c) depends on the dimensionality of the theory.
That is, flux 'ripples' and chains of single-winding vortices are
not stable against condensation into thicker vortices or heavier
cosmic strings. However they can be meta-stable for high enough
flux density. That meta-stability is stronger in the
superconducting model where a long-range magnetic interaction is
present (see next section). Figure \ref{figor}(a) shows a membrane
of magnetic flux in between two  superconducting blocks. The flux
membrane contains nine winding units of flux. They manifest in the
form of 'ripples' in figure \ref{figor}(b). The equilibrium of
such configuration is just maintained by the periodicity in the
direction $z$ along the junction. However, when we introduce a
small perturbation $\delta|\phi| \ll  |\phi|_{min}$ at $z=0$, the
periodicity is broken and the structure becomes unstable. The
vortices gather in groups of winding three (figure \ref{figor}(c))
and three thick vortices of winding three get formed (figure
\ref{figor}(d)).\\
\begin{figure}[htp!]
\begin{center}
\includegraphics[scale=0.72]{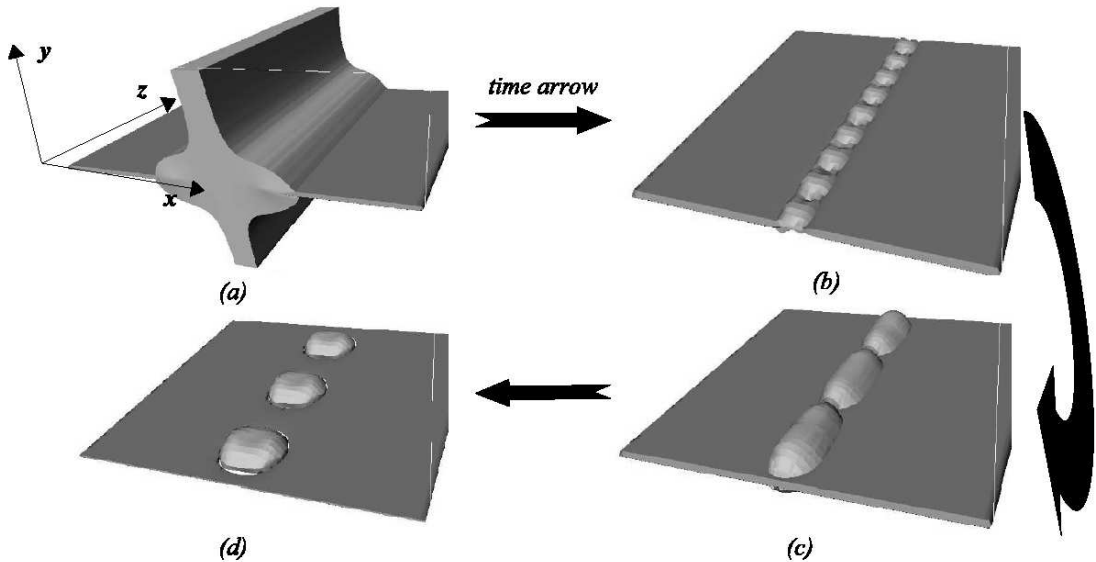}
\end{center}
\caption{\footnotesize{Superconducting model simulation. Snapshots
of the evolution of a membrane (\emph{light grey}) of magnetic flux
$18\pi/e$ between two blocks of broken phase (\emph{dark grey}).
Periodic boundary conditions  are imposed in the directions $y$,
$z$. The film thickness is $\mu=3$ in lattice spacing units. Light
grey is used to depict isosurfaces of equal magnetic energy density.
Breakdown of periodicity along the $z$-axis between (b) and (c)
leads to the formation of three vortices of winding three in
(d).}}\label{figor}
\end{figure}
\indent  The spontaneous phase difference $\alpha_{B}-\alpha_{A}$ is
indeed physical. It gives rise to a shift in the location of the
vortices along the junction according to (\ref{e2}). In a two-bubble
collision, whether a vortex forms at either side with respect to the
collision axis depends on this shift. In figure \ref{fig22} two
bubbles 'sandwich' a strip of magnetic flux as they coalesce. The
relative shift difference of value $3\pi/4$ between the cases
$(a)-(b)$ and $(c)-(d)$ gives rise to the production of one more
vortex along the junction in figure \ref{fig22}(b) with respect to
figure \ref{fig22}(d).
\begin{figure}[htp!]
\begin{center}
\includegraphics[scale=0.70]{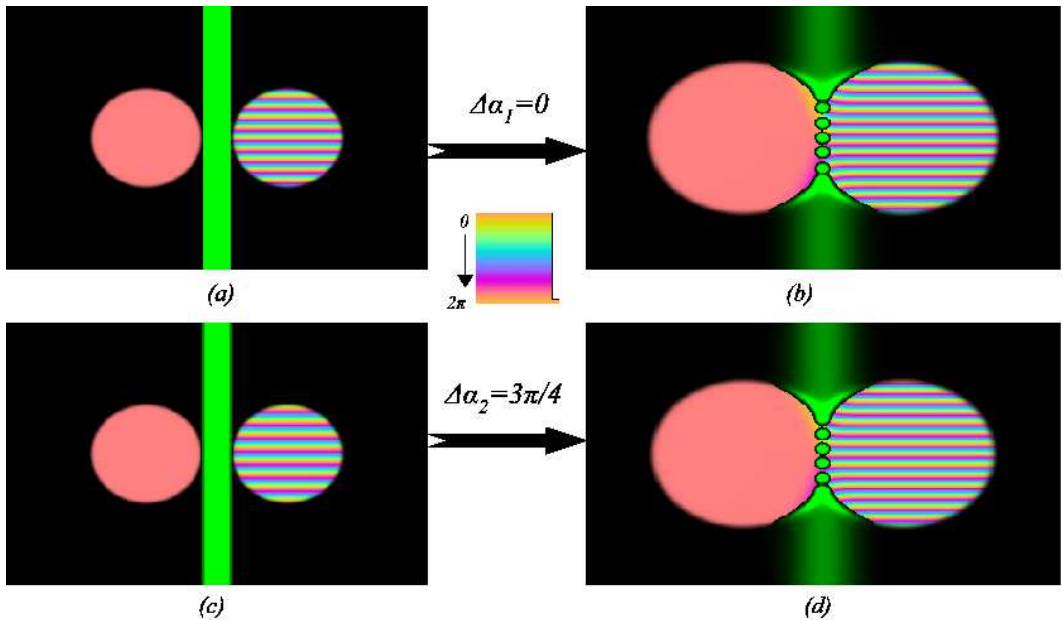}
\end{center}
\caption{\footnotesize{Fully two-dimensional model simulation. (a) A
strip of uniform magnetic flux $36\pi/e$ is placed between two
bubbles with spontaneous phase difference $\Delta\alpha_{1}=0$. (b)
After the collision of the bubbles in (a) a chain of five
single-winding vortices forms at the junction. (c) A strip of
uniform magnetic flux $36\pi/e$ is placed between two bubbles with
spontaneous phase difference $\Delta\alpha_{2}=3\pi/4$. (d)
 After the collision of the bubbles in (c) a
chain of four single-winding vortices forms at the
junction.}}\label{fig22}
\end{figure}
This is the main effect that SSB has in the total flux trapped in
the thick vortex which forms in a multi-bubble collision. The fact
that a single-winding piece ends up either forming part of the
thick vortex at the center or being expelled depends on this
shift.\\ \indent There is also another way in which SSB and flux
trapping mechanism interact. Let us consider the current
traversing the gap between the two bubbles in the presence of a
magnetic flux. The total current follows an interference pattern
(see formula (\ref{MSQUID}), section \ref{joseph}) and oscillates
as a function of the total flux. Its amplitude decreases notably
as the flux increases and it is zero if the total  flux in the
junction takes integral values of $2\pi/e$. As a result of that
current a fluxon ring would develop around the collision axis
giving rise to an extra contribution to the magnetic flux in the
thick vortex (similar to what happens in figure \ref{fig4} in
absence of magnetic field). However, because that current is
strongly suppressed for high values of the magnetic flux in the
junction, we expect that fluxon to be very weak. Consequently its
contribution to the flux in the thick vortex is negligible. For
instance, in the collision of the three bubbles of figure
\ref{fig6}, the probability for the final winding in the thick
vortex to be $\frac{e}{2\pi}\Phi_{F.T.}$ must be greater than the
value $3/4$ in (\ref{probas1}) due to this interference effect. We
conclude that he contributions from flux trapping and KZ mechanism
are not additive.
\section{The superconductor film model\label{sc}}
The model which describes a superconducting film differs from the
fully two-dimensional G-L model in the fact that the order
parameter $\phi$ is confined in a thin film while the
electromagnetic potential $A _{i}$ lives in the  three dimensional
bulk space. The magnetic field is not a pseudoscalar but a
three-component pseudovector. It propagates in the bulk and
induces a long-range interaction on the magnetic flux density in
the film which is absent in the fully two-dimensional case
\cite{Kibble03}.\\ \indent Let us take a thin superconductor film
of thickness $\mu$, placed at $y=0$ and parallel to the
$xz$-plane. If a net magnetic flux is present, it gets confined in
areas of symmetric phase. For a stationary configuration, the
electromagnetic energy reduces to the magnetic energy confined in
a volume $\mathcal{V}$ whose cross section at $y=0$ is the
symmetric phase area on the film,
$E^{E.M.}_{stat.}=\frac{1}{2}\int_{\mathcal{V}}\:d^{3}r\:B_{i}B^{i}$.
As $E^{E.M.}_{stat.}$ is quadratic in $B_{i}$, its most general
form in terms of the film magnetic flux density $B_{y}(\vec{r})$
reads
\begin{equation}\label{scA}
E^{E.M.}_{stat.}=\frac{1}{2}\int_{sym.}\int_{sym.}d^{2}\vec{r}d^{2}\vec{r}'\:B_{y}(\vec{r})V_{mag.}(\vec{r},\vec{r}')B_{y}(\vec{r}'),
\end{equation}
where two-dimensional vectors with components $x$ and $z$ in the
film are labeled with arrows on the top,
$V_{mag.}(\vec{r},\vec{r}')$ is the induced magnetostatic
potential interaction and the subindex \emph{sym.} stands for
'symmetric phase area'. The stationary configuration corresponds
to that for which (\ref{scA}) is minimum. That is,
$B_{y}(\vec{r})$ such that
\begin{eqnarray}\label{green1}
\frac{\delta E^{E.M.}_{stat.}}{\delta B_{y}(\vec{r})} & = &
0\nonumber\\& = &
\int_{sym.}d^{2}\vec{r}'\:V_{mag.}(\vec{r},\vec{r}')B_{y}(\vec{r}'),
\end{eqnarray}
subject to the presence of a fixed magnetic flux
$\Phi_{E.M.}=\int_{sym.}d^{2}\vec{r}\:B_{y}(\vec{r})$. We first
calculate $V_{mag.}(\vec{r},\vec{r}')$ from the stationary
Ampere's law,
\begin{equation}\label{Ampere}
(\nabla\times B)_{i}=j_{i}.
\end{equation}
Because the current lives in the superconducting film on the plane
$y=0$, only the components $x,z$ are non-zero in the vector
equation (\ref{Ampere}). On the other hand, because the magnetic
field is orthogonal to the film at $y=0$ and the flux gets trapped
in the symmetric phase regions, Ampere's law reads
\begin{equation}\label{Ampere2}
[\nabla\times B(r)]_{i}=[\vec{\nabla}\times
B_{y}(\vec{r})]_{i},\quad \vec{r}\in\:\Pi^{y=0}_{sym.},\quad
i=x,z,
\end{equation}
where $\Pi^{y=0}_{sym.}$ stands for 'symmetric phase area in the
plane $y=0$'.\\
 \indent As a first approximation we will consider the film infinitely
thin, that is, with zero thickness, $\mu=0$. That way, the
$y$-component of the magnetic field on the film can be written in
Fourier space as
\begin{equation}\label{greenyBmu0}
\tilde{B}_{y}^{\mu=0}(\vec{k})\equiv\int
d^{3}r\:B_{y}(r)\delta(r_{y})\exp{(i\vec{k}\cdot
\vec{r})},\quad\mu=0.
\end{equation}
\indent It is more convenient to write  (\ref{Ampere2}) in
function of the gauge field $A_{i}(r)$,
\begin{equation}\label{Ampere3}
[\nabla\times\nabla\times A(r)]_{i}=[\vec{\nabla}\times
B_{y}(\vec{r})]_{i},\quad \vec{r}\in\:\Pi^{y=0}_{sym.},\quad
i=x,z.
\end{equation}
In the Coulomb gauge, $\nabla_{i}A^{i}=0$,
\begin{equation}\label{Ampere4}
-\nabla^{2}A_{i}(r)=[\vec{\nabla}\times B_{y}(\vec{r})]_{i},\quad
\vec{r}\in\:\Pi^{y=0}_{sym.},\quad i=x,z.
\end{equation}
We solve for $A_{i}(r)$ using the Green's function of the operator
$\nabla^{2}$, that is, $G_{ij}^{\nabla^{2}}$ such that
\begin{equation}\label{Greenlapl1}
\nabla_{r}^{2}G_{ij}^{\nabla^{2}}(r,r')=\delta_{ij}\delta^{(3)}(r-r').
\end{equation}
In Fourier space,
\begin{equation}\label{Greenlapl2}
\tilde{G}_{ij}^{\nabla^{2}}(k)=\frac{\delta_{ij}}{k^{2}}.
\end{equation}
The gauge field can be computed as
\begin{equation}\label{Greenlap3}
A_{i}(r)=-\int d
r'_{y}\delta(r'_{y})\int_{sym.}d^{2}\vec{r}'G^{\nabla^{2}}_{ij}(r,r')[\vec{\nabla}_{\vec{r}'}\times
B_{y}(\vec{r}')]^{j}
\end{equation}
and likewise the magnetic field, making use of (\ref{Greenlapl2})
and expanding $B_{y}(\vec{r})$ in Fourier modes,
\begin{eqnarray}\label{GreenB1}
\fl B_{i}(r)&=&[\nabla\times A(r)]_{i}\nonumber\\ \fl &=& -\int d
r'_{y}\delta(r'_{y})\int_{sym.}d^{2}\vec{r}'\int
\frac{d^{3}k}{(2\pi)^{3}}\epsilon_{ijk}\nabla_{r}^{j}e^{ik\cdot(r-r')}\frac{\delta_{m}^{k}}{k^{2}}\int\frac{d^{2}\vec{p}}{(2\pi)^{2}}
\epsilon^{m}_{yn}\vec{\nabla}_{\vec{r}'}^{n}e^{i\vec{p}\cdot\vec{r}'}\tilde{B}_{y}(\vec{p})\nonumber\\
\fl &=&-\int d r'_{y}\delta(r'_{y})\int_{sym.}d^{2}\vec{r}'\int
\frac{d^{3}k}{(2\pi)^{3}}\frac{d^{2}\vec{p}}{(2\pi)^{2}}e^{ik\cdot(r-r')}e^{i\vec{p}\cdot\vec{r}'}\epsilon_{ijm}\epsilon^{m}_{yn}\frac{k^{j}p^{n}}{k^{2}}\tilde{B}_{y}
(\vec{p}),
\end{eqnarray}
where $\epsilon_{ijk}$ stands for the Levi-Civita tensor, the
subscript $y$ is fixed, $m,n$ take labels $\{x,z\}$ and $i,j,k$
run over $\{x,y,z\}$. Next, we define the 'pseudo-Green's
function'
\begin{equation}\label{pseudo1}
\mathcal{G}_{i}^{y}(r,\vec{r}')\equiv-\int d
r'_{y}\delta(r'_{y})\int
\frac{d^{3}k}{(2\pi)^{3}}\frac{d^{2}\vec{p}}{(2\pi)^{2}}e^{ik\cdot(r-r')}e^{i\vec{p}\cdot\vec{r}'}\epsilon_{ijm}\epsilon^{m}_{yn}\frac{k^{j}p^{n}}{k^{2}}
\end{equation}
and its corresponding Fourier transform
\begin{equation}\label{pseudo2}
\tilde{\mathcal{G}}_{i}^{y}(k,\vec{p})\equiv-\epsilon_{ijm}\epsilon^{m}_{yn}\frac{k^{j}p^{n}}{k^{2}},
\end{equation}
such that
\begin{equation}\label{forB}
B_{i}(r)=\int
d^{2}\vec{r}'\:\mathcal{G}_{i}^{y}(r,\vec{r}')B_{y}(\vec{r}').
\end{equation}
 \indent Therefore, we can write the magnetic energy as
\begin{eqnarray}\label{Eemstat}
 E^{E.M.}_{stat.}&=&\frac{1}{2}\int_{\mathcal{V}} d^{3}r\:B_{i}B^{i}\nonumber\\
&=&\frac{1}{2}\int_{sym.}d^{2}\vec{r}'\int_{sym.}d^{2}\vec{r}''\int
d^{3}r\:
B_{y}(\vec{r}')\mathcal{G}_{i}^{y\dag}(r,\vec{r}')\mathcal{G}^{yi}(r,\vec{r}'')B_{y}(\vec{r}''),
\end{eqnarray}
where $\dag$ denotes adjoint. From (\ref{scA}), one reads
immediately
\begin{equation}\label{identy}
V_{mag.}(\vec{r}',\vec{r}'')\equiv\int
d^{3}r\:\mathcal{G}_{i}^{y\dag}(r,\vec{r}')\mathcal{G}^{yi}(r,\vec{r}'').
\end{equation}
Again, we make use of the Fourier transforms to write, in the
infinitely thin film limit $\mu\rightarrow 0$,\\
\begin{eqnarray}\label{PotentialFour}
\fl E^{E.M.}_{stat.}&=&\frac{1}{2}\int
d^{2}\vec{r_{1}}''d^{2}\vec{r_{2}}''\Bigl[\int\frac{d^{2}\vec{p}}{(2\pi)^{2}}\frac{d^{2}\vec{s}}{(2\pi)^{2}}\int
 \tilde{B}_{y}(\vec{p})\int \tilde{B}_{y}(\vec{s})\nonumber\\ \fl
&\times&\int\frac{d^{3}k}{(2\pi)^{3}}\frac{d^{3}q}{(2\pi)^{3}}\:\int
d^{3}r d^{2}\vec{r_{1}}' d^{2}\vec{r_{2}}'\:
e^{ik\cdot(r-\vec{r_{1}}')}e^{i\vec{p}\cdot(\vec{r_{1}}'-\vec{r_{1}}'')}e^{-i\vec{s}\cdot(\vec{r_{2}}'-\vec{r_{2}}'')}e^{-iq\cdot(r-\vec{r_{2}}')}\nonumber\\
\fl &\times&
\tilde{\mathcal{G}}_{i}^{y\dag}(k,\vec{p})\tilde{\mathcal{G}}^{yi}(q,\vec{s})\Bigr],\quad\mu\rightarrow
0.
\end{eqnarray}\\
After the integration of $q,\vec{p},\vec{s}$ (which yields a
product of delta functions) we get to\\
\begin{equation}\label{Potential2}
\fl E^{E.M.}_{stat.}=\frac{1}{2}\int
d^{2}\vec{r_{1}}''d^{2}\vec{r_{2}}''\Bigl[\int\frac{d^{3}k}{(2\pi)^{3}}\tilde{B}_{y}(\vec{k})\tilde{B}_{y}(-\vec{k})e^{i\vec{k}\cdot(\vec{r_{1}}''-\vec{r_{2}}'')}
\epsilon_{ijr}\epsilon^{i}_{ab}\epsilon^{r}_{ym}\epsilon^{b}_{yd}\frac{k^{j}k^{m}k^{a}k^{d}}{k^{4}}\Bigr],
\end{equation}
where $k^{4}$ is the $4^{th}$ power of the norm of the
three-component wave vector $k$; $r,b,m,d$ take labels $\{x,z\}$
and $i,j,a$ run over $\{x,y,z\}$. By applying the identity
$\epsilon_{ijk}\epsilon^{klm}=\delta_{i}^{l}\delta_{j}^{m}-\delta_{i}^{m}\delta_{j}^{l}$
and integrating the mode $k_{y}$ , it is easy to get to
\begin{equation}\label{Potential2'}
\fl E^{E.M.}_{stat.}=\frac{1}{2}\int
d^{2}\vec{r_{1}}''d^{2}\vec{r_{2}}''\Bigl[\int\frac{d^{2}k}{(2\pi)^{2}}B_{y}(\vec{k})B_{y}(-\vec{k})e^{i\vec{k}(\vec{r_{1}}''-\vec{r_{2}}'')}\frac{1}{2}|\vec{k}|\Bigr],\quad\mu\rightarrow
0.
\end{equation}
Comparing (\ref{Potential2'}) with (\ref{Eemstat},\ref{identy}),
we identify
\begin{eqnarray}
V_{mag.}^{\mu=0}(\vec{r_{1}}'',\vec{r_{2}}'')&=&\int\frac{d^{2}k}{(2\pi)^{2}}e^{i\vec{k}(\vec{r_{1}}''-\vec{r_{2}}'')}(\frac{1}{2}|\vec{k}|)^{-1}\label{POT1}\\
&=&\frac{1}{2\pi}\frac{1}{|\vec{r_{1}}''-\vec{r_{2}}''|}\label{POT2}.
\end{eqnarray}\\
\indent It is remarkable that
$V_{mag.}^{\mu=0}(\vec{r_{1}}'',\vec{r_{2}}'')$ equals the inverse
of the two-point function $\langle
B_{y}(\vec{r_{1}}'')B_{y}(\vec{r_{2}}'')\rangle$ in the high
temperature field theory, normalized to $k_{B}T=1$
\cite{Kibble03}. The reason being that the long wavelength
spectrum of fluctuations of the gauge field at high temperature
--that is, the Rayleigh-Jeans spectrum of balckbody radiation--
equals the Green's function of the stationary Ampere's equation
(\ref{Ampere4}) in the symmetric phase.\\
 \indent Should we take into account the finiteness of the thickness of the
film, $\mu$, on the magnetic flux density
$\tilde{B}_{y}(\vec{k})$, we would integrate those modes for which
$k_{y}\leq 1/\mu$. We may use a Gaussian function to suppress
shorter wavelength modes. Replacing the delta function
$\delta(r_{y})$ in the integrand of (\ref{greenyBmu0}) and
thereafter with the Gaussian distribution
$\frac{1}{\sqrt{2\pi}\mu}\exp{(-r_{y}^{2}/2\mu^{2})}$, we define
the film magnetic flux density as
\begin{equation}\label{greenyB}
B_{y}^{\mu>0}(\vec{k})\equiv\int
d^{3}r\:\frac{1}{\sqrt{2\pi}\mu}\exp{(-r_{y}^{2}/2\mu^{2})}B_{y}(r)\exp{(i\vec{k}\cdot
\vec{r})}.
\end{equation}
This way, analog of (\ref{POT1}) with $\mu>0$ reads
\begin{equation}\label{Potentialmuno0}
V_{mag.}^{\mu>0}(\vec{r_{1}}'',\vec{r_{2}}'')=\int\frac{d^{2}k}{(2\pi)^{2}}e^{i\vec{k}\cdot(\vec{r_{1}}''-\vec{r_{2}}'')}[\frac{1}{2}\textrm{Erfc}(\mu|\vec{k}|)\:|\vec{k}|]^{-1},
\end{equation}
where Erfc stands for 'complementary error function'. If the
typical length of the symmetric phase area is much larger than the
film thickness, the long wavelength limit $|\vec{k}|\ll 1/\mu$ can
be taken in good approximation such that
$\textrm{Erfc}(\mu|\vec{k}|)\approx 1$ and the approximate
expression (\ref{POT1}) is recovered.\\
 \indent We end up with the result that, in a superconductor film, vortices
present a long-range interaction associated to their magnetic
flux. Also, the Coulombian-like interaction
$V_{mag.}(\vec{r},\vec{r}')$ implies that the magnetic flux
density on the symmetric phase areas of the film behaves as if it
were an electric charge density on the surface of an ordinary
metal conductor.\\ \indent When a magnetic flux is trapped in a
symmetric phase region, the stationary configuration corresponds
to  that for which (\ref{scA}) is minimum subject to the existence
of a total magnetic flux $\Phi_{E.M.}$. That is, $B_{y}(\vec{r})$
such that
\begin{equation}\label{scB}
\frac{\delta}{\delta
B_{y}(\vec{r})}\int\int_{sym.}d^{2}\vec{r}d^{2}\vec{r}'\Bigl[\frac{1}{2}B_{y}(\vec{r})
V_{mag.}(\vec{r},\vec{r}')B_{y}(\vec{r}')\:+\:H\:\delta^{2}(\vec{r}-\vec{r}')
B_{y}(\vec{r})\Bigr]=0,
\end{equation}
where $H$ is a lagrangian multiplier and
$V_{mag.}(\vec{r},\vec{r}')$ is given by (\ref{POT2}). The
solution of (\ref{scB}) is
\begin{equation}\label{scC}
B_{y}(\vec{r})=-H\int_{sym.}d^{2}\vec{r}'\:
V^{-1}_{mag.}(\vec{r},\vec{r}'),
\end{equation}
with the constraint
$\Phi_{E.M.}=\int\int_{sym.}d^{2}\vec{r}d^{2}\vec{r}'\:\delta^{2}(\vec{r}-\vec{r}')
B_{y}(\vec{r})$.

\subsection{Stability and meta-stability of thick vortices}
For the purpose of studying the dynamics of vortices, there are in
principle two length scales to be considered, the coherence length
$\xi$ and the London penetration depth $\lambda_{L}$. $\xi$ is
equal to the inverse of the mass of the order parameter at
$|\phi|_{min}$, that is, $\xi=1/m_{h}$, where
\begin{equation}\label{mh}
m_{h}^{2}\equiv\frac{\delta}{\delta|\phi|}\Bigl(|\phi|\frac{\delta
V(|\phi|)}{\delta|\phi|^{2}}\Bigr)|_{|\phi|=|\phi|_{min}}.
\end{equation}
For the potential in (\ref{a3}),
$m_{h}^{2}=m_{H}^{2}-3\kappa|\phi|_{min}+6\lambda|\phi|_{min}^{2}$.
Notice that even though $m_{h}$ is of the order of $m_{H}$ they
differ by an order one numerical factor. $\lambda_{L}$ is the
inverse of the photon mass,
$\lambda_{L}=1/m_{\gamma}=(\sqrt{2}e|\phi|_{min})^{-1}$. The
Ginzburg parameter, $\kappa_{G}\equiv\lambda_{L}/\xi = m_{h}/
m_{\gamma}$, depends on both the value of $|\phi|_{min}$ and the
curvature of the effective potential $m_{h}$ at $|\phi|_{min}$.
Finite temperature radiative corrections \cite{Hove02} and the
presence of an external magnetic field \cite{Ayala05} may give
rise to variations in $\kappa_{G}$.\\ \indent In a type-I
superconductor, where $\lambda_{L}<\xi$, there exists an
attractive vortex-vortex force at distances shorter than $\xi$
associated to the order parameter. At distances shorter than
$\lambda_{L}$ there is also a magnetic force interaction which is
repulsive for equal sign vortices. If no other force exists, two
vortices of the same sign separated by a distance less than
$1/m_{h}$ can gather to give rise to a stable thick vortex. That
is the case in the fully two and three-dimensional theories
\cite{Kajantie99}.
\\ \indent  In a real type-I superconductor film, the induced long-range
repulsive force (\ref{POT2}) between either equal sign vortices or
equal sign flux densities, modifies the picture above. In a thick
film, $\mu\gg\xi$, the long-range Coulombian interaction is
dominant at very long distances. Therefore, a meta-stable cluster
of vortices of the same sign form. If the film thickness is of the
order of $\xi$ and still greater than $\lambda_{L}$, it happens
that high winding vortices are stable up to some critical value of
the winding number. For higher winding they can be meta-stable
. Finally, if the film is thin,
$\mu\lesssim\lambda_{L}$, the Coulombian repulsive interaction
acts at distances  $r\gtrsim \lambda_{L}$ and so only
single-winding
 vortices are stable and an Abrikosov lattice form in an external magnetic field . The last case corresponds
 to the situation in
which a thin type-I bulk superconductor film behaves as type-II in
a perpendicular external magnetic field
\cite{Tinkham63,Chang66,Miller68}.\\ \indent Let us concentrate on
the case where both stable and meta-stable thick vortices can
form. The winding number of the stable vortices cannot be
arbitrarily high. Beyond some critical winding value, they break
up into smaller pieces. The argument for this to happen is as
follows. The total energy of a stationary winding $N_{w}$ vortex
contains three terms,
\begin{equation}\label{vA}
E(N_{w},r_{v})=C(\mu/r_{v})N_{w}^{2}\:r_{v}^{-1} + \mu
V\:r_{v}^{2} + T\:r_{v},
\end{equation}
where $C(\mu/r_{v})$ is a geometrical factor which depends on the
exact form of $V_{mag.}(\vec{r},\vec{r}')$ through $\mu/r_{v}$ --
formula (\ref{POT2}) is only exact in the thin film limit
$\mu/r_{v}\approx 0$ --, $V=\pi V_{min}$ and $T$ stands for $2\pi$
times the tension of the vortex wall. Stability requires
\begin{equation}\label{vB}
\frac{\partial E}{\partial r_{v}}|_{r^{stab.}_{v}}=-\frac{\partial
C}{\partial\mu}\mu N_{w}^{2}r_{v}^{-2}-C\:N_{w}^{2}r_{v}^{-2}+2\mu
V\:r_{v} + T = 0.
\end{equation}
Considering the film thickness fixed and the vortex radius large
in comparison to both $\mu$ and $\lambda_{L}$ so that formula
(\ref{POT2}) is a good approximation and the energy associated to
the vortex wall tension is negligible (i.e. $T r_{v}\ll C
N_{w}^{2}\:r_{v}^{-1},\mu V r_{v}^{2}$), the first and fourth
terms in (\ref{vB}) can be ignored and
\begin{eqnarray}\label{vC}
r^{stab.}_{v} & \approx & \frac{C^{1/3}}{(2\mu
V)^{1/3}}N_{w}^{2/3},\nonumber\\ E^{stab.}(N_{w})& \approx &
(2^{1/3}+2^{-2/3})C^{2/3}\mu^{1/3}V^{1/3}N_{w}^{4/3}.
\end{eqnarray}
Consequently,
\begin{equation}\label{vD}
E^{stab.}(N_{w}) + E^{stab.}(1) <
E^{stab.}(N_{w}+1)\quad\forall\:N_{w}
\end{equation}
and multiple winding vortices are not stable.\\ \indent Let us
illustrate this phenomenon by analyzing the stationary
configuration of the magnetic flux density in a thick vortex. For
sufficiently large values of $N_{w}$, the vortex radius scales as
$r_{v}\propto N_{w}^{2/3}$ and the magnetic energy is dominated by
the long-range interaction of the flux density in (\ref{POT2}). In
that case the minimum-energy equilibrium configuration of the
magnetic flux $\Phi_{E.M.}=\frac{2\pi}{e}N_{w}$ in a  vortex
approaches the distribution of an electric charge placed on an
ordinary conducting surface of circular shape and radius $r_{v}$
\cite{Jackson62}. Solving for $B_{y}(\vec{r})$ in (\ref{scC}),
\begin{eqnarray}\label{g6}
B_{y}(r) =\left\{
\begin{array}{ll}
 \frac{N_{w}}{e\:
r_{v}}\frac{1}{\sqrt{r^{2}_{v}-r^{2}}} & \textrm{  for
$r<r_{v}$}\\
 0 & \textrm{  for $r\geq r_{v}$},
\end{array}\right.
\end{eqnarray}
where $r$ is the distance to the center of the disc. Formula
(\ref{g6}) gets maximum along the periphery of the circle rather
than at the center as would be the case in a fully-two dimensional
vortex. As $N_{w}$ is increased, $r_{v}$ goes to infinity while
$B_{y}|_{r=0}$ tends to zero. Such a tendency maximizes the
interaction energy  between the magnetic field and the scalar
particles through (\ref{ddd0'}). The magnetic flux density $B_{y}$
goes to zero at $r=0$ where $\phi$ is zero while $B_{y}$ increases
as $r$ goes to $r_{v}$ where $|\phi|_{r=r_{v}}\approx |\phi|_{min}$,
making the configuration energetically unfavourable.\\ \indent
However, given a pair of values  $C(\mu/r_{v})$, $\mu$ in
(\ref{vB}), the energy associated to the vortex wall tension may
compensate the energy difference in the inequality (\ref{vD}) for
small values of $N_{w}$ and $r_{v}$. If that is the case,
inequality (\ref{vD}) can be inverted. Figures \ref{Split} and
\ref{N1} show this phenomenon in numerical simulations. In figure
\ref{Split} the magnetic flux is $6\pi/e$ and $\lambda_{L}$ is
approximately one in lattice spacing units. When the
superconducting film is thick ($\mu=21$ in lattice spacing units)
the magnetic flux profile presents a maximum at the very center
and falls off exponentially towards the periphery of the vortex.
As the thickness decreases, the profile of $|\phi|$ does not
change sensibly. However the magnetic flux profile does approach
the form of formula (\ref{g6}) and the maximum magnetic flux
density migrates towards the periphery of the vortex. Finally, for
a thickness $\mu=1\approx\lambda_{L}$ the winding-three vortex
splits up in three single-winding pieces. In figure \ref{Split}
only the profile of one of them is plotted. While the
winding-three vortex is stable for $\mu=21$, it is meta-stable for
$\mu=9,5,3$ and unstable for $\mu=1$.
\begin{figure*}
\begin{center}
\begin{tabular}{cc}
\includegraphics[angle=-90,scale=0.39]{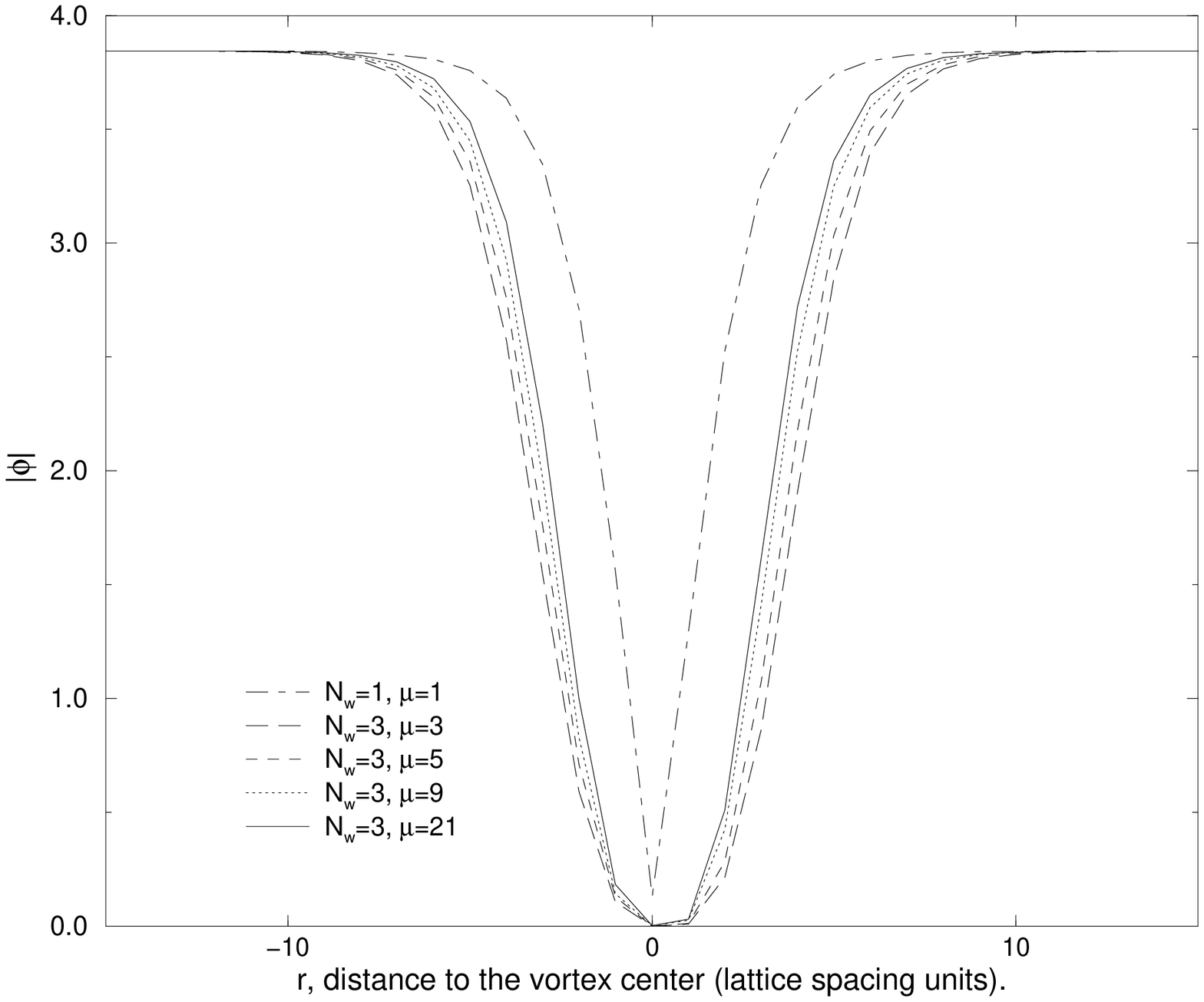}\qquad &
\includegraphics[angle=-90,scale=0.39]{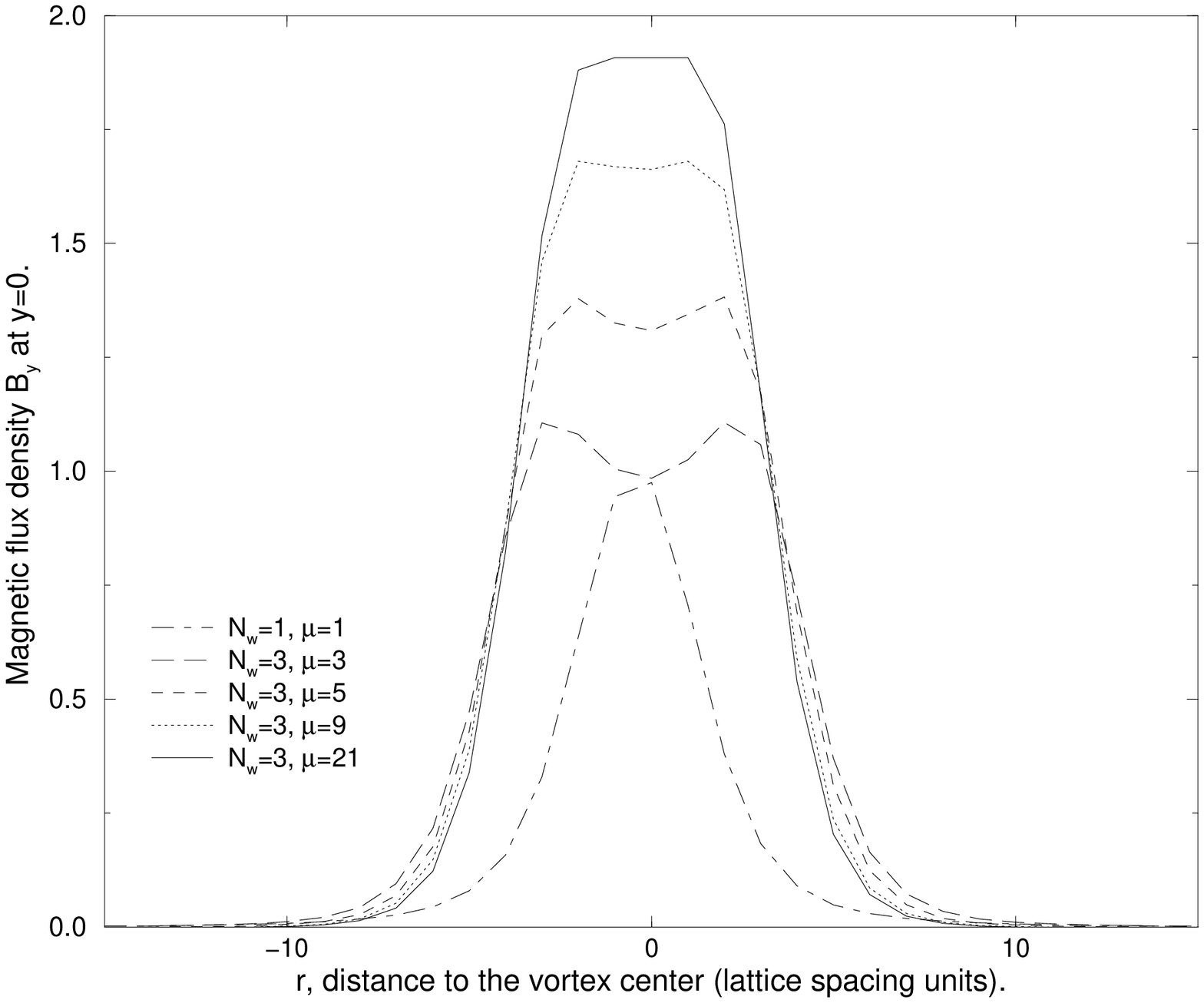}\cr
$(a)$&$(b)$
\end{tabular}
\end{center}
\caption{\footnotesize{Superconducting film simulation. (a) Profile
of the scalar field.} \footnotesize{(b) Profile of the magnetic flux
density at the meridional plane of the film $y=0$.
$\Phi_{E.M.}=6\pi/e$ for film thickness $\mu=3,5,9,21$.
$\Phi_{E.M.}=2\pi/e$ for the curve $\mu=1$.}}\label{Split}
\end{figure*}
Figure \ref{N1} shows how the energy per winding number of thick
vortices  $E(N_{w})/N_{w}$ scales with the winding number $N_{w}$
in comparison to the energy of a single-winding vortex $E(1)$. For
$\mu=1$ only single-windings are stable. For $15>\mu>1$  only
single-windings are still stable but numerical simulations show
that winding two and three vortices are meta-stable. It is
difficult to test how much meta-stable higher windings are because
the number of possible combinations and geometries in which a
thick vortex can decay increases quickly with its winding. For
$\mu=15$ a winding-two vortex is stable against decay into
single-winding pieces. For $\mu=21$, winding-four vortices are
stable and, at least up to $N_{w}=6$, thick vortices do not split
up in single-winding pieces. A winding-six is however meta-stable
and can decay into a pair of stable winding-three vortices.\\
\begin{figure}[htp!]
\begin{center}
\includegraphics[angle=-90,scale=0.50]{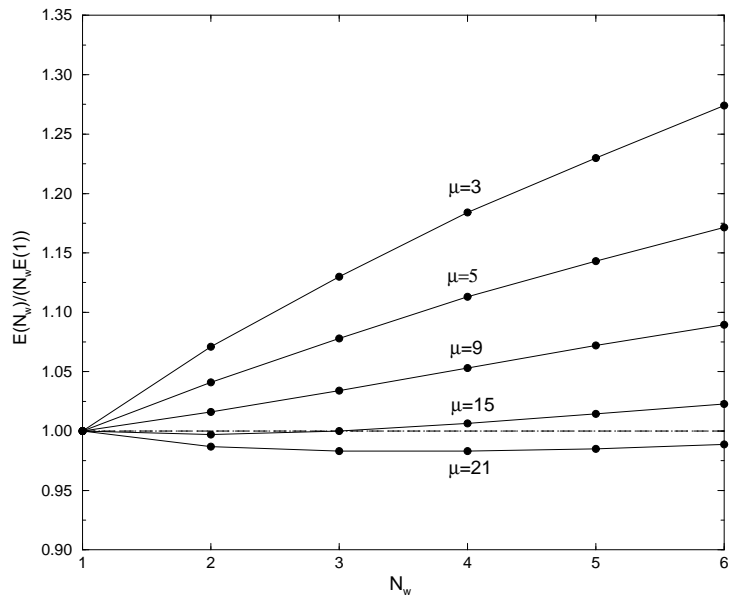}
\end{center}
\caption{\footnotesize{Superconducting film simulation. Scaling of
the energy per winding number of thick vortices with the winding
number. $E(1)$ stands for the energy of a single-winding vortex.
$\mu$ is the film thickness in lattice spacing units $\delta x$.
Because in our simulations $\lambda_{L}\approx\delta x$,
$\mu$-values are approximately in units of $\lambda_{L}$. The dashed
line marks stability against decay into single-winding pieces.
Vortex stability is guaranteed at points of non-positive
slope.}}\label{N1}
\end{figure}
\indent We expect that, by the mechanism explained in previous
sections, a variety of stable and meta-stable vortices form during
the phase transition of a type-I superconductor film in an
external magnetic field. Because of the inhomogeneous nature of a
first order phase transition, the primordial lattice can be very
different to the one corresponding to the minimum energy
configuration. Its evolution in the first stages must be dominated
by the long-range magnetic interaction. It is also very likely
that, by the enormous amount of meta-stable intermediate states,
the evolution of the vortex lattice gets stuck to some meta-stable
configuration. It is only for very thin films that the only stable
vortices are single-winding and the final configuration is a
stable Abrikosov lattice. The Abrikosov lattice configuration is
unique up to the fixing of its center of mass. The center of mass
consists of two degrees of freedom which correspond to the
two-dimensional analogue of the spontaneous phase difference
$\Delta\alpha_{ab}$ that determines the position of the
single-winding vortices in a finite vortex chain. Thus, they are
also vestiges of the spontaneous breakdown of the gauge symmetry.
\subsection{Meta-stability of flux strips}
The strips of magnetic flux described in section \ref{trapping}
are generally unstable in the fully two and three-dimensional
models. For thick superconducting films like the one in figure
\ref{figor} where $\mu\approx 3\lambda_{L}$ they may be unstable
as well. They can however be meta-stable for sufficiently thin
superconducting films and strong enough magnetic field. That is so
even if their width is much less than $2/m_{h}$ and short distance
effects become relevant. Numerical simulations show that, for a
fixed value of the linear magnetic flux density higher than
$\Phi^{chain,max}_{E.M.}/2\pi R$ ($2\pi R$ is the length of the
strip), the resultant strip is meta-stable for a film thickness
less than some critical value. Conversely, given a fixed value for
the film thickness, there exists a critical magnetic flux linear
density for which the strip is meta-stable. As an example let us
take the same setup of figure \ref{figor} for a superconductor
film of finite thickness $\mu$. Using the parameter values in
\ref{appendC}, if $\mu$ is set to $3$ in lattice spacing units,
the resultant strip is unstable for $N_{w}=9$ and three thick
vortices of winding three form as in figure \ref{figor}. However,
if we increase the total magnetic flux up to $12\frac{2\pi}{e}$,
the flux strip configuration turns out to be meta-stable.
Likewise, fixing the value of the total magnetic flux at
$12\frac{2\pi}{e}$, instability takes place when increasing the
thickness up to $\mu=7$. As a result, two thick vortices of
winding six form. Figure \ref{METAST} shows both configurations.
Similar structures of intermediate states have been observed
experimentally \cite{Schawlow56,Alers57}. However the ones
described in \cite{Schawlow56} and \cite{Alers57} do not arise as
a consequence of a phase transition in the presence of an external
magnetic field but as a result of the restoration of the symmetry
in a type-I superconducting film as a uniform external auxiliary
magnetic field $H$ is increased.
\\ \indent In order to test the
meta-stability/instability of a flux strip, we proceed as follows.
We first let the system relax to some relative minimum energy
state compatible with periodic boundary conditions along the
$z$-axis. That gives rise to a non-straight strip like the one in
figure \ref{figor}(b). Both the electric field and the scalar
field momentum are zero at that moment. Afterwards we implement
small perturbations $\delta|\phi|\ll |\phi|_{min}$ randomly along
the strip, once at a time. The size of the fluctuations in the
transverse direction $x$ vary from one lattice spacing up to the
width of the strip. In the longitudinal direction $z$ the size
takes values from one to $\pi R$. If the perturbation does not
grow the system goes back to the original strip structure. That a
strip is meta-stable and not stable can be easily seen by
comparing its energy with the one in which the same amount of flux
is confined in separated stable vortices. The latter is
energetically favourable.
\begin{figure*}
\begin{center}
\begin{tabular}{cc}
\includegraphics[angle=-90,scale=0.39]{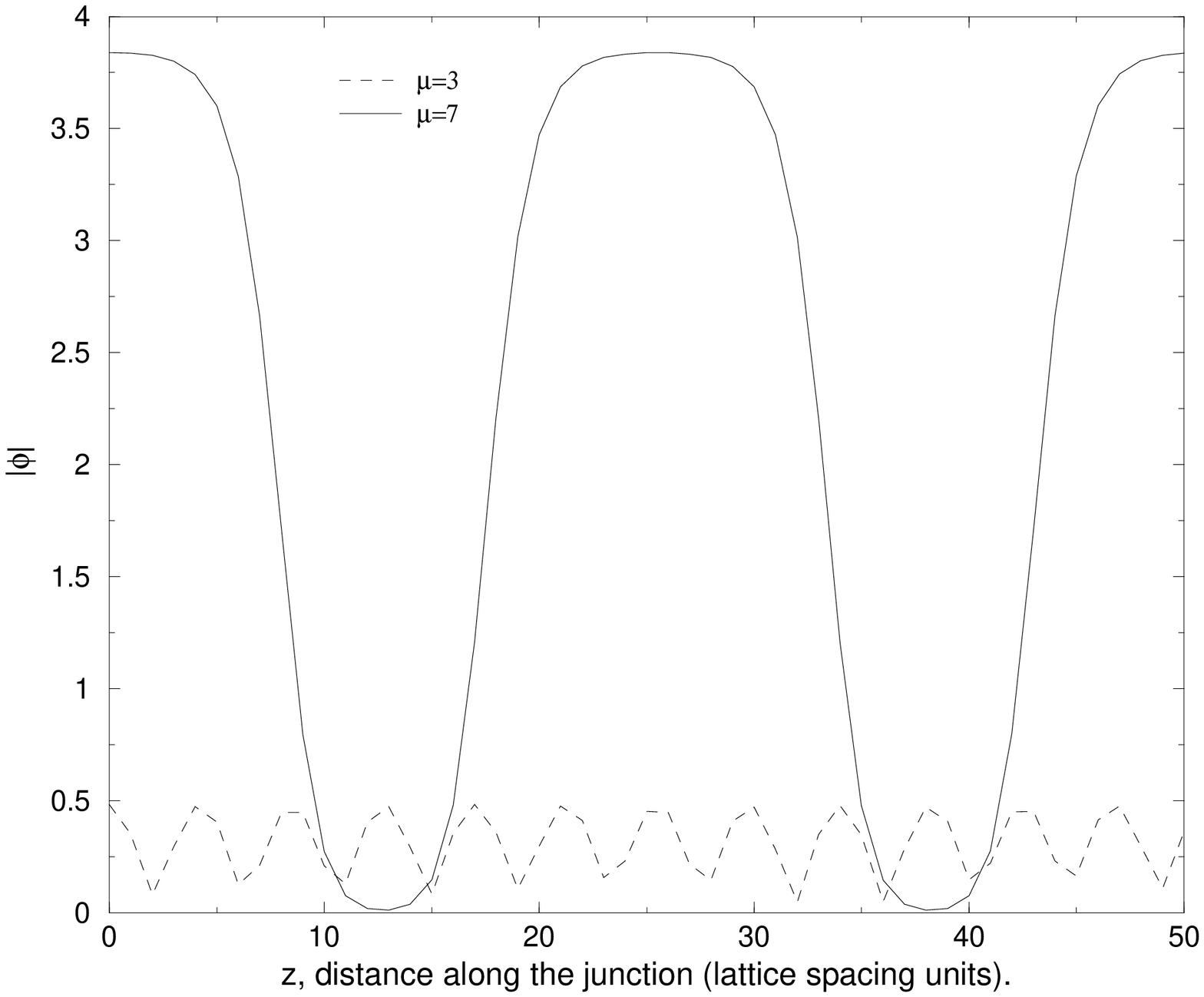}\qquad &
\includegraphics[angle=-90,scale=0.39]{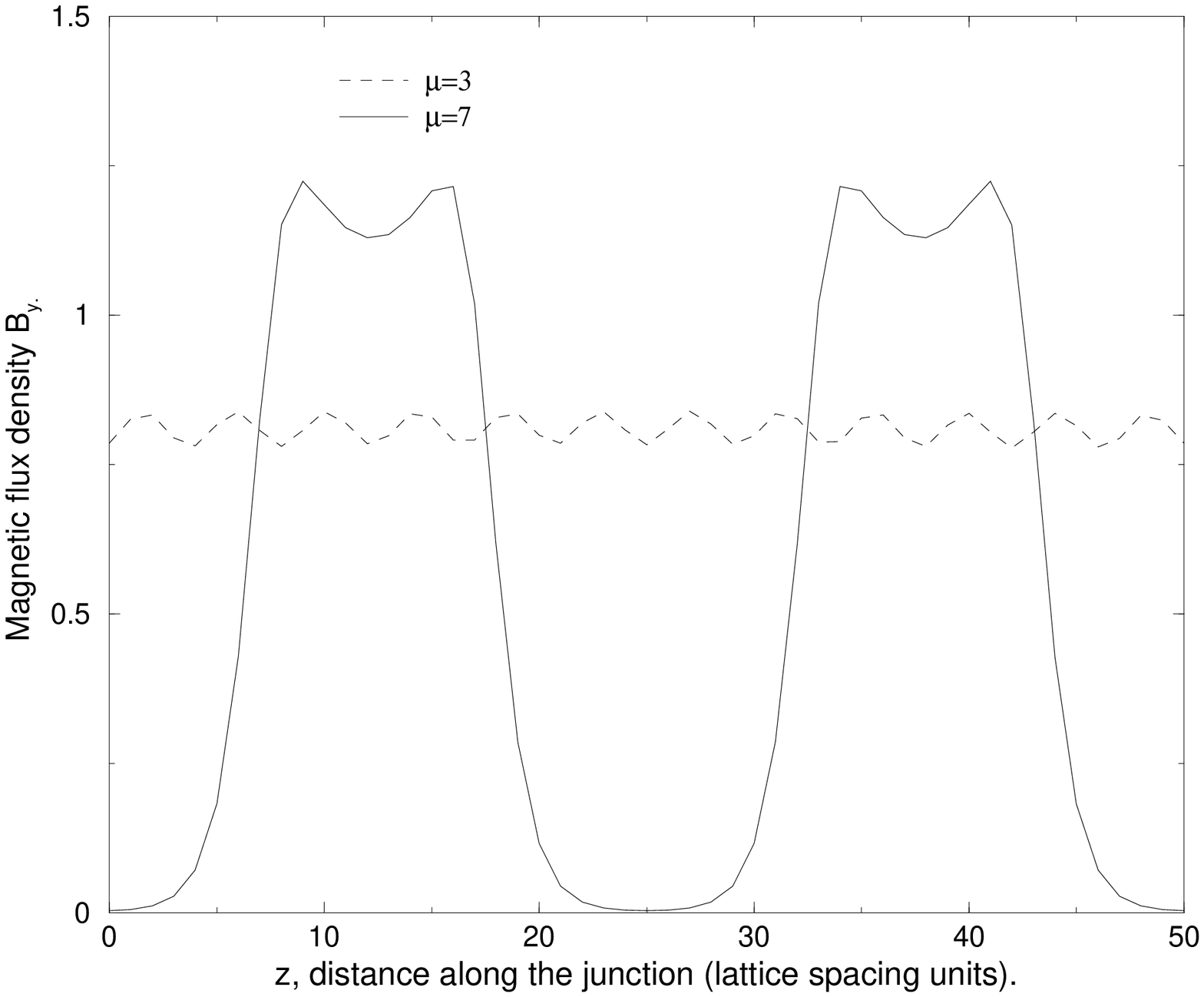}\cr
$(a)$&$(b)$
\end{tabular}
\end{center}
\caption{\footnotesize{Superconducting film simulation.}
\footnotesize{(a) Meta-stable profile of the scalar field along
the junction at the meridional plane of the film $y=0$.}
\footnotesize{(b) Meta-stable profile of the magnetic flux density
along the strip at $x=l_{x}/2$. $\Phi_{E.M.}=24\pi/e$. For
$\mu=3$, twelve minima in $|\phi|$ and twelve maxima in $B_{y}$
coexist, though no vortex forms. For $\mu=7$, two vortices of
winding six form with the magnetic flux profile of equation
(\ref{g6}).}}\label{METAST}
\end{figure*}
\section{Formation of spontaneous topological defects\label{spontaneous}}
In previous sections we have shown how local vortices and cosmic
strings form in the collision of bubbles by trapping of an
external magnetic flux. A similar mechanism applies to the
formation of topological defects when, instead of an external
magnetic field, magnetic flux is provided in the form of thermal
fluctuations. One can think of the magnetic field long wavelength
fluctuations as a uniform external magnetic field.\\ \indent The
actual magnetic field fluctuations and nucleation of the actual
bubbles need of a Monte Carlo simulation like the one performed in
\cite{Donaire05}. The analytical computation is a difficult task.
In the finite-temperature field theory a first order phase
transition effective potential of the form (\ref{a3}) comes from
radiative corrections to the zero-temperature potential. In the
particular case of the $U(1)$-invariant renormalizable theory,
\begin{equation}\label{h000}
\mathcal{L}_{T=0}=-\frac{1}{4}F_{\mu\nu}F^{\mu\nu}+D_{\mu}\phi
D^{\mu}\phi^{*} +\lambda(\phi\phi^{*} - \eta^{2})^{2},\qquad\eta ,\:
\lambda\quad\textrm{constants.}
\end{equation}
The first order transition effective potential has been calculated
in the type-I regime ($\lambda/e^{2}<1$) in several contexts. Some
of them are concerned with quantum corrections \cite{Coleman73}
while others deal with thermal perturbations
\cite{Kirzhnits74,Dolan74,Halperin74,Arttu03,Larkin05}. Because
$e^{2}>\lambda$, radiative corrections due to the gauge field are
dominant. The nucleation rate and the bubble structure are
affected by the interaction of the fluctuation of both the scalar
field and the gauge field \cite{Strumia99,Moore01}. In particular,
this implies that the bounce solution and the nucleation rate as
calculated in \cite{Coleman77,Linde83} are only approximations. We
will therefore simplify the problem by considering that the
nucleation rate $\Gamma$, the bubble expansion rate $v$ and the
typical bubble nucleation radius $R_{N}$ are free parameters which
must be ultimately given by the actual field theory.\\ \indent
Thus, the finite-temperature vacuum manifold consists of the set
of minima of a potential like
$-\lambda(\phi\phi^{*}-\eta^{2})^{2}$ radiatively corrected.
 The equilibrium state corresponds to the Gibbs ensemble.
Thermal fluctuations of the gauge field are strongly suppressed by
Meissner effect in the broken phase where the photon is heavy.
 They are however present in the large areas of symmetric phase
between bubbles. Therefore, with regards to the production of
spontaneous magnetic field, we will concentrate on the
long-wavelength fluctuations of $A_{i}$.
\subsection{Estimation of the typical winding number of thermal
topological defects\label{typical}} We will review here the
arguments and predictions presented in \cite{Rajantie02} and
\cite{Donaire05}.\\ \indent Let us assume that the dynamics of the
whole system is adiabatic. Therefore, bubbles expand slowly enough
such that the magnetic field can always be considered in thermal
equilibrium in the symmetric phase areas between bubbles.
 Thus, the spectrum of the wavelength fluctuations corresponds to the
long wavelength spectrum obtained in the thermal ensemble where
the saddle point approximation is accurate enough given that the
fluctuations are nearly gaussian. The spectrum  is so the
two-point correlation function of the gauge field in the symmetric
phase. We have already calculated the functional form of that
spectrum in the context of the zero-temperature field theory in
section \ref{sc}. In the thermal ensemble it is the Boltzmann
factor $\exp{(-F/T)}$ what determines the weight of a
configuration. Therefore, the spectrum of fluctuations is the
Green's function of the equation of $A_{i}$ for an extreme of the
function $F/T$, that is:
$\nabla_{n}^{2}G_{ij}=T\delta^{n}(r-r')\delta_{ij}$.
\footnote{This is a good approximation as long as only long wave
length fluctuations are considered. It corresponds to the
classical Rayleigh-Jeans spectrum of balckbody radiation and so
valid for $|\vec{k}|<T$. In the case in hands we are considering
$1/|\vec{k}|\gtrsim 2/m_{h}\gtrsim 2/T$. For the Lagrangian
(\ref{h000}), the strength of the last inequality depends on the
coupling $\lambda$ and the v.e.v. of the scalar field.} The label
$n$ denotes the number of space dimensions and so the functional
form of the magnetic field fluctuations depends on the spatial
dimensionality:
\begin{equation}\label{gg1}
\langle B(k)B(q)\rangle^{2D}\approx T(2\pi)^{2}\delta^{2}(k-q),
\end{equation}
\begin{equation}\label{gg2}
\langle B_{y}(\vec{k})\: B_{y}(\vec{q})\rangle^{SC}\approx
T(2\pi)^{2}\delta^{2}(\vec{k}-\vec{q})\frac{1}{2}|\vec{k}|,
\end{equation}
\begin{equation}\label{gg3}
\langle B_{i}(k)\: B_{j}(q)\rangle^{3D}\approx
T(2\pi)^{3}\delta^{3}(k+q)(\delta_{ij}-\frac{k_{i}k_{j}}{k^{2}}).
\end{equation}
Formulas (\ref{gg1}), (\ref{gg2}) and (\ref{gg3}) correspond to the
fully two-dimensional, superconducting and fully three-dimensional
models respectively. The superconducting one (\ref{gg2}) was derived
in \cite{Kibble03} following similar steps to those we followed in
the computation of (\ref{POT2}) in the context of the
zero-temperature theory. Again, top-arrows label two-dimensional
vectors living in a superconductor film which is embedded in a three
dimensional space and $B_{y}$ stands for the component of the
magnetic field orthogonal to the film. Generally, we will write the
formulas above as
\begin{equation}\label{h0}
\langle B_{i}(k)B_{j}(q)\rangle^{m} = T
(2\pi)^{n}\delta^{n}(k+q)G^{m}_{ij}(k),
\end{equation}
where the label $m$ denotes the model to which the function
corresponds and the indices $i$, $j$ and the integer $n$ take
their values accordingly.\\ \indent As bubbles expand the magnetic
flux associated to those fluctuations is expelled by Meissner
effect. When several bubbles meet, a simply connected region of
symmetric phase arises at the center of the collision. The
magnetic flux in that region gets trapped by the mechanism
explained in section \ref{gauge} and a thick vortex or heavy
cosmic string form. The typical amount of flux trapped can be
estimated under adiabatic conditions. As the free energy $F^{m}$
is quadratic in the magnetic field components $B_{i}(r)$,
 the magnetic energy $F^{m}_{mag}/T= \frac{1}{2T}\int
B_{i}(r)B^{i}(r)d^{n}r$ can be written in thermal equilibrium as
\begin{equation}\label{h01}
F^{m}_{mag}= \frac{1}{2}\int d^{n}r \int d^{n}r'
B^{i}(r)(G^{m})^{-1}_{ij}(r,r')B^{j}(r'),
\end{equation}
where $(G^{m})^{-1}_{ij}(r,r')$ is the inverse of $G^{m}_{ij}(k)$
defined in  (\ref{h0}) and so the two-point interaction potential.
In the fully two-dimensional model the thermal fluctuations are
not correlated, $(G^{2D})^{-1}(r,r')=\delta^{2}(r-r')$. In that
case, the equilibrium value of (\ref{h01}) can be exactly computed
on a surface of size $S$ in function of the total magnetic flux
$\Phi_{E.M.}$. That is
\begin{equation}\label{h02}
F^{2D,S}_{mag} =\frac{\Phi_{E.M.}^{2}}{2S}.
\end{equation}
In the superconducting model (SC) and the fully-three dimensional
model, because the thermal fluctuations are correlated, the integral
in (\ref{h01}) does depend on the shape of the surface $S$.
Nevertheless the magnetic energy is a quadratic function of the
total flux in both cases:
\begin{eqnarray}\label{h03}
F^{3D,S}_{mag} \propto\frac{\Phi_{E.M.}^{2}}{2\sqrt{S}},\\
F^{SC,S}_{mag} \propto\frac{\Phi_{E.M.}^{2}}{2\sqrt{S}}.
\end{eqnarray}
The factors of proportionality which are missing are geometrical
coefficients of order one.\\ \indent As the free energy is
quadratic in the magnetic flux in thermal equilibrium, the thermal
fluctuations of $\Phi_{E.M.}$ adjust to the gaussian distribution
$\sim \exp{(-\frac{\Phi^{2}_{E.M.}}{2TS})}$ in two-dimensions and
to $\sim \exp{(-\frac{\Phi^{2}_{E.M.}}{2T\sqrt{S}})}$ in three
dimensions. Correspondingly, the typical value of the thermal flux
in a surface of area $S$ is the root-mean-square of $\Phi_{E.M.}$:
\begin{equation}\label{h04}
\Phi_{E.M.}^{rms,2D}=\sqrt{TS},
\end{equation}
\begin{equation}\label{h05}
\Phi_{E.M.}^{rms,3D}\propto\sqrt{T}S^{1/4}.
\end{equation}
The typical size $S$ of a region of symmetric phase enclosed by
bubbles can be estimated using dimensional arguments. The
characteristic length of that region must be of the order of the
bubble radius. If the transition is strong first order, the radius
is determined by the nucleation rate $\Gamma$ and the expansion
rate $v$. $S\sim (v/\Gamma)^{2/3}$ holds in the fully
two-dimensional and the superconducting models while $S\sim
(v/\Gamma)^{1/2}$ does in the fully three-dimensional model.
\\ \indent Making use of the estimations above we can compute the
typical winding number of vortices and cosmic strings in each
case:
\begin{equation}\label{h07}
N_{w,typ.}^{2D} \propto\frac{e}{2\pi}T^{1/2}(v/\Gamma)^{1/3},
\end{equation}
\begin{equation}\label{h08}
N_{w,typ.}^{SC} \propto\frac{e}{2\pi}T^{1/2}(v/\Gamma)^{1/6},
\end{equation}
\begin{equation}\label{h09}
N_{w,typ.}^{3D} \propto\frac{e}{2\pi}T^{1/2}(v/\Gamma)^{1/8}.
\end{equation}
A variety of windings is expected in the actual distribution. The
properties of the winding numbers distribution function are to be
reported in \cite{Me2}.
\subsection{Flux trapping vs. Kibble mechanism\label{vs}}
As it was mentioned in section \ref{gauge} (see also section
\ref{geodesic}), Kibble mechanism can be used in the form
explained there to explain the formation of local topological
defects in absence of magnetic field. On the other hand, because
the typical winding predicted by Kibble mechanism is of the order
of the square root of the number of bubbles involved in a typical
collision, it must be of order one. In particular, if the width of
the bubble wall is negligible, it is exactly one. That is so
because the probability for more than three bubbles to collide at
a single point is zero. Therefore the trapping of thermal flux
will be dominant over Kibble mechanism in the formation of
vortices as $N_{w,typ.}>1$. In order to estimate when that is so
we have to find out a relation between the parameters involved in
the equations (\ref{h07}), (\ref{h08}) and (\ref{h09}). First, we
will assume that the electromagnetic coupling $e$ is of order one.
That is a reasonable assumption in both GUTs and
electromagnetism.\footnote[1]{It is however still missing in this
calculation the effect of the magnetic permeability in the
superconducting model.} Then, we have to compare the value of the
temperature $T$ to the value of some power of $\frac{v}{\Gamma}$
which is equivalent to some power of the radius of the bubbles. As
mentioned at the beginning of this section, $\Gamma$ is difficult
to estimate and so is $v$. We will take a more conservative
approach. If the transition is not weak $\Gamma$ is exponentially
suppressed either by $\exp{[-F(\phi_{T})/T]}$ if the transition is
temperature driven or by $\exp{[-S_{E}(\phi_{0})]}$ if it is a
quantum transition. $S_{E}(\phi_{0})$ stands for the Euclidean
action of the bounce solution at zero-temperature while
$F(\phi_{T})$ does for the free energy of the bounce solution at
finite-temperature $T$. Also, the size of the bubbles when they
meet must be larger than or equal to the initial nucleation radius
at the transition temperature $R_{N}(T)$. Following Linde's
arguments \cite{Linde83},
\begin{eqnarray}\label{h10}
\fl\Gamma & \sim & \exp{[-F(\phi_{T})/T]}\quad\textrm{if}\quad
F(\phi_{T})/T<S_{E}(\phi_{0}),\\
\fl\Gamma & \sim & \exp{[-S_{E}(\phi_{0})]}\qquad\textrm{if}\quad
S_{E}(\phi_{0})<F(\phi_{T})/T,\quad R_{N}(0)<(2T)^{-1}.
\end{eqnarray}
If the transition is temperature driven, $R_{N}\geq T^{-1}$.
Therefore, the above conditions together with our estimates in
equations (\ref{h07}), (\ref{h08}) and (\ref{h09}) imply that
$N_{w}>1$ is guaranteed if the transition is strong first order
and temperature-driven. For the theory in (\ref{h000}) that is the
case in the couplings range $e^{2}>\lambda>e^{4}$. Otherwise not
only the nucleation process but also the bubble expansion rate
must be considered.
\section{Josephson-like currents at the junctions and the geodesic rule in gauge theories}
\subsection{The gauge-invariant phase difference\label{phase}}
In section \ref{choice} it was argued that it is $D_{\mu}\theta$
the field to be tracked in the gauge theory in order to describe
the formation of topological defects.\\ \indent The definition of
$D_{\mu}\theta$ in (\ref{a40}) can be inferred from the
application of the differential operator $D_{\mu}$ onto the
product $\exp{(i\theta)}\cdot|\phi|$. Following the Leibnitz's
rule,
\begin{equation}\label{D1}
D_{\mu}(\exp{(i\theta)}\cdot|\phi|)=\exp{(i\theta)}D_{\mu}|\phi| +
i\exp{(i\theta)}|\phi|D_{\mu}\theta.
\end{equation}
On the other hand, the $LHS$ of (\ref{D1}) can also be written as
\begin{equation}\label{D2}
D_{\mu}(\exp{(i\theta)}\cdot|\phi|) =
(\partial_{\mu}+ieA_{\mu})|\phi|\exp{(i\theta)}.
\end{equation}
By expanding the $RHS$ of (\ref{D2}) and using the fact that the
covariant derivative of a neutral function like $|\phi|$ is
equivalent to its ordinary derivative in (\ref{D1}), identity
(\ref{a40}) holds.\\ \indent The local observables can be written
in terms of the gauge-invariant quantities $|\phi|$ and
$D_{\mu}\theta$ as:
\begin{eqnarray}
\textrm{Cooper pairs density}\qquad & |\phi|^{2}, &\\
\textrm{magnetic field} &\vec{B} = &
(1/e)\:\vec{\nabla}\times\vec{D}\theta,\\ \textrm{electric field}
&\vec{E} = & (1/e)\:(\partial_{0}\vec{D}\theta -
\vec{\nabla}D_{0}\theta),\\ \textrm{charge density} & q_{e} = &
2e|\phi|^{2}D_{0}\theta,\\ \textrm{electrostatic potential } &
V_{el.} = & (1/e)\:D_{0}\theta,\\ \textrm{Noether current density}
& \vec{j} = & -2e|\phi|^{2}\vec{D}\theta,
\end{eqnarray}
where in this case arrows denote spatial components vectors.\\
\indent Likewise, as $\partial_{\mu}\partial_{\nu}\theta =
\partial_{\nu}\partial_{\mu}\theta$, it is possible to write down
the equations of motion in terms of $|\phi|$ and $D_{\mu}\theta$
alone:
\begin{equation}\label{a10}
[\partial^{2}_{t} + \sigma\partial_{t}-\nabla^{2}-D^{\mu}\theta
D_{\mu}\theta+ \frac{\delta V(\phi)}{\delta \phi^{2}}]|\phi|=0,
\end{equation}
\begin{equation}\label{a11}
\partial^{\nu}\partial_{\nu}D_{\mu}\theta-\partial^{\nu}\partial_{\mu}D_{\nu}\theta +
2e^{2}|\phi|^{2}D_{\mu}\theta + \sigma\partial_{t} D_{i}\theta = 0,
\end{equation}
where in the last equation we can identify $\:e
\partial^{\nu}F_{\nu\mu}=\partial^{\nu}\partial_{\nu}D_{\mu}\theta-\partial^{\nu}\partial_{\mu}D_{\nu}\theta\:$ and
$\:-e(j_{\mu} + j_{i}^{ohm})=2e^{2}|\phi|^{2}D_{\mu}\theta +
\sigma\partial_{t} D_{i}\theta$.\\ \indent In the global theory
$D_{\mu}\theta$ reduces to $\partial_{\mu}\theta$ and the phase
difference $\int_{a}^{b}\partial_{x}\theta\:
dx=\alpha_{B}-\alpha_{A}\equiv\Delta\alpha_{ab}$ is meaningful. In
the gauge theory, when no magnetic field is present and the gauge
$A_{x}=0$ can be chosen, $\gamma_{ab}=\int_{a}^{b}D_{x}\theta\:
dx=\Delta\alpha_{ab}$ as well. In this case, the only reason for
the existence of non-zero $D_{\mu}\theta$ is the spatial overlap
between two different vacuum domains $A$ and $B$ as it occurs in
the junction between two colliding bubbles. In absence of magnetic
field, $D_{\mu}\theta$
 can be interpreted as an induced \emph{spontaneous connection} as
 a result of the coupling of the light ('fast') degrees of
freedom living in the junction to the heavy ('slow') degrees of
freedom living deep in the broken phase where
$m_{\gamma}=\sqrt{2}e|\phi|_{min}$. Even in the actual gauge
theory this spontaneous connection cannot be gauged away. One can
nevertheless perform the gauge transformation
\begin{eqnarray}\label{ddd5}
A_{x} & \rightarrow & (1/e)\:\partial_{x}\theta,\nonumber\\
\partial_{x}\theta & \rightarrow & 0
\end{eqnarray}
in such a way that $D_{x}\theta$ remains invariant and so does
$\gamma_{ab}$.\\ \indent If a magnetic field $B_{i}$ does exist
there is an extra divergenceless piece of connection induced by
$B_{i}$. It is always possible to write $D_{i}\theta$ as a sum of
an exact derivative $\partial_{i}\theta$ plus a divergenceless
component $eA_{i}$ in the Coulomb gauge:
\begin{equation}\label{ddd11}
D_{i}\theta=\partial_{i}\theta + eA_{i}\quad\textrm{such that
$\partial^{i}A_{i}=0$}.
\end{equation}
As $dD\theta=edA=eF$, $\partial_{i}\theta$ is the locally flat
component of the connection while $A_{i}$ carries the curvature in
the gauge (\ref{ddd11}).\\ \indent In summary, the local
connection $D_{i}\theta$ is composed of two parts:\\ $\bullet$A
spontaneous, flat component $\partial_{i}\theta$ which, in the
Coulomb gauge, carries the longitudinal modes of $D_{i}\theta$.
Its origin is in the breakdown of the $U(1)$ symmetry. In the
global theory, its interaction with the neutral scalar particles
gives rise to the geodesic rule.\\ $\bullet$An induced, curved
component $eA_{i}$ which, in the Coulomb gauge, carries the
transverse modes of $D_{i}\theta$. Its origin is in a net magnetic
field. In the gauge theory, its interaction with the charged
scalar particles gives rise to the Aharonov-Bohm effect.\\ \indent
Both the local quantities $\partial_{i}\theta$ and $eA_{i}$ and
their integrals $\int \partial_{i}\theta\:dx^{i}$ and $e\int
A_{i}\:dx^{i}$ come to be respectively equal (up to a sign) deep
in the broken phase where $D_{i}\theta=0$. As one approaches the
center of a vortex of winding $N_{w}$ at $r=0$, $D_{i}\theta$ is
not zero any more for $r\lesssim r_{v}\approx
\sqrt{N_{w}}\lambda_{L}$, where the magnetic flux is confined. The
holonomy of any loop $C$ of radius $r$ enclosing the center of the
vortex, $\oint_{C}\partial_{i}\theta\:dx^{i}=2\pi N_{w}$, does
however remain unaltered as $r$ goes to zero  by continuity. It is
the discrepancy between the holonomy and the magnetic flux in the
circle of radius $r_{v}$ what gives rise to the closed steady
screening currents in the core of the vortex.

\subsection{DC Josephson-like effect across bubbles junctions and flux strips \label{joseph}}
To our knowledge  Weinberg   was the first to show the periodicity
of the Josephson currents using field theory arguments
\cite{Weinberg86}. This will be our starting point to study the
electric current traversing the junction between two bubbles as
they collide.\\ \indent We will consider stationary situations and
so neglect time oscillations, i.e. $D_{0}\theta\approx 0$,
$\partial_{t}|\phi|\approx 0$. Gauge invariance implies that,
assuming no gradients along the junction between two
superconducting pieces separated by a gap of width $|b-a|$, the
value of the current self-interaction energy through the junction
is an even function of $\gamma_{ab}=\int_{a}^{b}D_{x}\theta\: dx$:
\begin{equation}\label{F}
\varepsilon_{ab}=\sum^{\infty}_{m=0}c_{m}\cos{(m\gamma_{ab})},
\end{equation}
where $\{c_{m}\}$ are constants which incorporate the degrees of
freedom of $|\phi|$ in $a<x<b$. If  stationary conditions hold,
the current density is \cite{Weinberg86}
\begin{equation}\label{j}
j_{ab}^{x}=
-e\frac{\partial\varepsilon_{ab}}{\partial\gamma_{ab}}.
\end{equation}
If the current flow is parametrised with the coordinate $x$, a
uniform steady current turns up in $a<x<b$ \cite{Me} if linear
superposition of the scalar fields is valid, that is, if
$2/m_{h}\ll |b-a|$. Uniformity of $j_{x}$ along the integration
path $a<x<b$ implies that the main contribution to formulas
(\ref{F},\ref{j}) comes from the terms $m=0$ and $m=1$. Uniformity
of the reduced charge-carriers density $|\phi|_{0}^{2}$ implies
that $c_{0}\approx c_{1}\approx -2|\phi|_{0}^{2}/\chi$. Therefore,
under these conditions, (\ref{F}) and (\ref{j}) read
respectively\footnote{A more precise computation of this equation
can be found in \cite{Me}.}
\begin{equation}\label{var1}
\varepsilon_{ab}=\frac{-2|\phi|_{0}^{2}}{\chi}[\cos{(\gamma_{ab})}
- 1],
\end{equation}
\begin{equation}\label{var2}
j_{ab}^{x} = \frac{-2e}{\chi}|\phi|_{0}^{2}\sin{(\gamma_{ab})},
\end{equation}
where $|\phi|_{0}$ is given in (\ref{phinot}). The potential
(\ref{var1}) is a Josephson coupling and (\ref{var2}) is
 a direct Josephson-like current though its origin is not in
quantum-tunneling \cite{Ambegaokar63}. Its amplitude,
\begin{equation}\label{nnn1}
j_{c}\equiv\frac{2e}{\chi}|\phi|_{0}^{2},
\end{equation}
is instead a function of the reduced charge-carriers density and
the length scale $\chi$.\\ \indent In the actual collision of two
superconducting bubbles or blocks,  steady and uniformity
conditions approximately hold at the middle of the junction. The
charge carriers density $|\phi|^{2}$ is roughly uniform at the
very middle of the gap, $x=0$, where $|\phi|$ presents a minimum
and so $\partial_{x}|\phi|_{x=0}=0$ (figure \ref{fig11}(b)). The
non-zero value of $D_{x}\theta$ concentrates also in a very narrow
strip along the junction with center at $x=0$ and width
$\chi$.\footnote[3]{In (\ref{gordon}), $\chi$ is what the authors
of \cite{Kibble95} identify with the width of the 'wave packet'
$D_{x}\theta$. In \cite{Me} it is shown that $\chi$ is of the
order of $m_{h}^{-1}$.} When $\partial_{z}|\phi|$ is negligible
the current flows across the junction in the $x$-direction. Figure
\ref{fig9}(a) shows the shape of the transverse current density
$j_{ab}^{x}$ in an actual simulation. It is in good agreement with
expression (\ref{var2}). The current is measured at the middle of
the gap ($x=0$) between the two superconducting blocks of figure
\ref{fig199}. There is a magnetic flux of value $6\pi/e$ uniformly
distributed in the gap so that the gradient of $\gamma_{ab}$ is
uniform in the direction $z$ along the junction. Figure
\ref{fig9}(b) shows the corresponding current self-interaction
energy as a function of $\gamma_{ab}$ normalized to
$\frac{2}{\chi}|\phi|_{0}^{2}=1$.\\
\begin{figure*}
\begin{center}
\begin{tabular}{cc}
\includegraphics[angle=-90,scale=0.39]{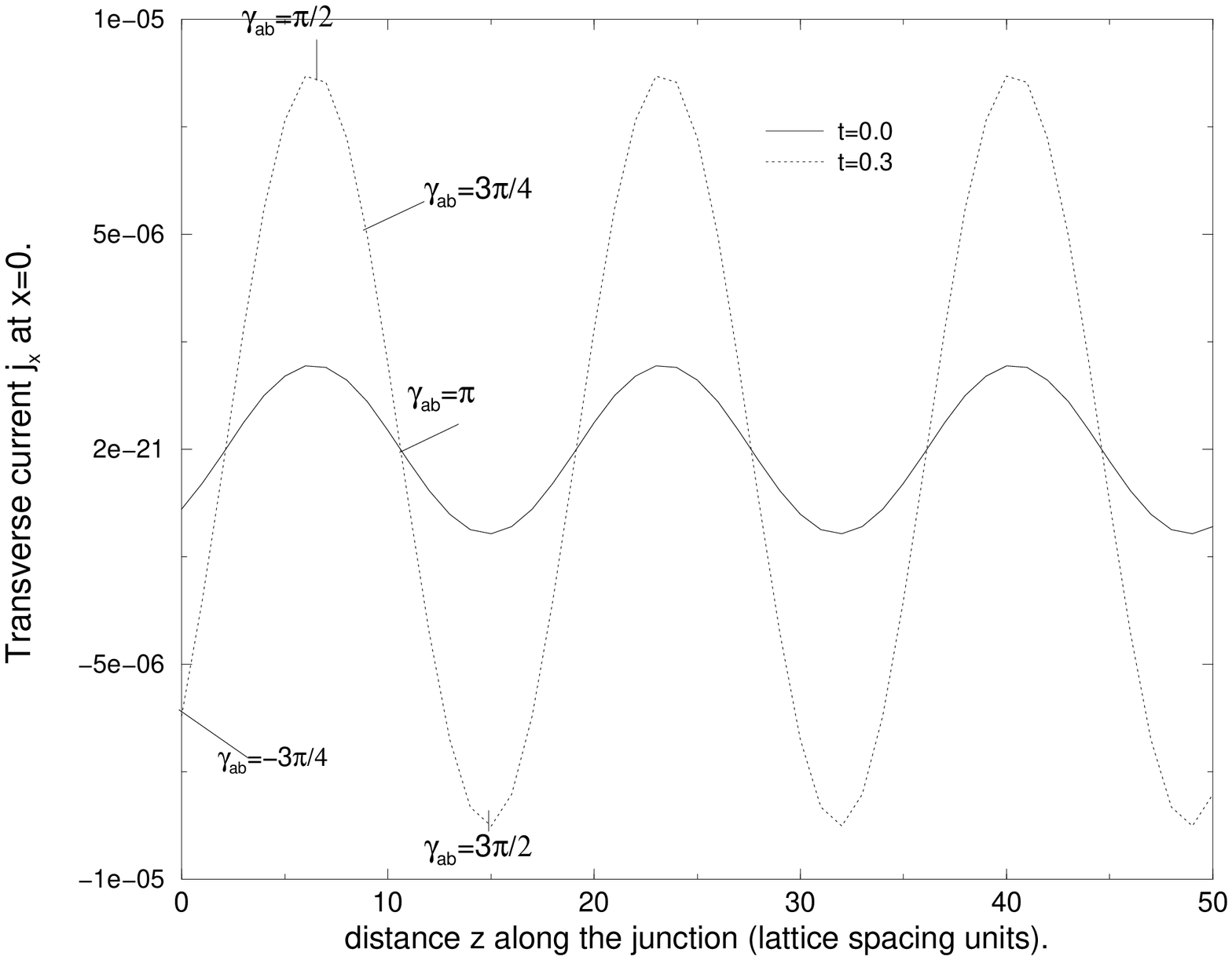}\qquad &
\includegraphics[angle=-90,scale=0.39]{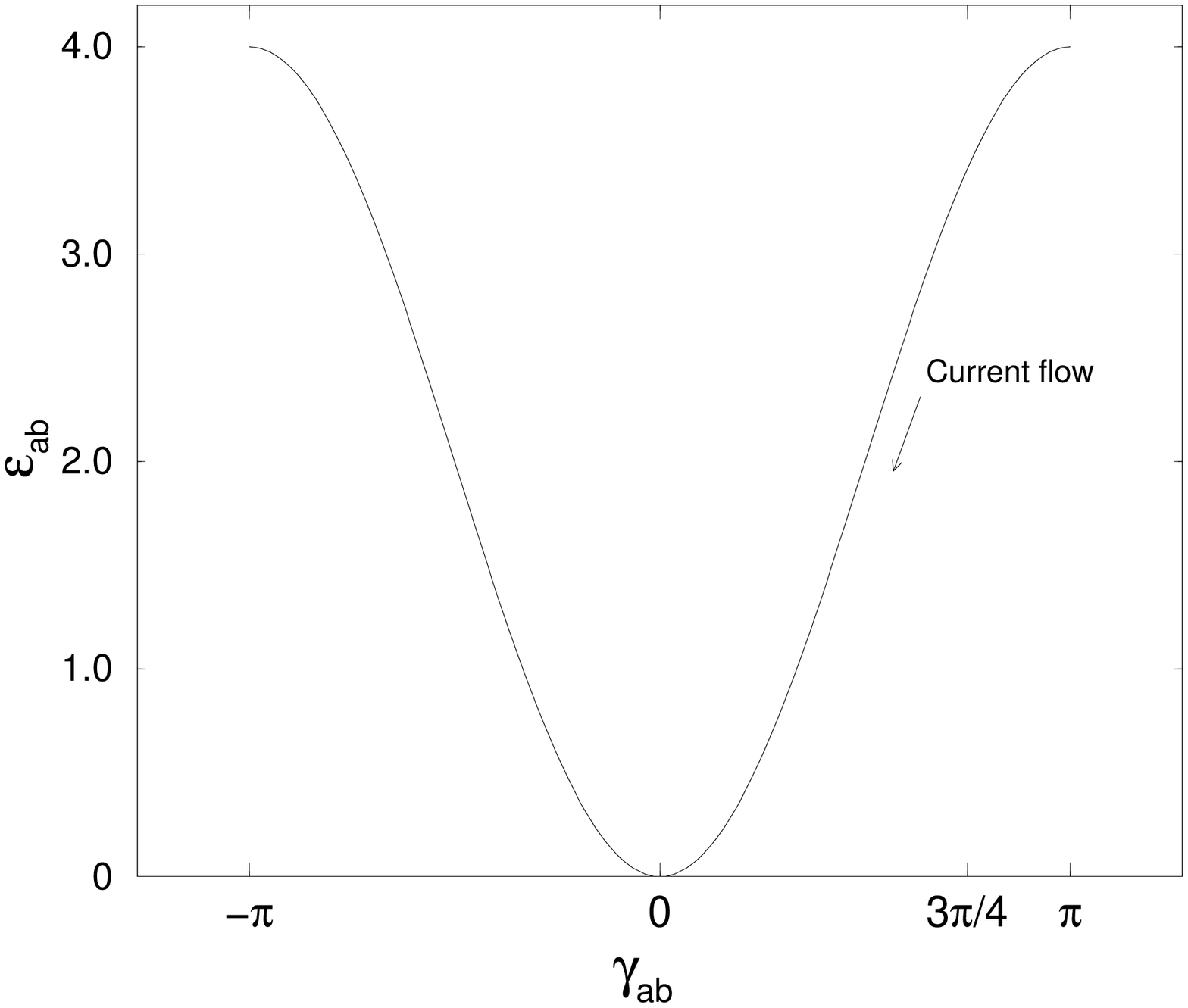}\cr
$(a)$&$(b)$
\end{tabular}
\end{center}
\caption{\footnotesize{Fully two-dimensional model simulation. (a)
Transverse current density $j_{x}$ along the junction at the
middle of the gap between two superconducting blocks. The setup
corresponds to that of figure \ref{fig199} with
$\alpha_{b}-\alpha_{a}=-3\pi/2$. Time $t$ is set to zero at the
moment the blocks get in touch. The current amplitude grows with
time as the blocks overlap and $|\phi|_{x=0}$ increases. The
current profile adjust to that in (\ref{var2}).} \footnotesize{(b)
Current self-interaction energy $\varepsilon_{ab}$ as a function
of $\gamma_{ab}$ (equation (\ref{var1})). $\varepsilon_{ab}$ is
normalized to $\frac{2}{\chi}|\phi|_{0}^{2}=1$.}}\label{fig9}
\end{figure*}
\indent An interesting scenario in which an actual steady
superconducting current can be found is a meta-stable flux strip.
In general, neither the density of charge carriers nor the
magnetic flux are uniform in a meta-stable strip like the one in
figure \ref{METAST} with $\mu=3$. The gradients of $|\phi|$ and
$D_{x}\theta$ across the strip make equations
(\ref{var1},\ref{var2}) inapplicable. Instead, the general
expression (\ref{F}) must be used. In addition, the gradients
along the strip must be incorporated in the coefficients of the
sum. Therefore, $\varepsilon_{ab}(z)$ takes the general form
\begin{equation}\label{ez}
\varepsilon_{ab}(z)=\sum_{m=0}^{\infty}C_{m}(z)\cos{[m\gamma_{ab}(z)]}.
\end{equation}
Correspondingly, the transverse current density is
\begin{equation}\label{curri1}
j_{ab}^{x}(z)=e\sum_{m=1}^{\infty}m\:C_{m}(z)\sin{[m\gamma_{ab}(z)]}.
\end{equation}
The functions $\{C_{m}(z)\}$ are not completely arbitrary.  All
the physical quantities are periodic along a meta-stable strip. If
the length of the strip is $L_{z}$ and it contains approximately
$N_{w}$ winding units of magnetic flux, their period is $\Delta
z\approx L_{z}/N_{w}$. Consequently, both the functions
$\{C_{m}(z)\}$ and $\gamma_{ab}(z)$ have period $\Delta z$ along
the junction (modulo $2\pi$ in the case of $\gamma_{ab}$). In
terms of the total magnetic flux $\Phi$, $\Delta z=\frac{2\pi
L_{z}}{e\Phi}$.\\ \indent A first consequence of the current
density distribution (\ref{curri1}) along a flux strip is an
interference pattern of the transverse total current different to
that in an actual Josephson junction. Taking into account the
symmetry and periodicity properties of the functions in
$j_{ab}^{x}(z)$, it can be formally integrated  along the junction
of a meta-stable strip in a superconductor film of thickness $\mu$
and total flux $\Phi$. That is
\begin{eqnarray}\label{MSQUID}
I_{ab}^{x}& = &
\int_{-L_{z}/2}^{L_{z}/2}\mu\:j_{ab}^{x}(z)\:dz\nonumber\\ & = &
\sum_{m=1}^{\infty}C'_{m}\sin{(m\Delta\alpha_{ab})}\Bigl[\frac{\sin{(m
e\Phi/2\pi)}}{m e\Phi/2\pi}\Bigr],
\end{eqnarray}
where the coefficients ${C'_{m}}$ are constants and
$\Delta\alpha_{ab}$ stands for the spontaneous phase difference.
The thickness $\mu$ is taken large enough such that $j_{ab}^{x}$
is approximately uniform in the direction normal to the film.
Figure \ref{mSQUID} shows the functional form of the first three
harmonics.\\ \indent The meta-stability of a flux strip
 requires the gap width $|b-a|$ to
be relatively broad. Broad means in the first place that $|b-a|$
is greater than $1/m_{h}$ so that the edges of the strip are
approximately straight and the linear magnetic flux density is
nearly uniform. Broad also implies that
$\int_{\chi}\partial_{x}|\phi|\:dx\ll
\chi^{-1}\int_{\chi}|\phi|\:dx$, $\chi\ll|b-a|$. If these
conditions are satisfied the density of charge carriers is nearly
uniform in the range of interest. Therefore the first harmonics
$m=0$, $m=1$ are the dominant terms in the sums
(\ref{ez}-\ref{MSQUID}).\\
\begin{figure}[htp!]
\begin{center}
\includegraphics[angle=-90,scale=0.46]{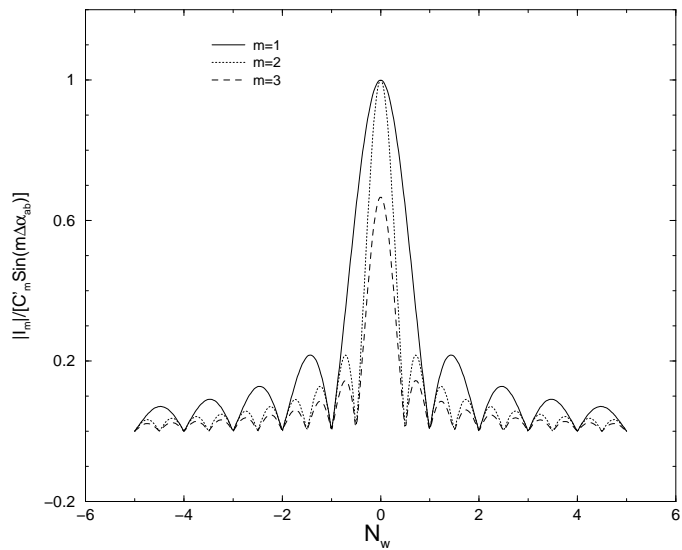}
\end{center}
\caption{\footnotesize{Interference pattern of the current across
a flux strip. $I_{m}$ denotes the $m^{th}$ term of the series
(\ref{MSQUID}).}}\label{mSQUID}
\end{figure}
\indent Because the total current only vanishes for integral
values of the total flux in units of $2\pi/e$, there must be a net
current passing through the gap during the process of formation of
a meta-stable flux strip. That total current stops when the  flux
in the junction gets quantized. It is worth noting that the
remaining current density $j^{x}_{ab}(z)$ along the strip does not
radiate because of
 topological reasons. It is the flux quantization rather than the existence of a mass gap
 what prevents this current from dissipating.\\ \indent The agreement between the
 lattice simulation and
the Josephson-like current in (\ref{var2}) deserves some comments.
At the moment the blocks get in contact
$\int_{-\chi/2}^{\chi/2}D_{x}\theta\: dx$ is of order $\pi$ and,
as we mentioned in section \ref{trapping}, the dynamics of
$\gamma_{ab}$ is governed by a sine-Gordon equation. In the
numerical simulation, this apparent problem is overcome naturally
with the use of the non-compact $U(1)$ lattice formulation
\cite{Kajantie}. As described in \ref{appendB}, it is the use of
the link field $\mathbf{U}_{i}=\exp\left(ie\delta x
\mathbf{A}_{i}\right)$ that incorporates the non-linear effects.
Hence, the Josephson-like coupling $-\frac{2}{\delta
x^2}\sum_{i}[{\rm Re}\phi_{(\mathbf x)}^* \mathbf{U}_{i,(\mathbf
x)}\phi_{(\mathbf x+\hat{i})}-|\phi_{(\mathbf x)}|^2]$ in
(\ref{Aa2}) and the sinusoidal functions of figure \ref{fig9}(a).
\subsection{The geodesic rule in gauge theories\label{geodesic}}
In \cite{Rudaz93} it was argued that the geodesic rule in a gauge
theory holds by equilibrium arguments, i.e. as a result of the
minimization of the energy. In \cite{Hindmarsh94} it was shown that
if no magnetic field is present the geodesic rule holds in the gauge
theory by dynamical arguments, i.e. as a result of the equations of
motion. A more detailed study of the bubble collision dynamics was
carried out in \cite{Kibble95} in terms of gauge-invariant
quantities. We will show with the aid of numerical simulations that
both energetic \cite{Rudaz93} and dynamical \cite{Kibble95}
arguments hold and agree with  the geodesic-magnetic rule presented
in section \ref{gauge}. The global geodesic rule as proved in
\cite{Hindmarsh94} is only satisfied if some particular
prescription for the gauge choice can be given (see below).\\
\indent Let us consider two superconducting bubbles $A$ and $B$
facing each other. Once more, if the radius of the bubbles is much
larger than the bubble-wall thickness the contact area can be well
approximated by two superconducting blocks with a gap in between.
Block $A$ is nucleated at the point $\alpha_{A}$ while block $B$ is
at the point $\alpha_{B}$ of the vacuum manifold. The gap in between
is initially in the symmetric phase and no magnetic flux at all is
in it. Let us take as an example
$\Delta\alpha_{ab}=\alpha_{B}-\alpha_{A}=3\pi/4$ and let the two
blocks expand and coalesce.
\begin{figure*}
\begin{center}
\begin{tabular}{cc}
\includegraphics[angle=-90,scale=0.37]{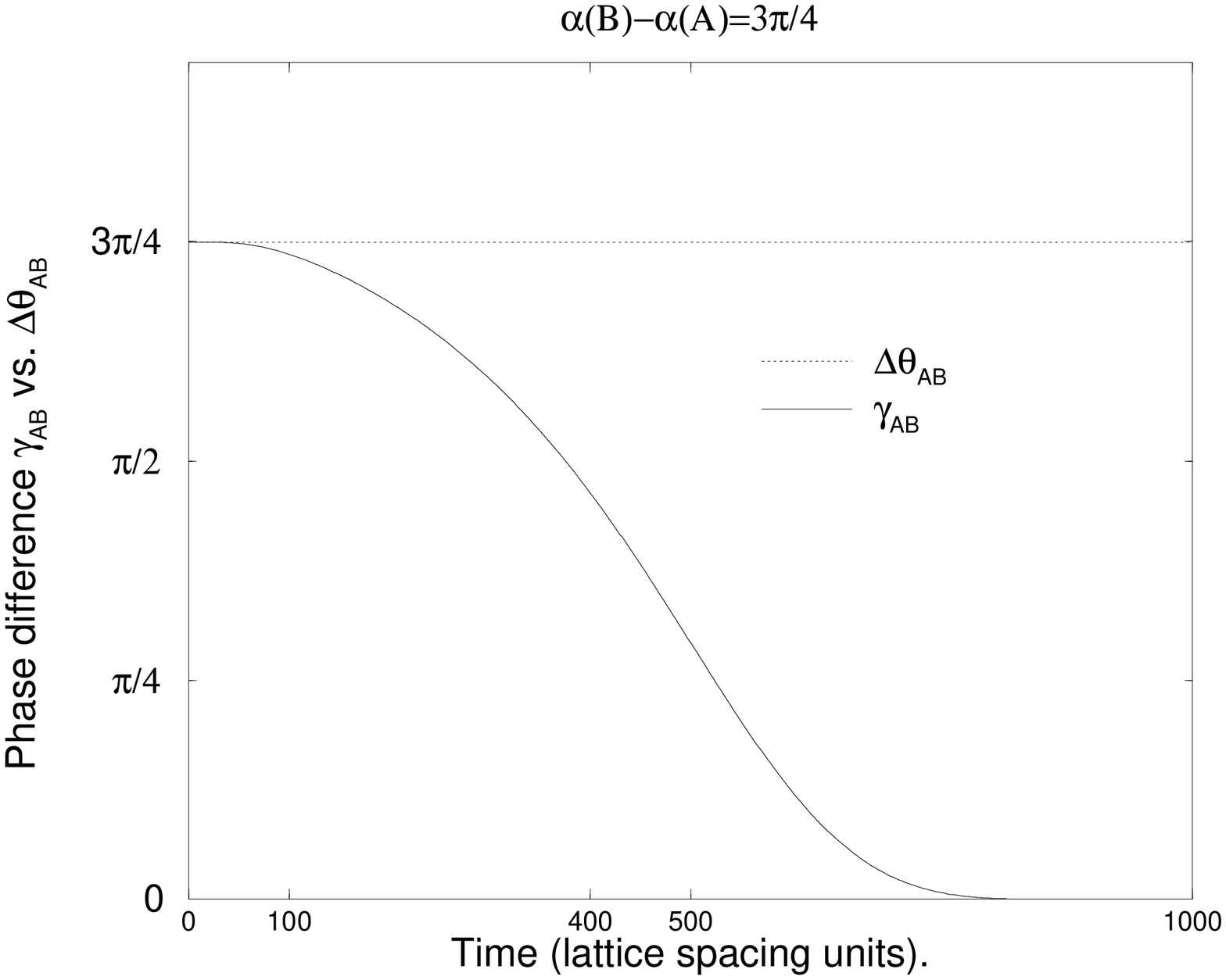}\qquad &
\includegraphics[angle=-90,scale=0.37]{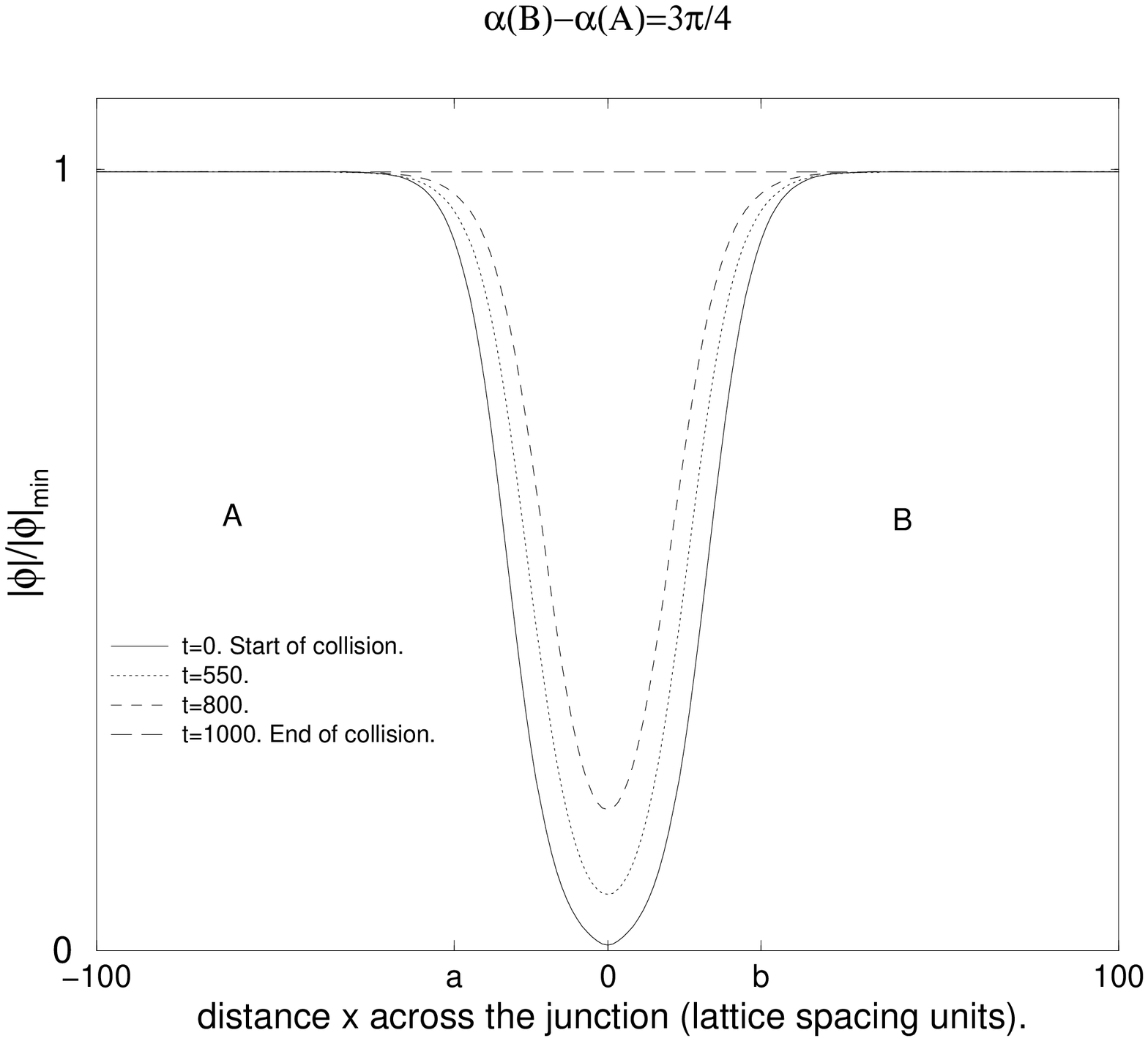}\cr
$(a)$&$(b)$
\end{tabular}
\end{center}
\caption{\footnotesize{Fully two-dimensional model simulation.}
\footnotesize{(a) Time evolution of the gauge-invariant phase
difference $\gamma_{ab}$ through the junction $a<x<b$ between two
superconducting blocks with spontaneous phase difference
$\Delta\alpha_{ab}=3\pi/4$.} \footnotesize{(b) Time evolution of
the scalar field profile $|\phi|$ across the junction
$a<x<b$.}}\label{fig11}
\end{figure*}
Figure \ref{fig11}(a) shows the time evolution of both $\gamma_{ab}$
and $\Delta\theta_{ab}$ across the junction $a<x<b$. In agreement
with the geodesic-magnetic rule, $\gamma_{ab}$ decreases in time
until vanishing. In the contrary, $\Delta\theta_{ab}$ remains
constant and equal to $3\pi/4$. The complex phase $\theta$ does
interpolate between $0$ and $3\pi/4$ as stated by the geodesic rule
because the gauge $A_{\mu}=0$ has been chosen initially. However, at
the moment the blocks meet we could have made the following gauge
transformation everywhere:
\begin{eqnarray}\label{ddd50}
A_{x} & \rightarrow & (1/e)\:\partial_{x}\theta,\nonumber\\
\partial_{x}\theta & \rightarrow & 0.
\end{eqnarray}
The complex phase would have been uniform in both the blocks and
the junction. In that case the evolution of $\gamma_{ab}$ would
have still been the one shown in figure \ref{fig11}(a) while
$\Delta\theta_{ab}$ would have remained equal to zero. This shows
that the physical modes are those of $D_{x}\theta$, neither the
Goldstone modes $\partial_{x}\theta$ nor the longitudinal modes of
the gauge field $A_{x}$. When the temporal gauge $A_{0}$ is
chosen, the evolution of $\partial_{x}\theta$ and $A_{x}$ just
depends on the initial fixing of the remaining gauge degrees of
freedom. Thus, the geodesic rule as applied in the global theory
is a gauge artifact in the gauge theory. It holds in the gauge
theory when the gauge choice $A_{\mu}=0$ can be made
\cite{Hindmarsh94}, which is only possible in the absence of a
magnetic field. One can however consider that in the absence of a
magnetic field topological defects form by Kibble mechanism. After
all, in the gauge theory it is also the spontaneous breakdown of
the symmetry that gives rise to the $D_{i}\theta$ modes.\\ \indent
In energetic terms what is happening is that, as bubbles coalesce,
the current tends to vanish as $\gamma_{ab}$ goes to zero. The
current could also vanish if $\gamma_{ab}$ evolved towards $\pi$.
However $\pi$ corresponds to a maximum and so unstable point of
the current self-interaction energy $\varepsilon_{ab}$ (figure
\ref{fig9}(b)). The current self-interaction potential
$\varepsilon_{ab}$ does decrease as $\gamma_{ab}$ evolves toward
its minimum at $\gamma_{ab}=0$.\\ \indent In dynamical terms, the
Maxwell equation (\ref{a11}) becomes a generalized Klein-Gordon
equation for $D_{\mu}\theta$ in the gauge
\begin{equation}\label{ddd15}
\partial^{\nu}D_{\nu}\theta=\textrm{const.},
\end{equation}
\begin{equation}\label{ddd16}
[\partial^{\nu}\partial_{\nu} + \sigma\partial_{t} +
2e^{2}|\phi|^{2}]D_{\mu}\theta=0.
\end{equation}
Integrating the resultant equation for $D_{x}\theta$ across the
junction (using the rough approximation $|\phi|\sim$ uniform), the
equation for $\gamma_{ab}$ is
\begin{equation}\label{ddd17}
[\partial_{tt} + \sigma\partial_{t} +
2e^{2}|\phi|^{2}]\gamma_{ab}=0.
\end{equation}
\begin{figure}[htp!]
\begin{center}
\includegraphics[angle=-90,scale=0.46]{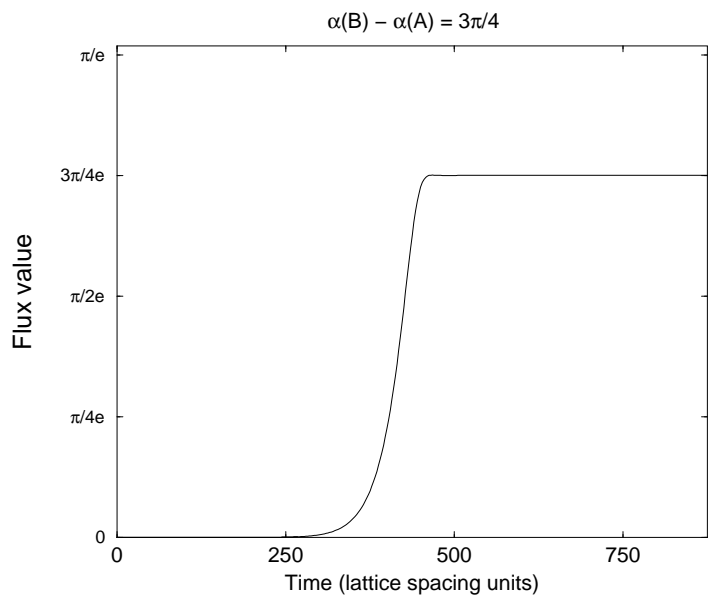}
\end{center}
\caption{\footnotesize{Fully two-dimensional model simulation. Time
evolution of the magnetic flux associated to the fluxon ring around
the collision axis of two bubbles. The spontaneous phase difference
between the bubbles is $\Delta\alpha_{ab}=3\pi/4$.}}\label{fig15}
\end{figure}
The study of the time-evolution of $\gamma_{ab}$ was carried out
in detail in \cite{Kibble95} taking $|\phi|\approx|\phi|_{min}$,
$\partial_{x} |\phi|\approx 0$. However, because the typical phase
difference between any two bubbles in a multi-bubble collision is
of the order of $\pi$ for a vortex to form, (\ref{ddd17}) is not a
good approximation. It is instead the sine-Gordon equation
(\ref{gordon}) what should be used in the first stages of the
collision. It is only when $\gamma_{ab}$ relaxes to
$\gamma_{ab}\ll \pi$ that (\ref{ddd17}) is valid and $\gamma_{ab}$
goes roughly as
$\sim(\alpha_{B}-\alpha_{A})e^{-\frac{2e^{2}|\phi|^{2}}{\sigma}t}$.
The numerical simulation of figure \ref{fig11}(a) so
confirms.\footnote[2]{In the simulation of figure \ref{fig11}(a)
the dynamics of $\gamma_{ab}$ is overdamped so that oscillations
in time are suppressed.}\\ \indent Figure \ref{fig15} shows the
evolution of the magnetic fluxon ring generated around the
collision axis when two bubbles meet with
$\alpha_{B}-\alpha_{A}=3\pi/4$. Because the collision there is
soft,\footnote[4]{If the collision is violent, the highly
non-equilibrium dynamics of the scalar field can lead to the
formation of unit-winding vortex-anti vortex pairs
\cite{Copelandpp} giving rise to an apparent violation of the
geodesic rule.} the fluxon can spread out in the directions
orthogonal to the collision axis. That is the reason for its
constant asymptotic value.
\section{Conclusions and outlook}
In the Abelian Higgs model we have shown that it is the local
dynamics of the gauge-invariant field
$D_{i}\theta=\partial_{i}\theta + eA_{i}$ and the global dynamics
of the gauge-invariant phase difference
$\gamma=\int\:D_{i}\theta\:dx^{i}$ that determine the formation of
vortices in superconductors and hypothetical cosmic strings in
grand unified theories.\\ \indent In the mean-field theory
approximation we have found that, when a first order phase
transition takes place in an external magnetic field, a variety of
objects of different winding form. These are meta-stable strips
and membranes of magnetic flux, chains of single-winding vortices,
thick vortices and heavy strings. Thick vortices turn out to be
stable in thin superconducting films up to some critical winding
value. The cause of the stability and/or meta-stability of
vortices and strips is a Coulomb-like long-range magnetic
interaction induced by the gauge field which propagates out of the
film. It might determine the order of the phase transition of a
thin superconducting film. A Monte Carlo simulation similar to the
one in \cite{Donaire05} with an external magnetic field is needed
to investigate this matter.\\ \indent The spontaneous breakdown of
the $U(1)$ gauge symmetry manifests in the location of the center
of mass of single-winding vortex chains and finite size Abrikosov
lattices.\\ \indent Under steady conditions we have derived a
direct Josephson-like current traversing the meta-stable flux
strips. For that calculation we have invoked non-local effects
within a length scale $\chi$ which we identify with the
correlation length of $D_{i}\theta$.\\ \indent Thermal
long-wavelength fluctuations of the gauge fields get confined in a
variety of winding topological defects in a first order phase
transition. In the particular case of the fully three-dimensional
Abelian Higgs model, a multi-tension cosmic string network arise
\cite{Donaire05}. The typical winding of the strings is $\sim
\frac{e}{2\pi}T^{1/2}(v/\Gamma)^{1/8}$ if the transition is strong
first order, $e^{2}>\lambda>e^{4}$. For a thin superconducting
film the prediction is $\sim
\frac{e}{2\pi}T^{1/2}(v/\Gamma)^{1/6}$. Even though the latter is
not easy to test experimentally, it is our belief that trapping of
an external magnetic flux in the form of strips, vortex chains and
thick vortices can be achieved experimentally by inducing
nucleation of bubbles in a type-I superconducting film sample.\\
\indent In the context of brane-world cosmology and extra
dimensions phenomenology, it would be interesting to study the
hypothetical non-Abelian currents of grand unified theories
traversing a $3+1$ dimensional brane (with or without fluxes) in
between two bulk
disconnected -- or weakly overlapped -- vacuum domains.
\ack I would like to thank very specially Arttu Rajantie for his
useful advice, discussion and corrections throughout all this
paper. I thank Tom Girard for his comments on section \ref{sc} and
discussion about experimental issues in superconductivity. I am
also grateful to the referee for pointing out an important error
in Section $4$.
\appendix
\section{Breakdown of the translational invariance of the setup in figure \ref{fig199}\label{appendA}}
Let us consider the fully two-dimensional theory and take two
parallel superconducting blocks  separated by a finite gap as in
figure \ref{fig199}. The blocks are semi-infinite in the
transverse direction $x$ and infinite along $z$. A non-zero
magnetic field exists in the gap, $x\in[-d/2,d/2]$, and it is
invariant along the $z$-axis. Before the blocks get in contact,
neglecting the tension of the walls, the initial equilibrium
configuration for the magnetic flux corresponds to that of a
uniform magnetic field $B$. If either the walls tension were
considered or we were working in the framework of the
superconducting model, translation invariance along the junction
would also hold but the magnetic field strength would show a
gradient across the gap.\\ We aim to show that, regardless of the
choice of gauge: a) local translation invariance along the
junction does hold as long as a finite width gap in which
$|\phi|=0$ $\forall z$ exists between the two blocks, b) local
translation invariance breaks down as the superconducting blocks
get in contact. Condition a) is  necessary for both SSB and for
the stability of the straight strip in figure \ref{fig19}(a).
Condition b) is  necessary  for the formation of the structures
presented in figures \ref{fig19}(b) and \ref{fig19}(c).\\\\
\emph{Proof of part a)}.\\\\ The initial conditions of the local
physical quantities are:\\ $1)$ Translation-invariance along the
$z$ direction:
\begin{eqnarray}\label{A1''}
\partial_{z}|\phi|_{t=0} & = & 0\quad\forall x,\\
\partial_{z}B|_{t=0} & = & 0\quad\forall x.
\end{eqnarray}
The above equations also imply:
\begin{eqnarray}\label{A1'}
\partial_{z}j^{i}|_{t=0} & = & 0,\quad
\textrm{at $x$ such that $\partial_{z}|\phi|_{x,t=0}=0$},\nonumber\\
\partial_{z}D^{i}\theta|_{t=0} & = & 0,\quad
\textrm{at $x$ such that $\partial_{z}|\phi|_{x,t=0}=0$ and
$|\phi|>0$}.
\end{eqnarray}
$2)$ The system is initially at rest:
\begin{eqnarray}\label{A2}
\partial_{t}\phi|_{t=0} & = & 0\quad\forall x,\nonumber\\
E_{i}|_{t=0} & = & 0\quad\forall x,\nonumber\\
\partial_{t}j^{i}|_{t=0} & = & 0\quad\forall x.
\end{eqnarray}
$3)$ Asymptotic vacuum state conditions:
\begin{eqnarray}\label{A4}
|\phi|_{t=0} & = & |\phi|_{min}\quad\textrm{for $x\leq -d/2$ or $x\geq d/2$,}\nonumber\\
D_{i}\theta|_{t=0} & = & 0,\qquad\:\textrm{for $x\leq -d/2$ or
$x\geq d/2$}.
\end{eqnarray}
$4)$ Existence of a narrow gap of symmetric phase at the center:
$x\in(-\epsilon, \epsilon)$ $\forall z$, $0<\epsilon\ll d$.
\begin{equation}\label{A4'}
|\phi|_{t=0}  =  0,\quad \textrm{for $-\epsilon<x<\epsilon$}.
\end{equation}
Bearing in mind that all the fields are evaluated at $t=0$, we
will drop the subscript $t=0$. Breakdown of translational
invariance along $z$ requires that
\begin{equation}\label{A3}
\partial_{z}(\partial^{2}_{t}|\phi(x,z)|)\neq 0\quad\textrm{for some
$x\in[-d/2,d/2]$}.
\end{equation}
The equation of motion for $|\phi|$ in the interval $x\in[-d/2,d/2]$
 at $t=0$ subject to the above conditions is
\begin{equation}\label{A5}
[\partial^{2}_{t}-\nabla^{2} + D_{x}^{2}\theta + D_{z}^{2}\theta+
\frac{\delta V(\phi)}{\delta \phi^{2}}]|\phi|=0.
\end{equation}
Taking the derivative of (\ref{A5}) along the direction $z$,
\begin{equation}\label{A6}
\partial_{z}\partial^{2}_{t}|\phi| -
\partial^{2}_{x}\partial_{z}|\phi|- \partial^{3}_{z}|\phi| +
\frac{1}{2}\partial_{z}\frac{\delta V(|\phi|)}{\delta |\phi|} +
\partial_{z}(D_{x}^{2}\theta + D_{z}^{2}\theta)|\phi|=0.
\end{equation}
The second, third and fourth terms in the $LHS$ of equation
(\ref{A6}) vanish according to (\ref{A1''}) and we are left with
\begin{equation}\label{A7}
\partial_{z}\partial^{2}_{t}|\phi|=-\partial_{z}[(D_{x}^{2}\theta +
D_{z}^{2}\theta)|\phi|].
\end{equation}
In order to satisfy the inequality (\ref{A3}), the $RHS$ of
(\ref{A7}) must be different to zero at some point $x\in[-d/2,d/2]$.
The Maxwell equation (\ref{a11}) in the gauge of equation
(\ref{ddd15}) reads
\begin{equation}\label{A8}
[(\partial_{x}^{2} + \partial_{z}^{2}) -
m_{\gamma}^{2}(x)]D_{i}\theta = 0,\qquad \textrm{where}\quad
m_{\gamma}^{2}(x)\equiv 2e^{2}|\phi(x)|^{2}.
\end{equation}
Let us write the $O(2)$ massive field $D_{i}\theta$ in terms of the
 polar coordinates $\varrho$ and $\psi$:
\begin{eqnarray}\label{A10}
\varrho^{2} & = & D_{x}^{2}\theta + D_{z}^{2}\theta,\nonumber\\
D_{x}\theta & = & \varrho \cos{(\psi)},\nonumber\\ D_{z}\theta & =
& \varrho\sin{(\psi)},
\end{eqnarray}
so that the inequality (\ref{A3}) implies
\begin{eqnarray}\label{A14}
\partial_{z}\partial^{2}_{t}|\phi|& \neq & 0\nonumber\\ \Rightarrow
-\partial_{z}(\varrho^{2}|\phi|) & \neq & 0 \nonumber\\
\Rightarrow -2|\phi|\varrho\:\partial_{z}\varrho & \neq & 0.
\end{eqnarray}
In (\ref{A14}), $\varrho$ does not vanish in the junction because
there is a magnetic field. The value of $|\phi(x)|$ equals $0$ in
$x\in(-\epsilon, \epsilon)$ according to (\ref{A4'}). $z$-invariance
of both $|\phi|$ and $B$ implies that $\partial_{z}\varrho=0$ if
$|\phi|\neq 0$ (equations (\ref{A1''}), (\ref{A1'})). Hence, the
inequality (\ref{A14}) does not hold and $z$-invariance remains as
long as $\epsilon>0$. Figures \ref{fig20}(a) and \ref{fig20}(c)
illustrate this situation. It
is the case for a separation between blocks $d\gg 2/m_{h}$.\\\\
\emph{Proof of part b)}.\\\\ Let us assume now that the blocks are
not separated but they overlap instead in a region of width
$2\epsilon\ll d$. Thus, $|\phi|>0$ everywhere but, at most, at a
finite number of points in the junction. Likewise the existence of a
net magnetic field in the junction implies that $\varrho>0$ in
$-d/2<x<d/2$. Therefore, in order to satisfy (\ref{A14}) we are left
with the proof of $\partial_{z}\varrho\neq 0$ at some
$x\in[-d/2,d/2]$. $z$-invariance holds at least in the range
$x<-\epsilon,\: x>\epsilon $ and therefore (\ref{A1''}) reads now
\begin{equation}\label{A1'''}
\partial_{z}|\phi|_{t=0}  =  0\quad x<-\epsilon,\: x>\epsilon.
\end{equation}
We aim to show that, if a $z$-invariant magnetic field is present
in the junction, $z$-invariance of  $|\phi|$ in the interval
$x\in(-\epsilon, \epsilon)$ is not possible.
 $z-$invariance of the magnetic field can be expressed as a
constraint equation for $D_{i}\theta$,
\begin{equation}\label{A9}
\partial_{z}^{2}D_{x}\theta = \partial_{z}\partial_{x}D_{z}\theta,
\end{equation}
and (\ref{A8}) in the variables (\ref{A10}) reads
\begin{equation}\label{A13}
[\partial_{x}^{2} + \partial_{z}^{2} -(\partial_{x}\psi)^{2} -
(\partial_{z}\psi)^{2} - m_{\gamma}^2(x)]\varrho = 0.
\end{equation}
Let us write (\ref{A13}) as
\begin{equation}\label{A15}
[\partial_{x}^{2} + \partial_{z}^{2} - m_{\gamma}^{'2}(x)]\varrho =
0,
\end{equation}
where we have defined
$m_{\gamma}^{'2}(x)\equiv(\partial_{x}\psi)^{2}
+(\partial_{z}\psi)^{2} + m_{\gamma}^{2}(x)$, and let us integrate
(\ref{A15}) from $x\rightarrow -\infty$ to $x\rightarrow +\infty$.
Because $\varrho$  falls deep inside the superconducting blocks as
$x$ goes to $\pm\infty$,
\begin{equation}\label{b17}
\int _{-\infty}^{+\infty}\partial_{x}^{2}\varrho\: dx =
\partial_{x}\varrho|^{+\infty}_{-\infty}= 0.
\end{equation}
Therefore,
\begin{equation}\label{b18}
\int _{-\infty}^{+\infty}\partial_{z}^{2}\varrho\: dx =\int
_{-\infty}^{+\infty}m_{\gamma}^{'2}(x)\varrho\: dx.
\end{equation}
The $LHS$ of (\ref{b18}) is greater than zero because the
integrand of the $RHS$, $m_{\gamma}^{'2}(x)\varrho$, is positive
definite and both $m_{\gamma}^{'2}(x)$ and $\varrho$ are greater
than zero at some point in the gap for there is a net magnetic
field and blocks overlap. That implies that, correspondingly,
there must be at least some point $x_{0}$ in the interval
$[-d/2,d/2]$ at which $\partial_{z}^{2}\varrho_{|x_{0}}> 0$. If
this is so and $\partial_{z}\varrho_{|x_{0}} \neq 0$ as well, our
proof is done and breakdown of the translational invariance of
both $|\phi|$ and $\varrho$ is shown. If however
$\partial_{z}\varrho_{|x_{0}} = 0$, then $x_{0}$ is a critical
point. Because $\varrho$ is gauge-invariant it is continuous in
the broken phase. Therefore there must be a neighborhood centered
at $x_{0}$ denoted by $\Omega _{x_{0}}$ in which
$\partial_{z}\varrho_{|\Omega _{x_{0}}} \neq 0$. Thus, the proof
is complete.$\Box$\\ \indent Figure \ref{fig20} shows the
breakdown of the $z$-invariance of
 $|\phi|$ and  $j^{2}=j_{x}^{2}+j_{z}^{2}$ as the two blocks get in
contact and there is a magnetic flux of value $6\pi/e$ in the
junction.
\begin{figure}[htp!]
\begin{center}
\includegraphics[scale=0.74]{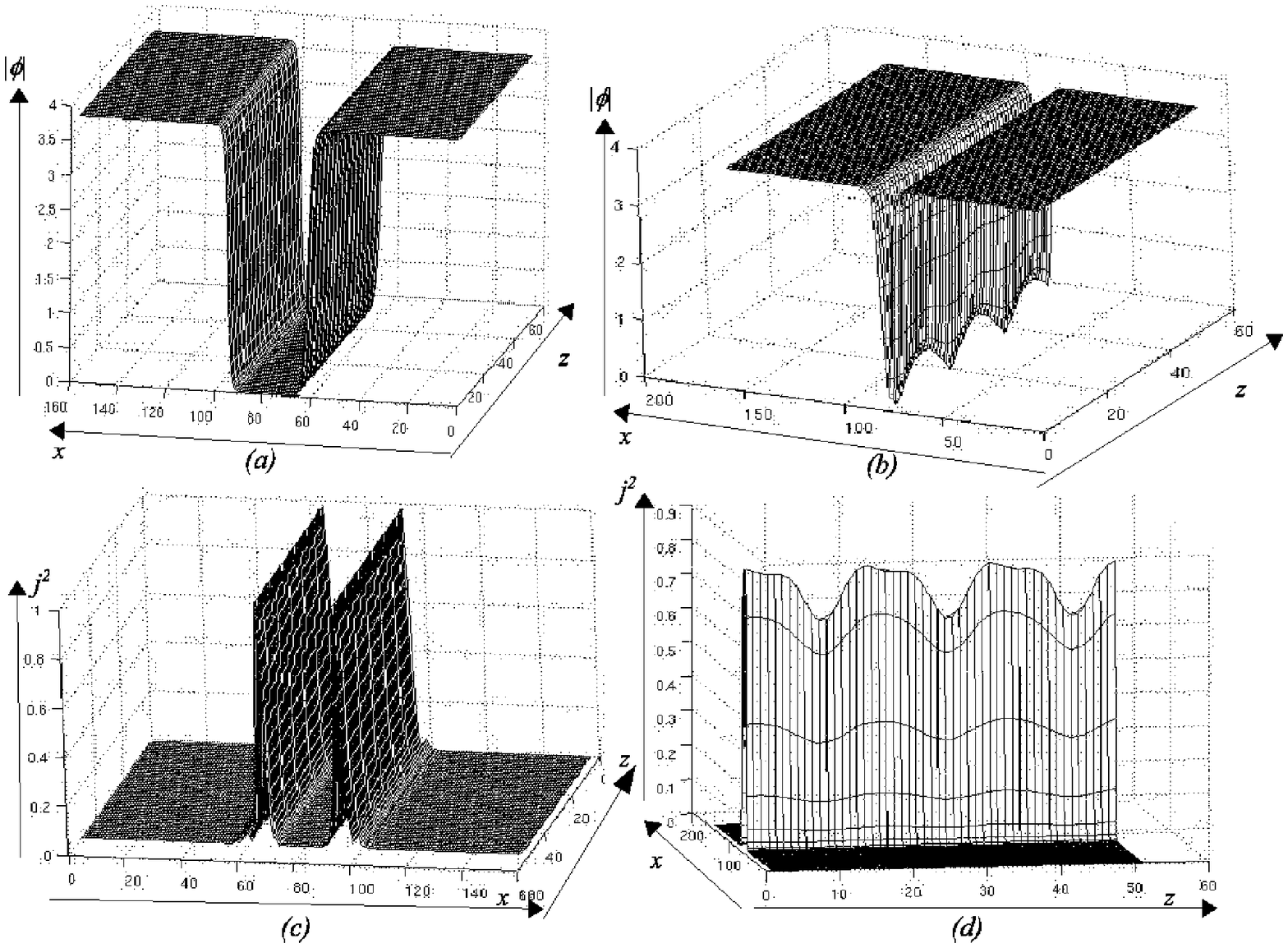}
\end{center}
\caption{\footnotesize{Fully two-dimensional model simulation. (a)
$t=0$. Initial profile of $|\phi|$ according to the setup of figure
(\ref{fig199}). Invariance along $z$ holds. (b) $t=30$.
$z$-invariance breaks down as blocks meet ($d\approx 2/m_{h}$) and
three minima ($=e\Phi_{E.M.}/2\pi$) show up. (c) $t=0$. Initial
profile of the square of the current density $j^{2}\equiv
j_{x}^{2}+j_{z}^{2}$. Invariance along $z$ holds. The current
density concentrates at the edges of the blocks where the magnetic
field initially migrates due to small structure effects. (d) $t=30$.
$z$-invariance breaks down as blocks meet. $j^{2}$ gets concentrated
around the minima of $|\phi|$.}}\label{fig20}
\end{figure}
\section{Lattice disretization\label{appendB}}
To write down the discretized equations of motion, we define the
forward and backward derivatives
\begin{equation}
\Delta^\pm_i f_{({\mathbf x})}= \pm\delta x^{-1}\left(f_{({\mathbf
x}\pm\hat{i})}- f_{({\mathbf x})}\right),
\end{equation}
where $\delta x$ is the lattice spacing, $\hat{i}$ is a unit
vector in the $i$ direction and vectors are denoted by bold-faced
letters. Similarly, we define the time derivative
\begin{equation}
\Delta_t f_{(t)} =\delta t^{-1}\left(f_{(t)}-f_{(t-\delta
t)}\right),
\end{equation}
where $\delta t$ is the time step. Using the link variable
\begin{equation}
\mathbf{U}_{i}=\exp\left(ie\delta x \mathbf{A}_{i}\right),
\end{equation}
we also define the corresponding covariant derivatives in the
superconductor bulk
\begin{eqnarray}
\mathbf{D}^+_i\phi_{(\mathbf{x})}&=&\delta x^{-1}\left(
\mathbf{U}_{i,(\mathbf{x})}\phi_{(\mathbf{x}+\hat{i})}
-\phi_{(\mathbf{x})}\right),\nonumber\\
\mathbf{D}^-_i\phi_{(\mathbf{x})}&=&\delta x^{-1}\left(
\phi_{(\mathbf{x})}-
\mathbf{U}^*_{i,(\mathbf{x}-\hat{i})}\phi_{(\mathbf{x}-\hat{i})}\right)
.\label{Aa0}
\end{eqnarray}
We make use of the Hamiltonian formalism to write the equations of
motion (\ref{a7}), (\ref{a8}) as first order time-differential
equations. $\pi_{(t,\mathbf{x})}$ is the conjugate momentum of the
scalar field $\phi_{(t,\mathbf{x})}$ and the electric field
${\mathbf E}_{i,(t,{\mathbf x})}$ is the conjugate momentum of the
gauge field ${\mathbf A}_{i,(t,{\mathbf x})}$. In the temporal
gauge ${\mathbf A}_{0,(t,{\mathbf x})}=0$, the equations of motion
in the fully two-dimensional model are
\begin{eqnarray}
\Delta_t {\mathbf A}_{i,(t,{\mathbf x})}&=&- {\mathbf
E}_{i,(t-\delta t,{\mathbf x})}, \nonumber\\
\Delta_t\phi_{(t,\mathbf{x})}&=&\pi_{(t-\delta t,\mathbf{x})},
\nonumber\\ \Delta_t {\mathbf E}_{i,(t,{\mathbf x})}&=& \sum_{jkl}
\epsilon_{ij}\epsilon_{kl} \Delta^-_j\Delta^+_k{\mathbf
A}_{l,(t,{\mathbf x})} -\sigma {\mathbf E}_{i,(t,{\mathbf
x})}\nonumber\\&& - 2 e
 {\rm Im}\phi^*_{(t,\mathbf{x})}\mathbf{D}^+_i\phi_{(t,\mathbf{x})},
\nonumber\\ \Delta_t\pi_{(t,\mathbf{x})}&=& \sum_i
\mathbf{D}_i^-\mathbf{D}_i^+\phi_{(t,\mathbf{x})}
-\sigma\pi_{(t,\mathbf{x})} \nonumber\\&&
-m_{H}^{2}\phi_{(t,\mathbf{x})}+\frac{3}{2}\:\kappa|\phi_{(t,\mathbf{x})}|\phi_{(t,\mathbf{x})}
-2\lambda|\:\phi_{(t,\mathbf{x})}|^2\phi_{(t,\mathbf{x})},
\label{Aaaaa1}
\end{eqnarray}
where $\epsilon_{ij}$ is the Levi-Civita tensor in 2D and the
indices run from $1$ to $2$ in the summations.\\ \indent
Analogously in 3D,
\begin{eqnarray}
\Delta_t {\mathbf A}_{i,(t,{\mathbf x})}&=&- {\mathbf
E}_{i,(t-\delta t,{\mathbf x})}, \nonumber\\
\Delta_t\phi_{(t,\mathbf{x})}&=&\pi_{(t-\delta t,\mathbf{x})},
\nonumber\\ \Delta_t {\mathbf E}_{i,(t,{\mathbf x})}&=&
\sum_{jklm} \epsilon_{ijk}\epsilon_{klm}
\Delta^-_j\Delta^+_l{\mathbf A}_{m,(t,{\mathbf x})} -\sigma
{\mathbf E}_{i,(t,{\mathbf x})}\nonumber\\&& - 2 e
 {\rm Im}\phi^*_{(t,\mathbf{x})}\mathbf{D}^+_i\phi_{(t,\mathbf{x})},
\nonumber\\ \Delta_t\pi_{(t,\mathbf{x})}&=& \sum_i
\mathbf{D}_i^-\mathbf{D}_i^+\phi_{(t,\mathbf{x})}
-\sigma\pi_{(t,\mathbf{x})} \nonumber\\&&
-m_{H}^{2}\phi_{(t,\mathbf{x})}+\frac{3}{2}\:\kappa|\phi_{(t,\mathbf{x})}|\phi_{(t,\mathbf{x})}
-2\lambda|\:\phi_{(t,\mathbf{x})}|^2\phi_{(t,\mathbf{x})},
\label{Aaaa1}
\end{eqnarray}
where $\epsilon_{ijk}$ is the Levi-Civita tensor in 3D and the
indices run from $1$ to $3$ in the summations.\\ \indent In the
superconducting model, the boundary conditions in (\ref{sccon})
are implemented modifying the equations for
$\Delta_t\pi_{(t,\mathbf{x})}$ and $\Delta_t {\mathbf
E}_{y,(t,{\mathbf x})}$ at the surface of the film. In the
equation for $\Delta_t\pi_{(t,\mathbf{x})}$ the index $i$ in the
summation only takes values $1$ and $3$ for $y\leq-\mu/2$ or
$y\geq\mu/2$. Likewise, the term $-2 e{\rm
Im}\phi^*_{(t,\mathbf{x})}\mathbf{D}^+_y\phi_{(t,\mathbf{x})}$ is
excluded from the equation of $\Delta_t {\mathbf E}_{y,(t,{\mathbf
x})}$ for $y\geq\mu/2$ or $y<-\mu/2$. The reason for these
modifications is that the covariant derivatives as defined in the
lattice, $\mathbf{D}^+_y\phi_{(\mathbf{x})}$ and
$\mathbf{D}^-_y\phi_{(\mathbf{x})}$, are ill-defined at
$y=\mu/2,-\mu/2-1$ and $y=-\mu/2,\mu/2+1$ respectively. They must
be discarded explicitly at those points because their vanishing is
not possible in the superconducting phase according to formula
(\ref{Aa0}).\\ \indent Once the above modifications have been
implemented and the conditions
\begin{eqnarray}
\phi_{(t=0,\mathbf{x})}& = & 0\qquad \textrm{for $y>\mu/2$ or
$y<-\mu/2$},\nonumber\\ \sigma & = & 0 \qquad \textrm{for
$y>\mu/2$ or $y<-\mu/2$ $\forall$ $t$}\nonumber
\end{eqnarray}
set up, the equations of motion in (\ref{Aaaa1}) take care of the
conditions in (\ref{sccon}).\\ \indent The lattice version of the
Hamiltonian is, in 2D,
\begin{eqnarray}
H_{tot}^{2D}&=&\frac{1}{2}\sum_{{\mathbf x},i}\delta x^{2}\left[
{\mathbf E}_{i}^2 +\sum_{j}\left(\epsilon_{ij}\Delta_i^+{\mathbf
A}_j\right)^2 \right]\nonumber\\ &&+\sum_{\mathbf x}\delta
x^{2}\Bigl[ \pi^*\pi -\frac{2}{\delta x^2}\sum_i[{\rm
Re}\phi_{(\mathbf x)}^* \mathbf{U}_{i,(\mathbf x)}\phi_{(\mathbf
x+\hat{i})}-|\phi_{(\mathbf x)}|^2]\nonumber\\ &&~~~~~~~~~~~ +
m^{2}_{H}|\phi_{(\mathbf x)}|^2-\kappa|\phi_{(\mathbf
x)}|^3+\lambda|\phi_{(\mathbf x)}|^4 \Bigr],\quad i,j=1,2.
\label{Aa2}
\end{eqnarray}
In three dimensions,
\begin{eqnarray}
H_{tot}^{3D}&=&\frac{1}{2}\sum_{{\mathbf x},i}\delta x^{3}\left[
{\mathbf E}_{i}^2 +\sum_{jk}\left(\epsilon_{ijk}\Delta_j^+{\mathbf
A}_k\right)^2 \right]\nonumber\\ &&+\sum_{\mathbf x}\delta
x^{3}\Bigl[ \pi^*\pi -\frac{2}{\delta x^2}\sum_i[{\rm
Re}\phi_{(\mathbf x)}^* \mathbf{U}_{i,(\mathbf x)}\phi_{(\mathbf
x+\hat{i})}-|\phi_{(\mathbf x)}|^2]\nonumber\\ &&~~~~~~~~~~~ +
m^{2}_{H}|\phi_{(\mathbf x)}|^2-\kappa|\phi_{(\mathbf
x)}|^3+\lambda|\phi_{(\mathbf x)}|^4 \Bigr],\quad i,j,k=1,2,3.
\label{Aa2}
\end{eqnarray}
 \indent When measuring the energy associated to a superconductor film, the
index $i$ in the last summation does not take the value $2$
associated to the $y$ direction orthogonal to the film at
$y=\mu/2$. It does however at $y=-\mu/2$.
\newpage
\begin{landscape}
\section{Table of parameter values in the numerical simulations\label{appendC}}
\begin{table}
\center \small \caption{Figure number, coupling constants, lattice
parameters and other setup values.}\label{tabfig4}
\begin{tabular}{lllllllllll}
\br fig.no. & $e$ &  $m_{H}$ & $\kappa$ &  $\lambda$
 & $l_{x}\times l_{z}\times l_{y}$ & $\delta x$ & $\delta t$ &  $\sigma$ & $B_{\rm ini}$ & other parameters\\
\hline \ref{fig4} & 0.2 & 0.656 & 0.295 & 0.043 & $384\times 384$ &
1.0 & 0.01 & 1.0 & 0.0 & \\
\hline \ref{fig6} & 0.2 & 0.656 & 0.295 & 0.043 & $384\times 384$ &
1.0 & 0.01 & 2.6 & $6\cdot10^{-3}/e$ & \\
\hline \ref{fig17} & 0.34 & 0.656 & 0.295 & 0.043 & $384\times
384\times 384(pbc)$ &
1.0 & 0.01 & 0.25 & $0.012/e$ & $d=143$\\
\hline \ref{magno} & 1.0 & 0.656 & 0.295 & 0.043 & $900\times 900$ &
1.0 & 0.01 & 0.0 & $7.76\cdot10^{-3}/e$ &  $d=300$ $R_{N}=14$\\
\hline \ref{fig19} & 0.1 & 0.656 & 0.295 & 0.043 & $160\times 2\pi
R_{a,b,c}$ &
1.0 & 0.01 & 1.0 & $\frac{3.0/e}{160\times R_{a,b,c}}$ &  $|b-a|=30$\\
\hline \ref{figor} & 0.1 & 0.656 & 0.295 & 0.043 &
$200\times50(pbc)\times100(pbc)$ &
1.0 & 0.01 & 0.5 & $0.0283/e$ &  $|b-a|=40$, $\mu=3$\\
\hline \ref{fig22} & 0.1 & 0.656 & 0.295 & 0.043 & $384\times384$ &
1.0 & 0.01 & 0.5 & $0.011/e$ &  $|b-a|=30$\\
\hline \ref{Split} & 0.2 & 0.656 & 0.295 & 0.043 &
$156\times80\times50(pbc)$ &
1.0 & 0.01 & 0.5 & $0.1885/e$ &  $d_{s}=10$\\
\hline \ref{N1} & 0.2 & 0.656 & 0.295 & 0.043 &
$156\times80\times50(pbc)$ &
1.0 & 0.01 & 0.5 & variable &  $d_{s},\mu$ variable\\
\hline \ref{METAST} & 0.2 & 0.656 & 0.295 & 0.043 &
$160\times50(pbc)\times50(pbc)$ &
1.0 & 0.01 & 0.5 & $5.03\cdot10^{-2}/e$ &  $|b-a|=30$\\
\hline \ref{fig11} & 0.1 & 0.2 & 0.0151 & 0.0014 & $300\times
60(pbc)$ &
1.0 & 0.01 & 1.2 & 0.0 & $|b-a|=30$\\
\hline \ref{fig15} & 0.2 & 0.656 & 0.295 & 0.043 & $384\times 384$ &
1.0 & 0.01 & 1.8 & 0.0 & $|b-a|=30$\\
\hline \ref{fig20} & 0.1 & 0.656 & 0.295 & 0.043 & $160\times
51(pbc)$ &
1.0 & 0.01 & 1.0 & $0.0123/e$ & $|b-a|=30$\\
\br
\end{tabular}
\end{table}
,where $l_{x}\times l_{z}\times l_{y}$ is the lattice size in
lattice spacing units; $(pbc)$ stands for 'periodic boundary
conditions'; $\delta x$ is the lattice spacing; $\delta t$ is the
time step; $\sigma$ is the damping rate; $B_{\rm ini}$ is the
initial magnetic flux density; $\mu$ is the film thickness in the
superconducting model; $d$ stands for bubble centers separation;
$d_{s}$ is the initial square side; $R_{N}$ is the initial bubble
radius; $|b-a|$ is the gap size between blocks or bubbles. The value
for $\sigma$ in fig.no. \ref{magno} is $0.0$ in the equation of the
gauge field and is  variable in the equation for $\phi$ as it
determines the bubble expansion rate.
\end{landscape}

\end{document}